\documentclass[longauth]{aa}  
\usepackage{graphicx}
\usepackage{float}
\usepackage{natbib}
\usepackage{txfonts}

\begin{document}

   \title{The Solar Orbiter mission}
   \subtitle{Science overview}
   
   \author{D.~M\"uller \inst{\ref{inst:estec}}, O.C. St.~Cyr \inst{\ref{inst:gsfc}}, I. Zouganelis \inst{\ref{inst:esac}}, H.R. Gilbert \inst{\ref{inst:gsfc}}, R. Marsden\inst{\ref{inst:estec}}, T. Nieves-Chinchilla\inst{\ref{inst:gsfc}}, E.~Antonucci\inst{\ref{inst:INAF-Torino}}, F.~Auch\`ere\inst{\ref{inst:ias}}, D.~Berghmans\inst{\ref{inst:rob}}, T.~Horbury\inst{\ref{inst:imp}}, R.A.~Howard\inst{\ref{inst:nrl}}, S.~Krucker\inst{\ref{inst:fhnw},\ref{inst:berk}}, M.~Maksimovic\inst{\ref{inst:obsm}}, C.J.~Owen\inst{\ref{inst:ucl}}, P.~Rochus\inst{\ref{inst:csl}}, J.~Rodriguez-Pacheco\inst{\ref{inst:alcala}}, M.~Romoli\inst{\ref{inst:unifi}}, S.K.~Solanki\inst{\ref{inst:mps},\ref{inst:KyungHee}}, R.~Bruno\inst{\ref{inst:INAF-Rome}}, M.Carlsson\inst{\ref{inst:Rosseland},\ref{inst:ITA}}, A.~Fludra\inst{\ref{inst:ral}}, L.~Harra\inst{\ref{inst:PMOD},\ref{inst:ETH}}, D.M.~Hassler\inst{\ref{inst:SwRI-Boulder}}, S.~Livi\inst{\ref{inst:SwRI-SanAntonio}}, P.~Louarn\inst{\ref{inst:IRAP}}, H.~Peter\inst{\ref{inst:mps}}, U.~Sch\"uhle\inst{\ref{inst:mps}}, L.~Teriaca\inst{\ref{inst:mps}}, J.C.~del Toro Iniesta\inst{\ref{inst:iaa}}, R.F.~Wimmer-Schweingruber\inst{\ref{inst:cau}}, E.~Marsch\inst{\ref{inst:cau}}, M. ~Velli\inst{\ref{inst:ucla}}, A.~De Groof\inst{\ref{inst:esac}}, A.~Walsh\inst{\ref{inst:esac}}, D.~Williams\inst{\ref{inst:esac}}
}
        
\authorrunning{M\"uller et al.}

\institute{
European Space Agency, ESTEC, P.O. Box 299, 2200 AG Noordwijk, The Netherlands\label{inst:estec}   
\and
NASA Goddard Space Flight Center, Greenbelt, MD, USA\label{inst:gsfc}
\and
European Space Agency, ESAC, Camino Bajo del Castillo s/n, Urb. Villafranca del Castillo, 28692 Villanueva de la Ca\~nada, Madrid, Spain\label{inst:esac} 
\and
INAF -- Astrophysical Observatory of Torino, Italy\label{inst:INAF-Torino}
\and
Institut d'Astrophysique Spatiale, CNRS, Univ. Paris-Sud, Universit\'e Paris-Saclay, B\^at. 121, 91405 Orsay, France\label{inst:ias}
\and
Royal Observatory of Belgium, Ringlaan -3- Av. Circulaire, 1180 Brussels, Belgium\label{inst:rob}
\and
Space and Atmospheric Physics, The Blackett Laboratory, Imperial College London, London, SW7 2AZ, UK\label{inst:imp}
\and
Naval Research Laboratory, Washington, DC, 20375, USA\label{inst:nrl}
\and
University of Applied Sciences Northwestern Switzerland\label{inst:fhnw}
\and
Space Sciences Laboratory, University of California, Berkeley, USA\label{inst:berk}
\and
LESIA, Observatoire de Paris, Universit\'e PSL, CNRS, Sorbonne Universit\'e, Univ. Paris Diderot, Sorbonne Paris Cit\'e, 5 place Jules Janssen, 92195 Meudon, France\label{inst:obsm}
\and
Mullard Space Science Laboratory, University College London, Holmbury St. Mary, Dorking, Surrey, RH5 6NT, UK\label{inst:ucl}
\and
Centre Spatial de Li\`ege, Universit\'e de Li\`ege, Av. du Pr\'e-Aily B29, 4031 Angleur, Belgium\label{inst:csl}
\and
Universidad de Alcal\'a, Space Research Group, 28805 Alcal\'a de Henares, Spain\label{inst:alcala}
\and
Dip. di Fisica e Astronomia, Universit\a di Firenze, Largo E.\ Fermi 2, 50125 Firenze, Italy, and INAF associated scientist\label{inst:unifi}
\and
Max-Planck-Institut f\"ur Sonnensystemforschung (MPS), Justus-von-Liebig-Weg 3, 37077 G\"ottingen, Germany\label{inst:mps}
\and
School of Space Research, Kyung Hee University, Yongin, Gyeonggi-Do, 446-701, Republic of Korea\label{inst:KyungHee}
\and
INAF -- Istituto di Astrofisica e Planetologia Spaziali, Rome, Italy\label{inst:INAF-Rome}
\and
Rosseland Centre for Solar Physics, University of Oslo, P.O. Box 1029 Blindern, NO-0315 Oslo, Norway\label{inst:Rosseland}
\and
Institute of Theoretical Astrophysics, University of Oslo, P.O. Box 1029 Blindern, NO-0315 Oslo, Norway\label{inst:ITA}
\and
RAL Space, STFC Rutherford Appleton Laboratory, Harwell, Didcot, OX11 0QX, UK\label{inst:ral}
\and
Physikalisch-Meteorologisches Observatorium Davos, World Radiation Center, 7260, Davos Dorf, Switzerland\label{inst:PMOD}
\and
ETH Z\"urich, H\"onggerberg Campus, Z\"urich, Switzerland\label{inst:ETH}
\and
Space Science \& Engineering Division, Southwest Research Institute, 1050 Walnut St., Suite 300, Boulder, CO  80302\label{inst:SwRI-Boulder}
\and
Southwest Research Institute, 6220 Culebra Road, San Antonio, TX 78238, USA\label{inst:SwRI-SanAntonio}
\and
Institut de Recherche en Astrophysique et Plan\'etologie, 9, avenue du Colonel Roche, B.P. 4346, 31028 Toulouse Cedex 4, France\label{inst:IRAP}
\and
Instituto de Astrof\'isica de Andaluc\'ia (IAA-CSIC), Apartado de Correos 3004, E-18080 Granada, Spain\label{inst:iaa}
\and
Division for Extraterrestrial Physics, Institute for Experimental and Applied Physics (IEAP), Christian Albrechts University at
Kiel, Leibnizstr. 11, 24118 Kiel, Germany\label{inst:cau}
\and
University of California Los Angeles, Los Angeles, CA 90095, USA\label{inst:ucla}
}

   \date{Received 22 May 2020 / Accepted 24 July 2020}

 
 %
 
  \abstract
{}
   {Solar Orbiter, the first mission of ESA's Cosmic Vision 2015--2025 programme and a mission of international collaboration between ESA and NASA, will explore the Sun and heliosphere from close up and out of the ecliptic plane. It was launched on 10 February 2020 04:03 UTC from Cape Canaveral and aims to address key questions of solar and heliospheric physics pertaining to how the Sun creates and controls the Heliosphere, and why solar activity changes with time. To answer these, the mission carries six remote-sensing instruments to observe the Sun and the solar corona, and four in-situ instruments to measure the solar wind, energetic particles, and electromagnetic fields.  In this paper, we describe the science objectives of the mission, and how these will be addressed by the joint observations of the instruments onboard.}
%
   {The paper first summarises the mission-level science objectives, followed by an overview of the spacecraft and payload. We report the observables and performance figures of each instrument, as well as the trajectory design. This is followed by a summary of the science operations concept. The paper concludes with a more detailed description of the science objectives.}
   {Solar Orbiter will combine in-situ measurements in the heliosphere with high-resolution remote-sensing observations of the Sun to address fundamental questions of solar and heliospheric physics. The performance of the Solar Orbiter payload meets the requirements derived from the mission's science objectives. Its science return will be augmented further by coordinated observations with other space missions and ground-based observatories.}
   {}


 \maketitle
%

\section{Introduction}
\label{sect-intro}
The extended atmosphere of the Sun, known as the heliosphere, is a domain of space where fundamental physical processes common to solar, astrophysical, and laboratory plasmas can be studied -- under conditions that are impossible to reproduce on Earth, and unfeasible to observe on stars from astronomical distances. 
Fundamentally, understanding the interrelation between the Sun and the heliosphere is key to understanding how our Solar System works.  

The Solar Orbiter mission, launched from Cape Canaveral on 10 February 2020 04:03\, UTC aboard a NASA-provided Atlas V 411 launch vehicle, has six remote-sensing instruments observing the Sun, the solar corona, and the inner heliosphere, and four in-situ instruments measuring the solar wind and the physical parameters of the heliosphere. 
Solar Orbiter is an ESA-led mission with strong NASA participation. The payload consists of ten experiments, some of which have several sensors for in-situ measurements or multiple telescope channels for remote imaging of the Sun and its atmosphere. Eight of the ten instruments have been provided by national agencies of the ESA member states, with Principal Investigators from the leading country. Another instrument, the SPICE imaging spectrograph, has been provided as a European facility experiment by a consortium supported by national funding agencies and ESA. As another European facility experiment, ESA has also funded the Suprathermal Ion Spectrograph (SIS), a sensor of the Energetic Particle Detector (EPD) suite. The Heliospheric Imager (SoloHI) and the Heavy Ion Sensor (HIS) of the Solar Wind Analyser suite (SWA) have been provided as NASA's contribution to the scientific payload, in addition to the launch vehicle.

Together, the ten Solar Orbiter instruments will enable correlative studies to provide a complete description of the plasma making up the solar wind -- its origin, transport processes, and elemental composition. Solar Orbiter data are expected to vastly improve on those of the Helios spacecraft \citep{Porsche:1977aa,Schwenn:1990aa,Schwenn:1991ab}, launched in 1974 and 1976, and the Ulysses mission \citep{Wenzel:1992aa}. 
Solar Orbiter's six remote-sensing instruments will provide novel imaging and spectroscopic observations of the Sun, its corona, and the inner heliosphere, and address key problems in heliophysics by measurement of the Sun's magnetic field, in particular at the solar poles, and of the plasma state of the corona. 
Solar Orbiter's out-of-ecliptic observations of the Sun from short distance will be unique, and will be further augmented by measurements of NASA's Parker Solar Probe mission \citep{2016SSRv..204....7F} at even lower solar distances.

In this paper, we present an updated mission science overview, expanding the earlier paper of \cite{Mueller:2013a}. Portions of the text have been reproduced with permission from \cite{Mueller:2013a} copyright by Springer.

\paragraph{Mission selection}
As described in the Solar Orbiter Assessment and Definition Study Reports \citep{Marsden:2009aa_TR,Marsden:2011aa_TR}, the Solar Orbiter mission has its origins in a proposal called `Messenger' that was submitted by Richter et al.\ in 1982 in response to an ESA call for mission ideas. At the meeting `Crossroads for European Solar and Heliospheric Physics' held on Tenerife in March 1998, the heliophysics community recommended to `launch an ESA Solar Orbiter as ESA's [next  flexible] mission, with possible international participation, [for launch] around 2007'. The kick-off meeting for a pre-assessment study of the `ESA Solar Orbiter' concept was held in March 1999.
Solar Orbiter was subsequently proposed in 2000 by E.\ Marsch et al., and was selected by ESA's Science Programme Committee (SPC) in October 2000 as a `flexible' mission for launch after ESA's BepiColombo mission to Mercury \citep[][launched in October 2018]{2010P&SS...58....2B}. 

Following a number of internal and industrial studies, SPC instructed the ESA Executive in 2007 to find ways to implement Solar Orbiter within a confined financial envelope. In response to this request, a Joint Science and Technology Definition Team (JSTDT), comprising scientists and engineers appointed by ESA and NASA, studied the benefits to be gained by combining ESA's Solar Orbiter mission and NASA's Solar Sentinels into a joint programme.
This led to the release of an ESA Announcement of Opportunity (AO) for the Solar Orbiter payload in September 2007 and a NASA Small Explorer Focused Opportunity for Solar Orbiter (SMEX/FOSO) AO in October 2007. Initially NASA selected two instruments and two sensors as parts of suites for Phase A study: SPICE (Phase-A PI: D.~Hassler), SoloHI (Phase-A PI: R.~Howard), SWA/HIS (Phase-A PI: S.~Livi), and EPD/SIS (Phase-A PI: G.~Mason).
 
 In early 2009, a joint ESA-NASA panel confirmed the validity of the previously selected payload in light of the major changes induced by the start of ESA's Cosmic Vision planning cycle, NASA's prioritisation of Solar Probe in its Living With a Star programme, and the strong science synergies between Solar Orbiter and Solar Probe. As a result, the instruments' selection, as originally recommended by the Payload Review Committee in 2008, was formally announced in March 2009, in coordination with NASA.

At its meeting in February 2010, ESA's SPC recommended that Solar Orbiter be one of the three M-class candidates to proceed into definition phase and made a further programmatic change by endorsing a `fast track' approach. In line with this approach, major industrial studies were kicked-off in February 2011, and the System Requirements Review was completed in mid-2011.
In March 2011, NASA announced a reduction of its contribution to the payload to one full instrument and one sensor, in addition to their provision of the launch vehicle. Given the scientific importance of the descoped investigations, SPICE and SIS, ESA's SPC decided that, should Solar Orbiter be selected, the SPICE and SIS measurement capabilities should be recovered through the inclusion of European-led instruments in the payload, procured under ESA's responsibility. Solar Orbiter was ultimately selected and adopted as the first M-class mission of ESA's Cosmic Vision programme by the SPC on 4 October 2011. The prime contract between ESA and Astrium UK, now Airbus Defence and Space, was signed in April 2012, and in March 2014, NASA selected United Launch Alliance (ULA) to launch Solar Orbiter onboard an Atlas V 411 rocket from Cape Canaveral.

\paragraph{Mission objectives}
The Solar Orbiter mission has been designed to address the overarching science objective `How does the Sun create and control the Heliosphere -- and why does solar activity change with time?' Responding to this objective will be made possible by the mission's unique combination of short distance to the Sun (minimum perihelion of 0.28\,AU), out-of-ecliptic vantage points (reaching $18^\circ$ heliographic latitude during its nominal mission phase, NMP, and above  $30^\circ$ during the extended mission phase, EMP), and its comprehensive suite of combined in-situ and remote-sensing instruments.

Since the launch of the Helios missions, the results of solar and heliospheric missions such as Voyager \citep{Stone:1977aa}, Solar Maximum Mission \citep{1980SoPh...65....5B}, Ulysses, Yohkoh \citep{Acton:1992aa}, SOHO \citep{Domingo:1995aa}, WIND \citep{1995SSRv...71...55O, 1997AdSpR..20..559O}, ACE \citep{1998SSRv...86....1S}, TRACE \citep{Handy:1999aa}, RHESSI \citep{Lin:2002aa}, Hinode \citep{Kosugi:2007aa}, STEREO \citep{Kaiser:2008aa}, SDO \citep{Pesnell:2012aa}, IRIS \citep{De-Pontieu:2014lr}, and, most recently, Parker Solar Probe \citep{2016SSRv..204....7F}, have formed the foundation of our understanding of the solar corona, the solar wind, and the heliosphere and greatly advanced our knowledge of solar physics. 

Nevertheless, two areas have not  yet been fully explored: (a) the inner heliosphere, where the nascent solar wind evolves and heliospheric structures are formed, and (b) the Sun's polar regions, which are key to understanding the solar dynamo processes that drive the Sun's activity cycle. 

Solar Orbiter will combine in-situ measurements with high-resolution remote-sensing observations of the Sun to resolve fundamental science problems. 
These problems include the question of the sources of the solar wind, the causes of eruptive releases of plasma and magnetic field from the Sun, often in the form of coronal mass ejections (CMEs), the evolution of CMEs and their interaction with the ambient solar wind flow, and the origins, acceleration mechanisms, and transport of solar energetic particles that may be hazardous to human explorers as well as robotic spacecraft that operate in the highly variable environment outside of Earth's magnetosphere.
The combination of in-situ and remote-sensing observations will, for example, enable us to trace solar wind structures back to their sources on the Sun (e.g. Helios~1 and 2  carried only a single, limited remote-sensing instrument, the Zodiacal Light Experiment; \cite{1975RF.....19..264L}).

Near perihelion, Solar Orbiter provides time series of ultraviolet (UV) and extreme ultraviolet (EUV) images of the Sun's corona at unprecedented spatial resolution, which allow us to study coronal heating processes at small scales and better understand the dynamic forcing of the heliosphere by the Sun's corona. So far, data of this kind have only been obtained for short time intervals during suborbital rocket flights \citep[e.g.\ by Hi-C,][]{2014SoPh..289.4393K,Rachmeler:2019ve}.
\cite{Williams_2020} analysed data from the Hi-C 2.1 rocket flight in 2018 and find that Hi-C 2.1 can resolve individual `strands' of coronal loops as small as $\sim200$\,km. This is significantly smaller than the spatial resolution of SDO/AIA \citep[down to $\approx 1.5$\,arcseconds, which corresponds to around 1000\,km;][]{Lemen:2012lr,Boerner:2012aa} and highlights the strong scientific rationale for high-resolution Extreme Ultraviolet Imager (EUI) observations by Solar Orbiter, which are expected to resolve spatial scales down to 100\,km at perihelion. 
High-resolution image sequences at high temporal cadence by the EUI High Resolution Imagers will also allow the imaging of transverse waves propagating in coronal loop systems. Based on the observation of higher wave harmonics, coronal seismology techniques will be used to infer key plasma properties.

Observations by the same telescopes might also provide evidence for nanoflares at different phases of the solar cycle and at different solar latitudes, and could shed light on the proposed  `magnetic field-line braiding' process that might play a key role in heating the Sun's corona \citep{Parker1988ApJ}.

Solar Orbiter is also providing the first-ever magnetograms from outside the Sun--Earth line, which will be key to improving global models of the Sun's extended magnetic field. In addition to other applications, these will help to advance models of particle transport in the inner heliosphere and to improve the foundations of space weather forecasting.

In the later phases of the mission, Solar Orbiter will reach significant heliographic latitudes. From this perspective, it will provide the first-ever images and detailed surface magnetic maps of the Sun's polar regions. This, in turn, will enable us to add new constraints on solar dynamo models, a key ingredient to better understand the Sun's 11-year activity cycle.
In addition, during this phase of the mission, the first ever in-situ observations in the inner heliosphere from outside the ecliptic will be made, providing new insights into the fast solar wind emanating from the polar coronal holes. In particular, the combination with the remote-sensing instruments will allow us to pinpoint the physical properties in the source region of the solar wind plasma detected \textit{in situ}, thus opening a new avenue to identifying the acceleration processes.   

Solar Orbiter's overarching scientific objective can be expanded into four interrelated top-level scientific questions \citep{Mueller:2013a}:

\begin{itemize}
\item What drives the solar wind and where does the coronal magnetic field originate?
\item How do solar transients drive heliospheric variability?
\item How do solar eruptions produce energetic particle radiation that fills the heliosphere?
\item How does the solar dynamo work and drive connections between the Sun and the heliosphere?
\end{itemize}

\noindent
Making in-situ measurements close to the Sun and observing the Sun remotely from outside the ecliptic plane are two fundamental drivers for the mission, which will approach the Sun to as close as 0.28\,AU and reach heliographic latitudes of up to 33$^\circ$.
The specific minimal distance is enabled by the usage of solar array technology from ESA's BepiColombo mission, and the maximum latitude is determined by fuel available for gravity assist manoeuvres (GAMs), as well as the fact that the latitude increase per GAM decreases in the extended mission phase.

The varying vantage point of Solar Orbiter will provide new perspectives for remote-sensing instruments. In addition to close-up imaging of the solar corona at unprecedented spatial resolution in UV and EUV wavelengths and obtaining the first images and surface magnetic field maps of the Sun's polar regions as mentioned above, this will also allow the imaging of CMEs from outside the ecliptic plane for the first time.

The following section provides an overview of the mission's science objectives followed by sections on the Solar Orbiter spacecraft, its instruments, the mission design and science operations. Table \ref{T-MissionSummary} gives a one-page mission summary.


\begin{table*}[ht!]
\caption{Solar Orbiter mission summary.}

\label{T-MissionSummary}
\begin{tabular}{lp{13cm}}
\hline              
\rule{0pt}{3ex} {\bf Top-level science questions} & 
\vspace{-6mm}
\begin{itemize}
\item What drives the solar wind and where does the coronal magnetic field originate?
\item How do solar transients drive heliospheric variability?
\item How do solar eruptions produce energetic particle radiation that fills the heliosphere?
\item How does the solar dynamo work and drive connections between the Sun and the heliosphere?
\end{itemize}
\vspace{-4mm}\\
\hline
\rule{0pt}{3ex} {\bf Science Payload} &
{\bf In-situ instruments:} 
\begin{itemize}
\item Energetic Particle Detector \citep[EPD,][]{Rodriguez2020a}
\item Magnetometer \citep[MAG,][]{Horbury2020}
\item Radio and Plasma Wave analyser  \citep[RPW,][]{Maksimovic2020a}
\item Solar Wind Analyser  \citep[SWA,][]{Owen2020a}
\end{itemize}
{\bf Remote-sensing instruments:}
\begin{itemize}
\item Extreme Ultraviolet Imager \citep[EUI,][]{Rochus2020a}
\item Visible light and UV Coronagraph \citep[Metis,][]{Antonucci2020a}
\item Polarimetric and Helioseismic Imager  \citep[SO/PHI,][]{Solanki2020a}
\item Heliospheric Imager  \citep[SoloHI,][]{Howard2020a}
\item EUV Imaging Spectrograph  \citep[SPICE,][]{SpiceConsortium2020}
\item X-ray Spectrometer/Telescope  \citep[STIX,][]{Krucker2020a}
\end{itemize}
\vspace{-4mm}\\
 \hline
\rule{0pt}{3ex} {\bf Mission Profile} & 
\vspace{-6mm}
\begin{itemize}
\item Launched on 10 February 2020 04:03 UTC, on NASA-provided Atlas V 411 
\item Interplanetary cruise with chemical propulsion and seven gravity assists at Venus and one at Earth to decrease perihelion distance
\item Venus resonance orbits with multiple GAMs  to increase inclination
\end{itemize}
\vspace{-4mm}\\
\hline
\rule{0pt}{3ex} {\bf Closest Perihelion} & 0.28\,AU (furthest aphelion: 1.02 \,AU)\\
\hline
\rule{0pt}{3ex} {\bf Orbital period} & $150-180$\,days\\
\hline
\rule{0pt}{3ex} {\bf Max. Heliolatitude} & 
\vspace{-6mm}
\begin{itemize}
\item 7$^\circ$ (cruise phase)
\item18$^\circ$ (nominal mission, reached first on 22 March 2025)
\item 24$^\circ$ (start of extended mission, reached first on 28 January 2027)
\item 33$^\circ$ (extended mission, reached first on 24 July 2029)
\end{itemize}
\vspace{-6mm}\\
\hline
\rule{0pt}{3ex} {\bf Spacecraft} &
Three-axis stabilised platform, heat shield, two adjustable, single-sided solar arrays of dimensions: 2.5\,m $\times$ 3.1\,m $\times$ 2.7\,m (launch configuration, i.e.\ with folded solar arrays), launch mass $\sim$1720\,kg, incl.\ 248.7\,kg fuel (149.0\,kg nitrogen tetroxide and 99.7\,kg monomethylhydrazine)
\\
 \hline
\rule{0pt}{3ex} {\bf Telemetry Band} & Dual X-band\\
 \hline
{\rule{0pt}{3ex} \bf Data Downlink} & 150 kbit/s at 1\,AU spacecraft--Earth distance (requirement; higher in-orbit performance)\\
 \hline
\rule{0pt}{3ex} {\bf Nominal Mission Duration} & 7 years (incl.\ cruise phase, nominal mission defined to start with the Earth GAM on 26 November 2021 and to end with the fifth Venus GAM on 24 December 2026) \\
 \hline
\rule{0pt}{3ex}  {\bf Extended Mission Duration} & 3 years\\
 \hline
\end{tabular}
\end{table*}


\section{Science objectives}
\label{sect-science}
The Sun is surrounded by a million-degree solar corona, and the physical mechanisms that heat the corona to these temperatures are still not fully understood today, even after many decades of observations and theoretical work. 
However, what is known is that coronal plasma continuously expands outwards and develops into the supersonic solar wind that creates the heliosphere \citep{Parker:1958aa}. The solar wind streams radially outward, evolving and interacting with itself, Earth, and other planets, and extends all the way to the boundary of the heliosphere, as measured by the Voyager spacecraft \citep{Stone:1977aa}. Thus, the solar wind fundamentally affects the Solar System planets as well as their environments, including, for example Earth's magnetosphere.

\begin{figure*}
\resizebox{\hsize}{!}{
 \includegraphics{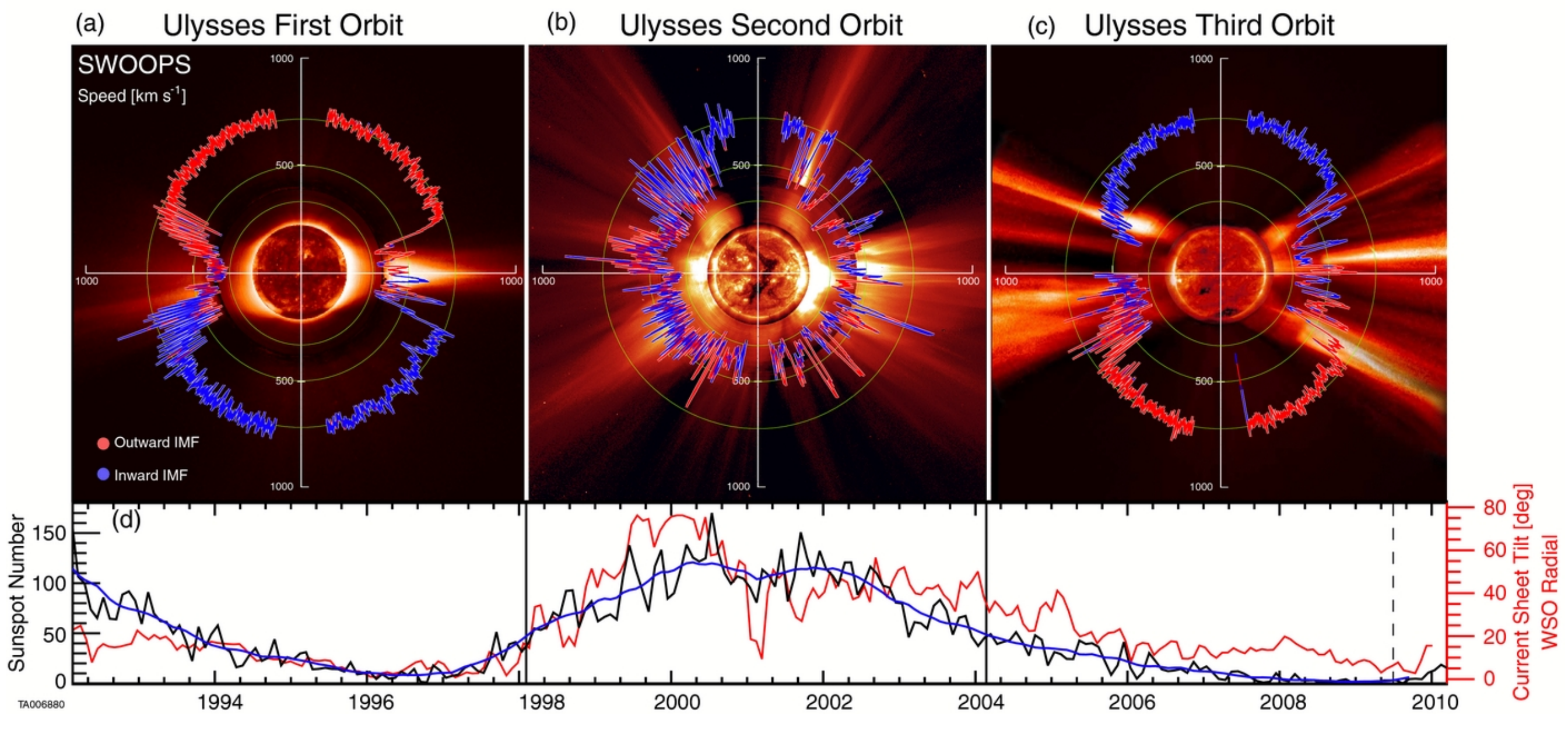}  
}
\caption{Graphic from \cite{2013ApJ...779....2M} showing polar plots of the solar wind speed as measured by the Ulysses mission. The interplanetary magnetic field
is color coded: outward (red) and inward (blue). For each polar plot, time progresses counterclockwise from the nine o'clock position. Matching time ranges in the bottom panel (d) are indicated by the vertical lines; this panel provides the sunspot number (black, see \cite{2016SoPh..291.2629C} for the recalibrated sunspot number), smoothed sunspot number (blue), and Wilcox Solar Observatory (WSO) calculated heliospheric current sheet tilt angle (red). As visual indicators of the general solar and coronal structure for these three orbits, the figure displays images taken on 1996 August 17, 2000 December 7, and 2006 March 28, respectively, combining (from the inside out) images from SOHO's Extreme ultraviolet Imaging Telescope (Fe XII at 19.5\,nm), the Mauna Loa K coronameter (700--950 nm), and SOHO's LASCO C2 white light coronagraph. (Reproduced with permission from the AAS.)}
              
             \label{F-McComas+al2013}
   \end{figure*}

The heliosphere is permeated by two classes of solar wind, `fast' and `slow', whose frequency distribution is modulated by the 11-year solar cycle (Fig.~\ref{F-McComas+al2013}). The fast solar wind ($\sim800$\,km/s) is comparatively steady and primarily emerges from polar coronal hole regions. However, it does also emanate from equatorial coronal holes; from a space-weather perspective, these are the most important ones.

The slow solar wind ($\sim300 - 500$\,km/s) is present in the ecliptic plane throughout the solar cycle, and so both types play a role in shaping the near-Earth space environment. The mass flux and composition of the slow solar wind are different from those of the fast wind, a result indicating that the plasma sources could be confined to small-scale coronal loops that open only intermittently.
In-situ plasma and magnetic field measurements from Parker Solar Probe during its first perihelion (36 to 54\,$R_\sun$) have provided evidence for a small low-latitude coronal hole being a source of slow solar wind \citep{BaleNature2019}. The same study also suggests that there is an impulsive mechanism associated with solar-wind energisation, and that micro-instabilities also contribute to the solar wind heating. 

In addition to the above, the heliosphere is also frequently experiencing transient events, such as CMEs, on a wide range of scales.
In the following sections, we discuss the four overarching science questions of Solar Orbiter in detail and specify how its data will contribute to answering them.

\subsection{What drives the solar wind and where does the coronal magnetic field originate?}
\label{S-Goal1}

Based on comet tail observations, \cite{Biermann1951b} deduced that the Sun continuously emits a flow of charged particles. A few years later, \cite{Parker:1958aa} realised that the latter is closely linked to the presence of a hot corona surrounding the Sun. While the detailed physical mechanisms that heat the Sun's corona to millions of degrees are still not fully understood, most coronal heating mechanisms proposed attribute key roles to the Sun's convective motions in the photosphere and its magnetic field, which can release energy via magnetic reconnection and/or guide waves that might dissipate energy in the corona \citep[for reviews, see][] {Klimchuk:2006aa,Reale:2010LRSP,2019ARA&A..57..157C}.

The hot plasma of the solar corona travels outward into interplanetary space to form the solar wind, thereby blowing a cavity in the interstellar medium that is known as the heliosphere. During solar minimum, large-scale regions of a single magnetic polarity in the Sun's atmosphere  --- polar coronal holes  --- open into space and are the source of high-speed solar wind flows that are comparatively steady (Fig.~\ref{F-McComas+al2013}). The origin of the slow wind, on the other hand, is not yet fully determined. It is thought to predominantly originate from magnetically complex regions at low latitudes near coronal hole boundaries and is highly variable in speed, composition, and charge state. 

Near maximum solar activity, this stable bimodal configuration transforms into a complex interplay of slow and fast solar wind streams. The fast wind from the polar coronal holes carries magnetic flux of opposite polarity into the heliosphere, separated by the heliospheric current sheet (HCS) embedded in the slow wind. {\it Ulysses} and {\it Wind} measurements of this boundary have shown that it is not symmetric around the Sun's equator \citep{Smith:2000aa}, at least during the period observed. \cite{Wang:2011aa} argue that this is due to Joy's law and the observed hemispheric sunspot number asymmetry in the sunspot numbers.\footnote{We note that in 2016 the sunspot number was recalibrated. Figure~\ref{F-HeliosphericFlux_2017} uses the new version, while Fig.~\ref{F-McComas+al2013} uses the old version. That is why the smoothed sunspot curve in Fig.~\ref{F-McComas+al2013} peaks mid-2000 below 125, while in Fig.~\ref{F-HeliosphericFlux_2017} it peaks above 170 mid-2000. The new version is available at {\tt http://sidc.be/silso/monthlyssnplot}, and the old version at {\tt http://sidc.be/silso/archivemonthlyssnplot}.}
In the following sections, we discuss three questions that flow down from the first top-level question.

\subsubsection{What are the source regions of the solar wind and the heliospheric magnetic field?}
\paragraph*{Current understanding.} 
The way the solar wind is structured at large scales and how the magnetic field in the heliosphere is connected to the global coronal magnetic field are reasonably well understood \citep[see e.g.][]{2007bsw..book.....M}. However, it becomes increasingly difficult to map the heliospheric magnetic field to the one in the lower corona, or even the photosphere, which is the only layer in which the Sun's magnetic field can be accurately measured to date.  The strong expansion of the Sun's magnetic field from the photosphere into the corona, in a domain that is governed by complex interactions between hydrodynamic and magnetic forces, combined with the lack of accurate magnetic field diagnostics in the optically thin solar corona, make it hard to extrapolate the magnetic field beyond just a few solar radii outwards. At the same time, plasma measured \textit{in situ} at larger distances from the Sun has evolved dynamically on its way out, which complicates matters further. 

\vspace{3mm}
\noindent
{\it(a) Source regions of the solar wind.} As a function of radial distance to the Sun, the solar wind speed has been shown to be anti-correlated with the modelled rate of magnetic field expansion \citep{Wang:2006aa,wang_small_2017}: The central areas of polar coronal holes exhibit the fastest solar wind streams, while progressively slower wind is observed close to coronal hole boundaries. Strong outflows inside coronal holes are correlated with the intense magnetic flux elements found at the intersections of supergranules \citep{Hassler1999Sci}; these expand into the corona as `funnels' \citep{Tu:2005aa}, preferentially from regions dominated by flux of the coronal hole polarity \citep{McIntosh:2006aa, panasenco_large-scale_2019}.
The solar wind speed in the corona has also been measured  directly with SOHO's UVCS instrument, both in the central areas of coronal holes and closer to the coronal hole boundary \citep[][]{2006SSRv..124...35A, 2012SSRv..172....5A,1999ApJ...511..481C,2000SoPh..197..115A, 2002ApJ...571.1008S,2005A&A...435..699A}.

\begin{figure}   
   \centerline{\includegraphics[width=\columnwidth,clip=true]{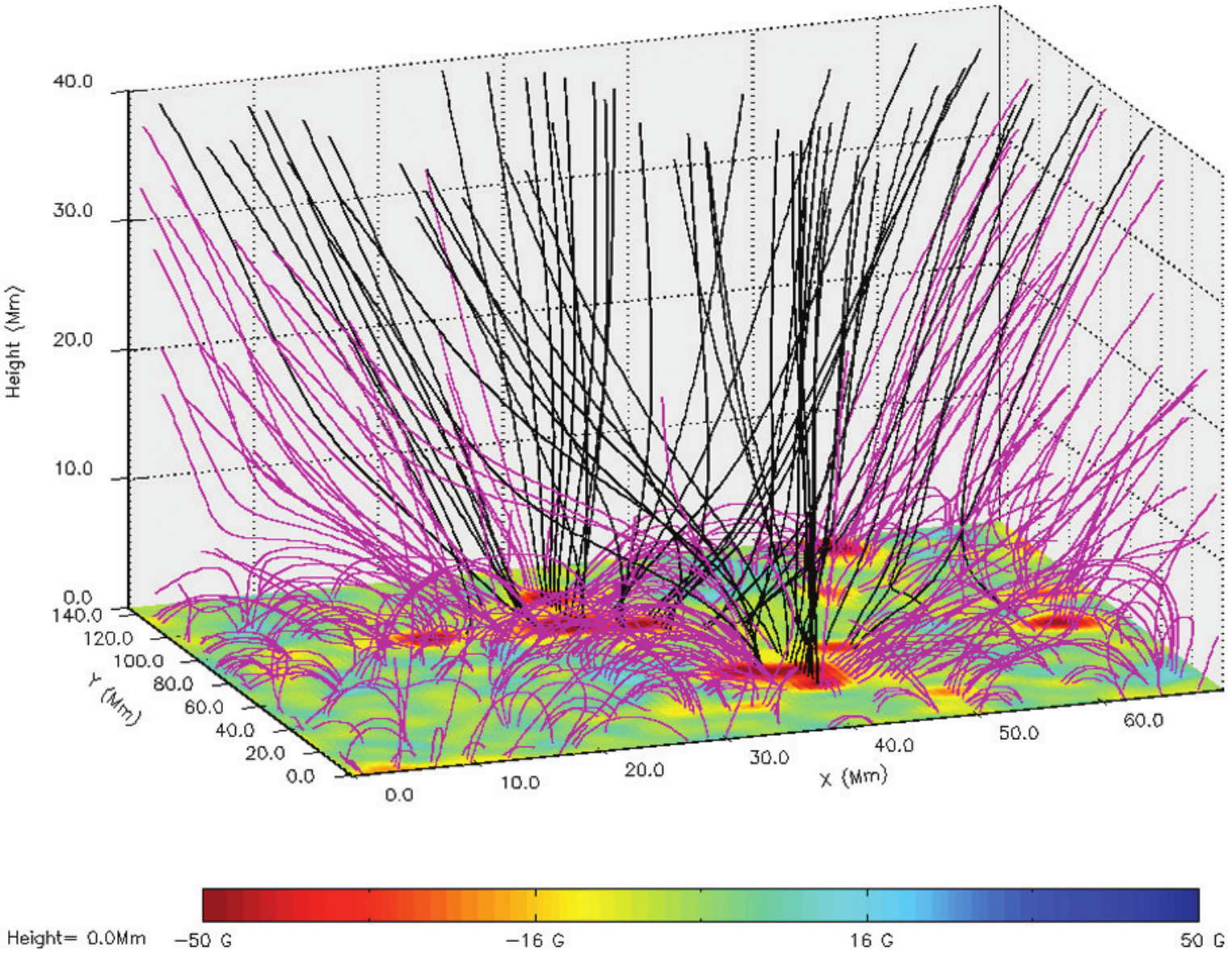}}
              \caption{Modelled magnetic field of the transition region and lower corona in a polar coronal hole based on measurements of the photospheric magnetic field. This figure illustrates the complex connections between the solar surface and the heliosphere: only the black field lines extend far from the surface. One of Solar Orbiter's key objectives is to determine the links between the observed solar wind streams and their source regions. \citep[From][]{Marsch:2006aa}}
     \label{F-Marsch+al2006}
   \end{figure}

The solar chromosphere and transition region, which are the atmospheric layers from which the solar wind originates \citep{Marsch:2006aa}, are highly structured by magnetic field and highly dynamic in space and time (Fig.~\ref{F-Marsch+al2006}). 
The solar chromosphere is permeated by spicules, cool and dense jets of chromospheric plasma (see e.g. \cite{pontieu_observations_2017}). In the past, spicules have been considered too slow and cold to contribute significantly to the solar wind, but a more dynamic type of spicule was discovered by Hinode \citep{2007PASJ...59S.655D,De-Pontieu:2007ab} that exhibits shorter lifetimes, higher velocities, and higher temperatures. Such spicules also support waves, which might transport sufficient energy to accelerate fast wind streams in coronal holes \citep{De-Pontieu:2009aa,De-Pontieu:2011aa,Mart_nez_Sykora_2018,chitta_hot_2019}.
Hinode has also observed the frequent occurrence of very small-scale X-ray and UV jets in polar coronal holes \citep{Cirtain:2007aa, 2015A&A...579A..96P}. These X-ray jets were first observed with the Yohkoh mission \citep{1992PASJ...44L.173S}.
 Their origin is thought to be magnetic reconnection of coronal field lines. At the reconnection site, Alfv\'en waves develop and produce outflow velocities up to $\sim800$\,km/s, while the energy released by the reconnection heats the plasma locally, generating mass motions with sonic speeds of $\sim200$\,km/s. Given the high velocities and frequency of these events, it has been suggested that they contribute to the fast solar wind. However, their relation to the photospheric magnetic field has not been established, as the high heliographic latitudes at which they occur make it difficult to distinguish
  their photospheric footpoints from the ecliptic plane (see e.g. \cite{tiwari_evidence_2018}). Other fine-scale ray-like structures, namely coronal plumes, permeate coronal holes and are correlated with small-scale bipolar structures inside the hole \citep[for a review, see][]{poletto_solar_2015}. Measurements in the UV show that these structures are cooler than the surrounding background, and have slower but denser outflows. Within the fast solar wind, in-situ measurements have revealed the existence of faster and slower microstreams \citep{Neugebauer:1995aa, 2012ApJ...750...50N}, as well as other fine-scale structures \citep{Thieme:1990aa}. However, these have not been unambiguously linked to coronal features. For more recent attempts with Parker Solar Probe data,
  see also \cite{BaleNature2019,badman_magnetic_2020, allen_solar_2020,panasenco_exploring_2020}. Observations with remote-sensing instruments aboard Solar Orbiter will allow us to understand the relation of these jets from a unique vantage point, providing comparable high-angular-resolution data simultaneously in the corona, chromosphere, and photosphere \citep[for an in-depth review of this topic, see ][]{2016SSRv..201....1R}.

\begin{figure}[ht]
\includegraphics[width=0.5\textwidth]{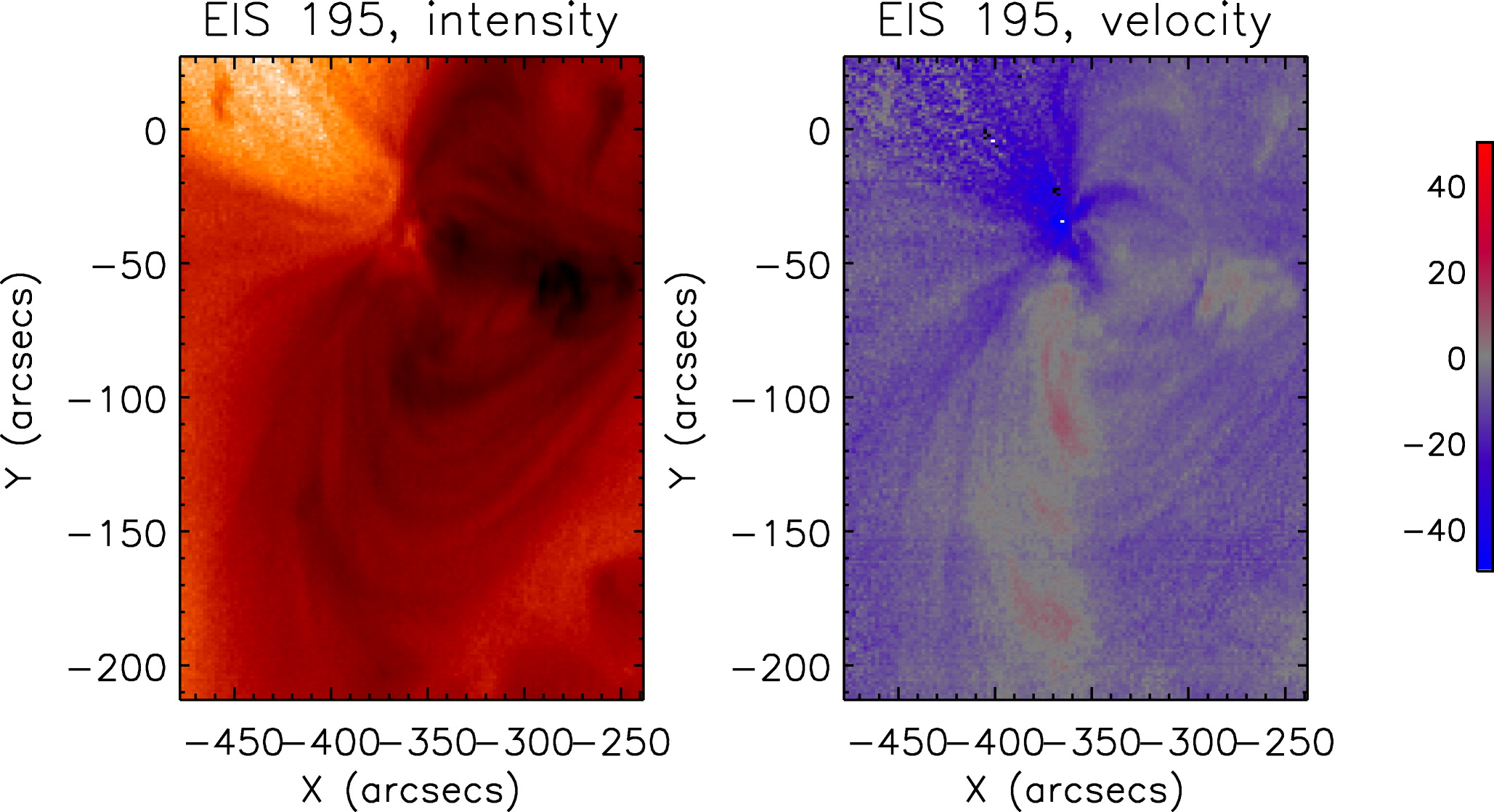}
\caption{Hinode/EIS raster scan of an active region. The left image shows the intensity of the Fe{\sc xii} emission line, and the right image shows the Doppler velocity determined from that line. The strongest blueshifts are occur in the lower-intensity region at the edge of the active region, seen on the left-hand side.  \citep[From][reproduced by permission of the AAS.]{Harra2008} \label{upflows}}
\end{figure}

The sources of the slow solar wind are less clear. Observations from the Hinode EUV Imaging Spectrometer (EIS) instrument consistently show evidence of blueshifted plasma at the edges of active regions \citep[e.g.][]{Harra2008}.  As an example, Fig.~\ref{upflows} shows a raster scan of an active region. The EIS instrument builds up a rastered image through moving a mirror. Each pixel strip in the $y$-direction is created through one exposure of the slit. The image is built up from right to left in time. Each pixel is therefore a spectrum, and this figure shows the Fe{\sc xii} intensity image which is derived by fitting each spectral emission line. The Doppler velocity map is also shown, demonstrating  the bulk flow shifts of the plasma where red is redshifted (away from the observer) and blue is blueshifted (towards the observer). In the context of Solar Orbiter science, we are particularly interested in any blueshifted plasma that may be able to make its way into the solar wind. 

Figure~\ref{upflows} shows that the closed loops in the active region tend to be redshifted, but at the left side of the active region there is strong blueshifted plasma that is located in the weakest-intensity region. The important aspect about this observation is that these upflows always exist in active regions. It may be dominantly on one side, or on both sides, but we are aware of no observations where there are no upflows. Different physical explanations have been put forward for the existence of these upflows, including waves, reconnection in the corona \citep{Baker2009}, and reconnection from below \citep{Bart2007}. The plasma is clearly moving outwards, but the question is whether or not this plasma can make it all the way into the solar wind which is a goal that Solar Orbiter will be able to address.

Modelling is an important aspect of understanding the flows, and work carried out by \citet{Edwards2016} aimed to determine whether or not the upflowing plasma can escape into the solar wind. In this latter work, seven active regions were studied, and the upflows and composition were measured. The structure of the magnetic field was determined from modelling to determine sites of open magnetic field where this plasma could freely flow into the solar wind. In most cases, these upflows did not, in fact, correspond to locations of open field, and so it cannot be assumed that because an upflow is observed, it will automatically lead into the solar wind. However, the sites of upflows were intersected by separatrix surfaces with null points located high in the corona. This indicates that these regions could be sites of reconnection that would have an impact on a large scale. It is clear that local and global magnetic field modelling will be used hand-in-hand with Solar Orbiter data. 

\begin{figure}[ht]
\includegraphics[width=0.5\textwidth]{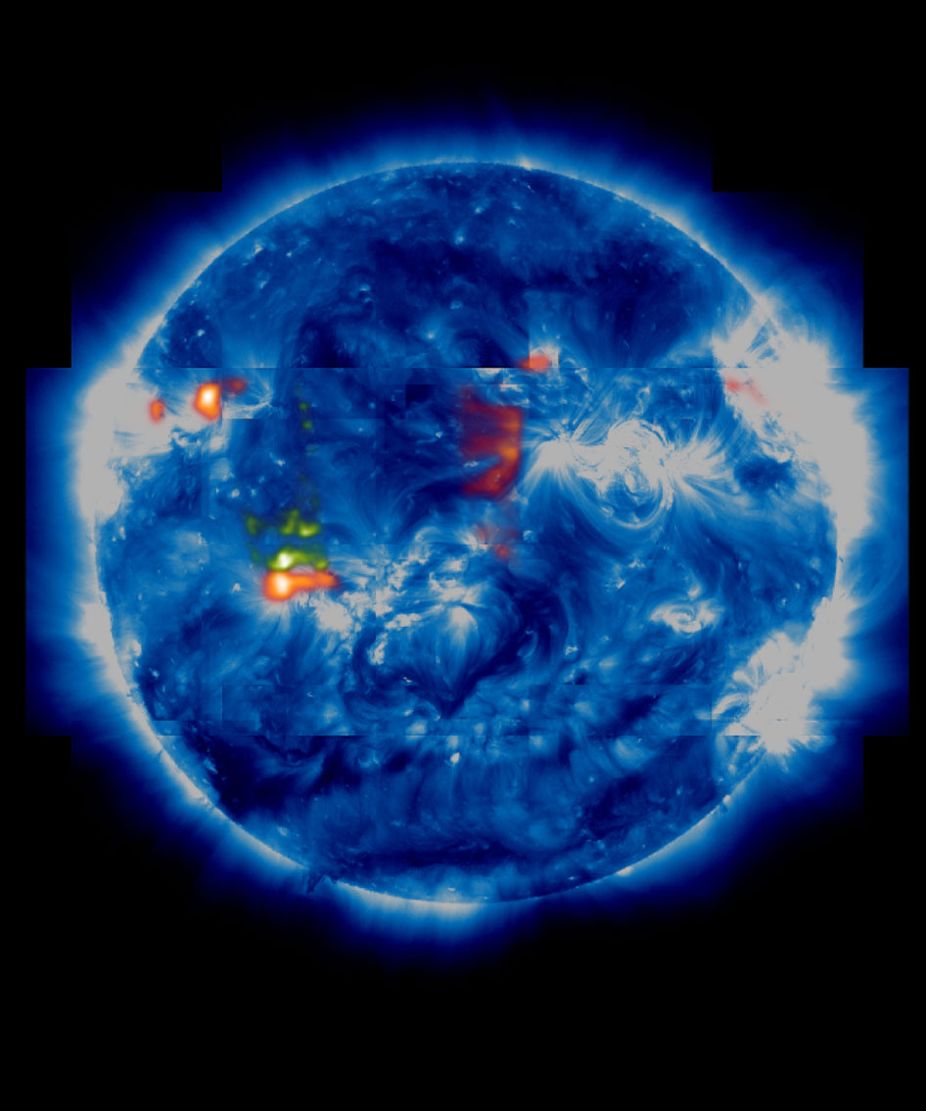}
\caption{Source regions of the slow solar wind. The background image is an SDO/AIA image in the 193\,\AA~band. The sources overlaid on this image are all potential regions where coronal plasma with slow-wind composition is outflowing on open field lines that reach close to the ecliptic plane. The red sources are the larger concentrations of such sources and the green regions are the weaker ones. \citep[From][figure under Creative Commons License, {\tt https://creativecommons.org/licenses/by/4.0/}]{2015NatCo...6.5947B} \label{fullsun}}
\end{figure}

An important way to determine possible sources of the solar wind is to compare the elemental composition determined spectroscopically with in-situ measurements. Imaging spectrometers have a small field of view due to the time it takes to build up an image (see Fig.~\ref{upflows} for the case of Hinode/EIS). However, solar wind can emanate from multiple regions, so a large field of view significantly increases the likelihood that one of them is magnetically connected to the spacecraft's location. One way to explore this is to perform full-Sun rasters, which are time consuming but provide spectroscopic information across the whole Sun. This takes a couple of days and is now carried out by both Hinode/EIS and the Interface Region Imaging Spectrometer (IRIS)  on a regular basis. \citet{2015NatCo...6.5947B} describe the analysis of such a dataset, from which they determined the Doppler velocity and the elemental composition, and combined these with global magnetic field models. This allowed sources of the solar wind to be determined, some of which are active regions as can be seen in Fig.~\ref{fullsun}. This method is a powerful tool, although observationally and analytically time-consuming. On Solar Orbiter,  we plan to perform similar full-Sun rasters using the SPICE instrument.

The anti-correlation of the slow solar wind's expansion and its speed suggests that it is accelerated along bundles of open magnetic field lines that have the greatest expansion rate of their cross-sections. This would correspond to the bright rays observed at the interfaces between coronal holes and streamers \citep[e.g.][]{Wang:2007aa} and to outflows from coronal hole boundaries \citep{Antonucci:2005aa, higginson_dynamics_2017}. However, it is not clear whether or not this interpretation is consistent with measurements of the chemical composition: a significant fractionation of elements is observed in the solar wind relative to the photosphere \citep[e.g.][]{Geiss1982SSRv}, which scales with the first ionisation potential (FIP) 
\citep[FIP][]{laming_element_2019}. The low-FIP metallic ions are more abundant in the solar wind than mid- or high-FIP elements, in contrast to conditions in the photosphere
 \citep{von-Steiger:1997aa, von_steiger_solar_2015, laming_fip_2015}. Ulysses has revealed that the degree of fractionation differs systematically between the two classes of solar wind. Fast wind associated with coronal holes has a composition similar to that of the photosphere, whereas the slow solar wind is characterised by a substantially larger degree of fractionation. In fact, the fast wind, which does not exhibit strong FIP enhancements, could come directly from the photosphere, from small, cool coronal loops and open magnetic funnels at the base of coronal holes or spicules, which also exhibit small FIP enhancements. Remote observations have revealed many cases of macrospicules undergoing reconnection and erupting within coronal holes \citep{loboda_what_2019}. Do they contribute to the fast solar wind streams? Polar plumes, as well
as polar regions within plumes \citep[interplume lanes;][]{2000ApJ...531L..79G,2003ApJ...588..566T,panesar_hi-c_2019,qi_relation_2019,wang_converging_2016}, have also long been suspected to be a significant source of fast solar wind \citep{1997SoPh..175..393D}. Micro-streams of plasma originating in the coronal holes may be related to polar plumes \citep{1995JGR...10023389N}, though evidence for this is controversial \citep[e.g.][]{1996A&A...316..368M}.  However, the relation could be difficult to observe since large-amplitude Alfv\'enic fluctuations generate micro-stream signals in the fast stream \citep{2014GeoRL..41..259M,stansby_diagnosing_2019, kasper_alfvenic_2019, horbury_sharp_2020}.
Direct measurements of the oxygen abundance in corona in the slow and  fast  wind regions are summarised in \cite{2006SSRv..124...35A} and \cite{2012SSRv..172....5A}.

The so-called `S-web' model of \cite{Antiochos:2011aa} can account both for the observed large angular width (up to $\approx 60^\circ$) of the slow wind and its FIP-enhanced coronal composition. According to this model, the most likely source of the slow wind is a network of narrow, possibly singular corridors of open magnetic field in the surrounding closed-field corona. These map to a web of separatrices (S-web) and quasi-separatrix layers in the heliosphere, and the model proposes that the process that releases coronal plasma into the solar wind would have to be either the opening of closed magnetic flux, or interchange reconnection between open and closed magnetic field lines \citep{karpen_reconnection-driven_2016,kepko_implications_2016,kumar_first_2019, owens_signatures_2020}. 
Closed magnetic field lines close to the Sun confine the plasma in loops where the compositional differentiation occurs, but these are continuously destroyed when neighbouring open field lines are advected into them. Interchange reconnection between the open and closed field allows the plasma to flow outwards into space.
This process is expected to predominantly occur at coronal hole boundaries, but may also be active in the intermediate areas of the quiet Sun \citep{bellot_rubio_quiet_2019}, and is the underlying mechanism invoked by Fisk and coworkers (\cite{Fisk:1998aa}; \cite{Fisk:2003aa}; \cite{Fisk:2006aa}; \cite{Fisk:2009aa, zhao_relation_2017}) in their model for the heliospheric magnetic field.

\begin{figure}   
   \centerline{\includegraphics[width=\columnwidth,clip=true]{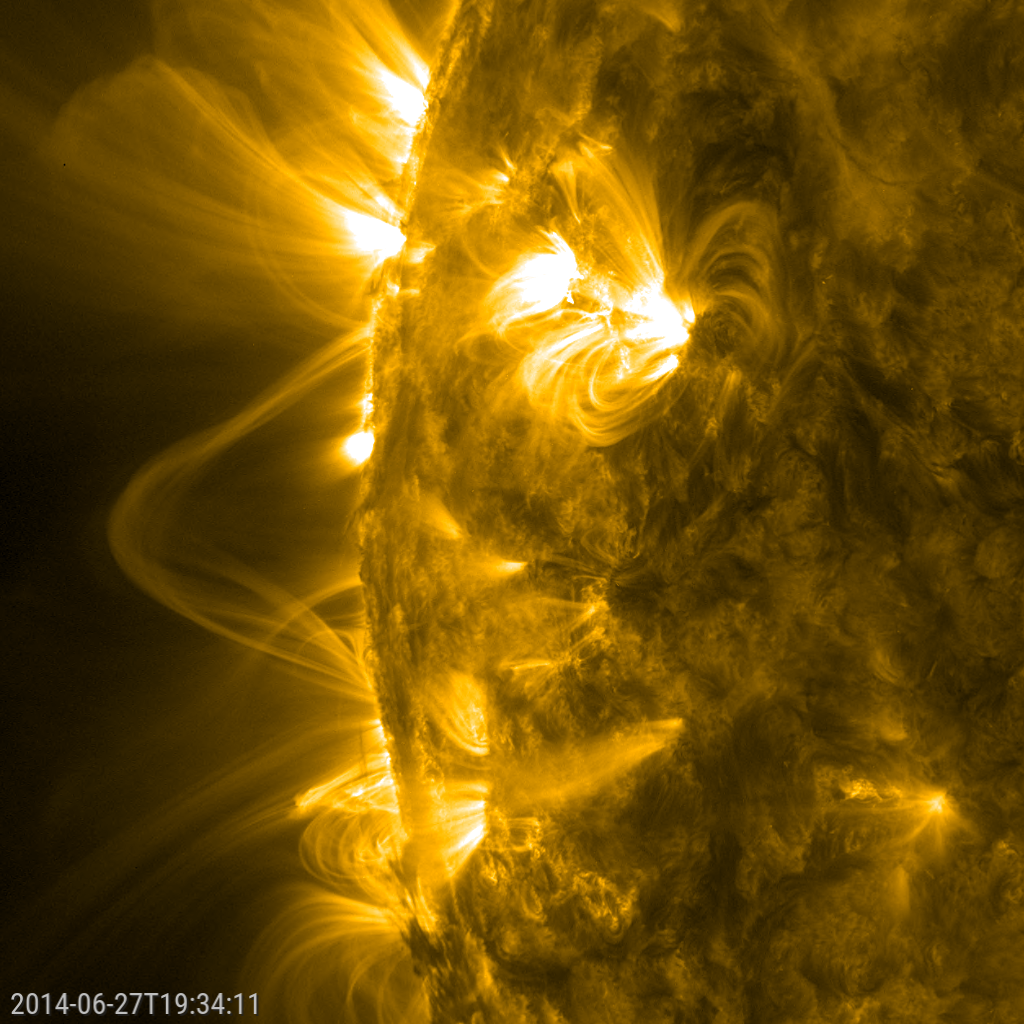}}
              \caption{Ultraviolet emission from plasma in the Sun's atmosphere, revealing some of the complex magnetic field structures around active regions (SDO/AIA 17.1\,nm image).
}
     \label{F-AR_AIA}
   \end{figure}

Additional contributions to the slow wind \citep{kilpua_sources_2016, deforest_highly_2018, sanchez-diaz_situ_2019} could arise from the opening of previously closed field lines in the middle and lower corona, from the tops of helmet streamers or the complex magnetic fields around active regions (Fig.~\ref{F-AR_AIA}, \cite{2007ApJ...658L..63N, abbo_coronal_2015, nieves-chinchilla_analysis_2020}), releasing plasmoids (discrete `blobs' of plasma) into the heliosphere. 

Such plasma blobs are observed in white-light coronagraphic images as the continual, episodic releases of plasma from the tips of helmet streamers; they are also observed \textit{in situ} at 1\,AU and tracked from the upper corona to 1\,AU, swept up and compressed by the fast solar wind from low-latitude coronal holes \citep{2010ApJ...715..300S}. These blobs are small interplanetary transient flux ropes (size of 0.05-0.12\,AU) and can be traced back to streamer events, but also to CMEs \citep{2009ApJ...694.1471S,2011ApJ...734....7R} with a rate of about four per day or approximately every 6 hours and observed with speeds of 150\,km/s at 5\,$R_\sun$ and 300\,km/s at 25\,$R_\sun$. They are thought to be released through either interchange reconnection and/or complete disconnection, and in either case, the reconnection takes place at high altitudes \citep{Wang:2000aa,2001JGR...10616001Z,2004JGRA..109.3108C,2009JGRA..114.4103S}. In addition, there are other, smaller periodic density structures, which are often not flux ropes \citep{2009GeoRL..3623102V} and are observed in 70-80\% of the slow solar wind and in much of the ecliptic fast wind \citep{2008JGRA..113.7101V, 2010SoPh..267..175V, 2015ApJ...807..176V}.

\vspace{3mm}
\noindent
{\it (b) Source regions of the heliospheric magnetic field.} 
The large-scale structure of the interplanetary or heliospheric magnetic field (IMF) is well known (see e.g.\ \cite{2007bsw..book.....M} and \cite{owens_heliospheric_2013}, \cite{balogh_heliospheric_2013} for reviews): the Sun's rotation winds up the magnetic field into the Parker spiral; compression and rarefaction of the plasma in co-rotating interaction regions (CIRs) produce increases and decreases of the field strength; the polarity of the solar source field is reflected in that of the IMF; and the field is pervaded by waves and turbulence over a wide range of scales. Over the Sun's 22-year magnetic cycle, the IMF reflects the changing character of the solar field, from an approximately dipolar to a much more complex, multi-polar structure. 
However, the mapping between solar and interplanetary fields is only known on relatively large scales and in a crude manner. 

Observations of the Sun's surface photospheric magnetic field combined with coronal observations or magnetohydrodynamics (MHD) models of the corona make it possible to estimate the mapping between the lower corona and the `source surface' \citep{Schatten:1969yq} at several solar radii. Nevertheless, many fundamental questions remain about how the Sun's magnetic field opens into space \citep[e.g.][]{2007ApJ...671..936A,wiegelmann_coronal_2017}, particularly with regard to the emergence of new coronal holes and the long-range connectivity of active regions, as well as how the IMF disconnects from the Sun. Distant observations by Ulysses over the Sun's poles have helped to constrain such mappings \citep[e.g.][]{1995SSRv...72..137H, 1997GeoRL..24.3101F}, but Solar Orbiter, being much closer to the Sun and hence eliminating many of the uncertainties caused by local stream--stream interactions, will dramatically improve the precision with which this can be constrained.

Our knowledge of the Sun's magnetic field and its extension into the solar atmosphere is based on polarimetric measurements in the Sun's photosphere. However, the vast majority of the magnetic flux from the Sun closes within the photosphere, chromosphere, and lower corona in small bipolar regions with strong local magnetic fields, and only a small fraction of magnetic flux tubes extend sufficiently high into the solar atmosphere to connect to the heliosphere. In addition, the magnetic field in the Sun's lower atmosphere is highly variable and dynamic at all spatial and temporal scales that can be resolved with current instrumentation. For example, high-resolution observations with the Sunrise balloon-borne telescope \citep{2013SoPh..283..253W} show that the connectivity of magnetic loops changes rapidly on timescales of around ten minutes in the photosphere, and only three minutes in the upper solar atmosphere. Spicules, as well as chromospheric and lower coronal jets, are further examples of highly dynamic phenomena in this domain. The magnetic connection between the solar wind and the Sun's surface therefore critically depends on understanding what determines the amount of open flux from the Sun, how open field lines are distributed on the solar surface, and how these field lines change their connectivity across the solar surface. These processes are thought to be controlled by interchange reconnection \citep{Wang:2000aa,Fisk:2001aa,crooker_interchange_2012,kong_observational_2018,wang_observations_2018}.

The HCS is embedded in the slow solar wind and shows structures on small spatial scales. Some of the first Parker Solar Probe results \citep{szabo_heliospheric_2020} suggest that these evolve significantly between the Sun and 1\,$AU$. As mentioned in Sect.~\ref{S-Goal1}, one of the most surprising results regarding the HCS is that it is not symmetric around the equator, but appeared to be displaced southward by around $10^{\circ}$\citep{Smith:2000aa,koskela_southward_2018} during the last few solar minima, which causes a difference in cosmic ray fluxes between hemispheres. Similar asymmetries exist in the Sun's polar magnetic fields \citep[e.g.]{nistic?_north-south_2015,bhowmik_polar_2019} and even sunspot numbers \citep{Wang:2011aa,iijima_effect_2019}, but the origin of the asymmetry of the HCS is not yet understood.
The north--south asymmetry in solar activity must be driven by the solar dynamo and the interaction of the sub-surface magnetic field with the flows present there. This question can therefore only be answered if our understanding of the solar dynamo is improved. The HCS is also of vital importance for the motion of cosmic rays throughout the heliosphere: depending on the polarity of the solar cycle, ions or electrons tend to migrate to low latitudes and along the HCS as they enter the Solar System.
 
The HCS is remarkably thin -- just a few thousand kilometers across \citep{2005ESASP.592..659Z} -- surrounded by the much thicker, denser heliospheric plasma sheet (HPS, \cite{wu_solar_2019}). It is not clear how thin the HCS and HPS are close to the Sun, which could provide clues to their origin (the HCS seems to become thinner with distance for example). Both the HCS and HPS are also highly variable, but the origin of this is unclear. Reconnection appears to occur here \citep{2006GeoRL..3317102G} and has already been measured by Parker Solar Probe in the inner heliosphere \citep{phan_parker_2020}, but it is still unclear how frequent this is close to the Sun, where the solar wind is seen to be more dynamic with the recently observed switchbacks \citep{kasper_alfvenic_2019}. There is also evidence for folds in the magnetic field \citep{tenerani_magnetic_2020}, from cross-helicity \citep{1999GeoRL..26..631B} and proton-alpha streaming data \citep{2004JGRA..109.3104Y}, but their origin is also unknown. Are they related to photospheric reconnection \citep[e.g.][]{rouppe_van_der_voort_reconnection_2016, ortiz_ellerman_2020}, chromospheric reconnection features such as jets \citep[e.g.][]{2007Sci...318.1591S,robustini_chromosphere_2018} or velocity shears \citep{landi_three-dimensional_2012}?

\paragraph*{How Solar Orbiter will address these questions.}
Detailed measurements of the solar wind plasma and its magnetic field, simultaneously acquired with remote-sensing measurements of the photosphere and corona, are key to deciphering the source regions of the solar wind and the heliospheric magnetic field. As outlined above, the composition of the solar wind, measured \textit{in situ}, can be compared with the spectroscopic signatures of coronal ions with differing charge-to-mass ratios and FIP. Magnetic connectivity can be inferred by measuring energetic electrons and the associated X-rays and radio emissions and using this information to trace magnetic field lines that have crossed the spacecraft's trajectory back to the Sun.
Extreme-ultraviolet spectroscopy and imaging can indicate magnetic reconnection in the solar transition region and corona, for example by the observation of plasma jets or of explosive events. The large overlap between Solar Orbiter's EUI's Full-Sun Imager and the Metis coronagraph will help to connect off-disc coronal structures to the lower corona.

However, from an operational point of view,  this is one of the most challenging science goals of the mission. In order to link remote-sensing observations with in-situ measurements, we need to have pre-selected the right region of the Sun that we will observe remotely in high resolution and that would later be magnetically connected to the spacecraft. In order to do so, we will be using precursor observational data (called low-latency data) and running connectivity models. The details and the adopted strategy are explained in the modelling working group paper \citep{Rouillard2020a}, in the Solar Orbiter operations paper \citep{Sanchez2020} and in the Science Activity Plan \citep{Zouganelis2020a}.

\subsubsection{What mechanisms heat the corona, and heat and accelerate the solar wind?}

\paragraph*{Current understanding.} 
As stated above, the physical mechanisms that heat the Sun's corona are still not fully understood \citep[see][for recent reviews]{2019ARA&A..57..157C, Reale:2014pi, cranmer_origins_2017, verscharen_multi-scale_2019}, but identifying them is of critical importance for both fundamental and applied solar physics. Understanding how the solar wind is accelerated is essentially linked to this question of how a small fraction of the energy contained in the flows of the Sun's convection zone is transformed into magnetic and thermal energy in the solar atmosphere.
Numerous coronal heating mechanisms have been proposed: these involve reconnection caused by the convectively driven braiding of magnetic field lines \citep{Parker1988ApJ,Parker1991Oslo}, dissipation of electric direct currents \citep[Joule heating,][]{Gudiksen+Nordlund2002}, heating by different types of waves (sound waves, magneto-acoustic waves, Alfv\'en waves; see, e.g. \cite{2011ApJ...736....3V,2014ApJ...782...81V} and references therein), and reconnection in the chromosphere \citep{2018A&A...615L...9C}.

Advances in numerical models have allowed simulations of the corona above active regions. Based on 3D magnetohydrodynamics, these models provide insight into the complex energy transport within the magnetically closed structures \cite[][]{Gudiksen+Nordlund2002,Gudiksen+Nordlund2005a,Gudiksen+Nordlund2005b,Bingert+Peter2011,Rempel2017} and can provide explanations for a number of observational findings, such as systematic Doppler shifts \cite[][]{Peter+al2004,Peter+al2006,Hansteen+al2010}, and the appearance of coronal loops \cite[][]{Peter+Binger2012,Warnecke+Peter2019} and flares \cite[][]{Cheung+al2019}. However, all these models are local, in the sense that their computational domain is restricted to a single active region. Consequently, magnetically open structures are not captured by these models. Instead, in such cases one can employ global models that encompass the whole sun \cite[][]{Rouillard2020a}. While this class of models is very useful for studying large-scale evolution, even the most sophisticated numerical models cannot simultaneously include all physical mechanisms at both large {and} small scales (i.e. at the kinetic scales) that are needed to understand the processes of heating and acceleration of the plasma.

Observationally, coronal rain has been proposed as a marker for coronal heating mechanisms by \cite{2010ApJ...716..154A}: Near active regions, small `blobs' of cool and dense plasma in the much hotter surrounding corona are often seen to fall towards the solar surface along coronal loops \citep{Leroy:1972rw,2001SoPh..198..325S,2005A&A...443..319D}. This phenomenon can be explained by a local loss of thermal equilibrium due to heating processes that predominantly deposit energy low in the corona. The loss of equilibrium initiates a runaway cooling process that results in cool, dense plasma in a gravitationally unstable position, which is subsequently seen to flow back towards the solar surface \citep{Karpen:2001fj,Mueller+al2003AA,Mueller+al2004a,Mueller+al2005a}. As both the occurrence and timescale of this phenomenon depend on the average distribution of coronal heating with height above the solar surface, detailed studies of coronal rain can help to constrain theoretical models and help differentiate between them \citep[see, e.g.][]{2018ApJ...855...52F,2018ApJ...865..111W}.

In general terms, energy that is deposited in the corona is subject to a number of physical processes: heat conduction, radiation, kinetic energy, and enthalpy fluxes.
A large part of the deposited energy is transported back to the chromosphere by heat conduction, where it is emitted as radiation. Some heat is conducted outwards, some energy is lost by radiation in the corona itself, and a fraction of it contributes to accelerating the solar wind plasma via kinetic energy and enthalpy flux \citep{2012SSRv..172...89H}.
While transition region pressure, coronal densities, temperature, and asymptotic solar wind speed sensitively depend on the details of the heating processes, the mass flux only depends on the total energy flux \citep{Hansteen:1995aa}. %
A basic observation that models struggle to account for is the fact that the fast wind originates from regions of low electron temperature and density, while the slow wind emanates from hotter parts of the corona. 
This anti-correlation is supported by Ulysses data showing a strong anti-correlation between solar wind speed and `freezing in' temperature of the different ionisation states of oxygen and magnesium in the solar wind \citep{Geiss:1995aa}. 
Based on Helios observations of the very high temperatures and anisotropies of solar wind helium ions \citep{1982JGR....87...35M} and protons \citep{1982JGR....87...52M}, it has been suggested that other processes such as magnetic mirror and wave-particle interactions might significantly contribute to the expansion of the fast wind \citep{Li:1998aa,Kohl:1997aa,Kohl:1998aa,Kohl:2006aa,Dodero:1998aa}.

All observations show that it is insufficient to reduce the question of coronal and solar-wind heating to a mere increase in temperature. Instead, it is necessary to determine the processes that create these observed kinetic features in order to understand coronal heating, solar-wind heating, and the acceleration of the solar wind. For an extensive review of the kinetic physics of the solar corona and solar wind, see \cite{Marsch:2006ab} and \cite{verscharen_multi-scale_2019}. 

In a plasma with low-to-medium collisionality like the solar wind, deviations from thermodynamic equilibrium can lead to instabilities \citep[see][]{Marsch:2006ab,2008JGRA..113.3103S,2013ApJ...773..163V,angeo-36-1607-2018} 
that create small-scale fluctuations in the electromagnetic fields \citep[see monograph by][]{1993tspm.book.....G}. Particles scatter on these fluctuations and thereby reduce the deviations from equilibrium that caused the instability in the first place \citep{angeo-29-2089-2011}.

These processes are thermodynamically relevant since they not only generate electromagnetic fluctuations but also equilibrate the plasma, change the temperatures of the plasma components, and regulate the heat flux in the system \citep{2013JGRA..118.5421H,2015ApJ...806..157V,2017ApJ...838..158H,2018ApJ...854..132R}. We expect that the relevance of the acting mechanisms depends on location, the source regions of the solar wind, the magnetic configuration and connectivity, and the solar cycle. It is therefore necessary to study the kinetic properties of the solar wind under different conditions in order to quantify the contributions of the relevant heating and acceleration mechanisms.

Models of fast solar wind acceleration broadly fall into three different categories. First, there are models in which the shuffling of magnetic flux tubes by convective motions in the photosphere induce wave-like fluctuations that propagate upwards in the solar atmosphere. These waves are partially reflected back towards the surface, dissipating over a range of heights in the process \citep[see][and references therein]{Cranmer:2007aa}

Secondly, there are interchange reconnection models. In these, the energy flux usually results from magnetic reconnection between open and closed magnetic flux systems. In these models, differences between fast and slow wind result from the different rates of magnetic flux emergence, reconnection, and coronal heating in different regions on the Sun \citep{Axford:1992aa,Fisk:1999aa,Schwadron:2003aa}.

Finally, a third class of kinetic self-consistent models referred to as exospheric are based on the velocity filtration mechanism \citep{1992ApJ...398..319S, scudder_thermal_2019} and the assumption that there are suprathermal electrons in the solar atmosphere (i.e.\ power-law tails that depart from the Maxwellian velocity distributions), as observed in the solar wind at different heliocentric distances. Despite this minimal assumption \citep[which may or may not be true for the Sun; see e.g. the model calculations of][]{2012ApJ...753...31S}, these models \citep[e.g.][]{zouganelis_transonic_2004, zouganelis_acceleration_2005} can successfully predict some aspects of the in-situ wind properties at 1\,AU and explain some of the electron properties as observed in the inner heliosphere by Helios \citep{bercic_scattering_2019} and Parker Solar Probe \citep{moncuquet_first_2020, martinovic_enhancement_2020, maksimovic_anticorrelation_2020, halekas_electrons_2020}. The velocity filtration mechanism, which is a physical phenomenon of the solar wind \citep[for a thorough discussion on controversies and uncontroversial mechanisms, see][]{cranmer_origins_2017}, like any other coronal heating theory requires converting some other form of energy (i.e.\ kinetic or magnetic) into thermal energy. The difference is that this conversion would have to occur down in the chromosphere, where a combination of Coulomb collisions and radiative losses would keep mostly the electrons cool.

\paragraph*{How Solar Orbiter will address the question.} Solar Orbiter's combination of high-resolution measurements of the photospheric magnetic field with SO/PHI together with UV and EUV images from EUI and spectra from SPICE will make it possible to identify plasma processes such as reconnection and shock formation and wave dissipation in rapidly varying surface features, observe Doppler shifts of the generated upflows, and determine compositional signatures. 
In particular, the high resolution at which time series of magnetic field and coronal emission will be measured around perihelia will finally allow us to determine the importance of the role played by the coronal heating mechanism proposed by \cite{2017ApJS..229....4C,2018A&A...615L...9C} and \cite{2018ApJ...862L..24P}.

Global maps of the hydrogen outflow velocity, obtained by applying the Doppler dimming technique to the resonantly scattered component of the most intense emission lines of the outer corona (H\,I Ly\,$\alpha$ 121.6\,nm) observed with the Metis coronagraph, will provide the contours of the maximum coronal acceleration for the major component of the solar wind, and the role of high-frequency cyclotron waves will be assessed by determining the height where the maximum gradient of outflow velocity occurs \citep{Telloni:2007aa,telloni_fast_2019}. 
In addition, SoloHI will measure the velocity, acceleration, and mass density of structures in the accelerating wind, which can subsequently be compared to predictions of the different types of solar wind models.

\subsubsection{What are the sources of turbulence in the solar wind and how does it evolve?}

\paragraph*{Current understanding.} Turbulence and instabilities are common features of the solar wind \citep[see reviews by][]{1995SSRv...73....1T,bruno_solar_nodate,matthaeus_who_2011,narita_spacetime_2018,verscharen_multi-scale_2019}. 
At large scales, the fast wind is dominated by anti-sunward-propagating Alfv{\'e}n waves, which are thought to be generated by photospheric motions. At smaller scales, these waves decay and generate a turbulent cascade with a Kolmogorov-type frequency dependence of $f^{-5/3}$ \citep[see, e.g.][]{carbone_scaling_2009,hadid_energy_2017}. The turbulence observed in the slow wind, contrarily, does not have a dominant Alfv{\'e}nic component, and is fully developed over all measured scales \citep[see e.g.][]{alexandrova_solar_2013,sahraoui_three_2010,bruno_solar_2017,perrone_fluid_2018,bruno_low-frequency_2019}. There is strong evidence that the cascade to smaller scales is anisotropic, but it is not known how the anisotropy is generated or driven \citep{chen_anisotropy_2010,wicks_anisotropy_2011,horbury_anisotropy_2012,verdini_3d_2018}. The question is to what extent information about the origin of the turbulence and of the solar wind itself can be deduced from the observed differences of the fast- and slow-wind turbulence. Mechanisms for the evolution of solar wind turbulence include slow-fast wind shears, the presence of fine-scale structures, and gradients \citep{Tu:1990aa,Breech:2008aa,pucci_generation_2018,borovsky_properties_2019}. 

The dissipation of energy in a turbulent cascade contributes to the heating of the solar wind plasma. However, while measurements of the properties of solar wind turbulence in near-Earth orbit largely agree with observed heating rates \citep{carbone_scaling_2009,matthaeus_turbulence_2016,van_ballegooijen_heating_2016}, the details are controversial and dependent on precise models of turbulent dynamics \citep[see e.g.][]{vech_nature_2017,verdini_turbulent_2019,isenberg_perpendicular_2019,kellogg_heating_2020}). 
Recent data from Parker Solar Probe in the inner heliosphere show similar features, but with turbulent energy levels increased by more than an order of magnitude \citep{chen_evolution_2020,bandyopadhyay_enhanced_2020}. This is consistent with models in which the solar wind is powered by turbulence (e.g.\ \cite{2011ApJ...743..197C,2014ApJ...782...81V}).
Statistical analysis of magnetic field fluctuations shows that fine-scale structures, like discontinuities, are commonly present in the solar wind. However, it is not clear whether these originate from complex coronal structuring in the form of advected strands of small-scale magnetic flux tubes \citep{Borovsky:2008aa,Bruno:2001aa,borovsky_plasma_2016} or  are generated locally by turbulent fluctuations. Parker Solar Probe shows very different turbulence properties inside and outside of the switchback structures (rapid polarity reversals; see e.g. \cite{tenerani_magnetic_2020}), indicative of an increasing complexity of the physical processes with decreasing distance form the Sun \citep{dudok_de_wit_switchbacks_2020,mozer_switchbacks_2020}.

At scales around and below the proton gyroradius, turbulent fluctuations interact directly with the ions of the solar wind, but the nature of the turbulent cascade at these scales is poorly understood. Below the electron gyroradius, conditions are even less certain and the partitioning of turbulent energy into electron or ion heating is still unknown. In addition, solar wind expansion constantly drives distribution functions toward kinetic instabilities \citep[e.g.][]{Marsch:2006aa,verscharen_multi-scale_2019}. 
Questions that Solar Orbiter will address include the role of kinetic effects at varying solar distance, the role of wave--particle interactions in accelerating the fast solar wind, and the contribution of minor ions to the solar wind's turbulent energy density close to the Sun.

\paragraph*{How Solar Orbiter will address the question.}
Solar Orbiter will measure waves and turbulence in the solar corona and solar wind over a wide range of latitudes and distances. By travelling over a range of distances, the in-situ instruments will measure how the turbulence evolves as it is swept outward by the solar wind. This will enable us to differentiate between competing theories of turbulent dissipation and heating mechanisms in a range of plasma environments, which in turn is required to advance our understanding of coronal heating.
Because Solar Orbiter is a three-axis stabilised spacecraft, it can continuously view the solar wind beam with 
SWA's Proton and Alpha Particle Sensor (PAS). By measuring how the distributions and waves change with solar distance and between solar wind streams with different plasma properties, Solar Orbiter will help to determine the relative contributions of instabilities and turbulence towards the heating of the solar wind.

\subsection{How do solar transients drive heliospheric variability?}
\label{S-Goal2}

The Sun exhibits many forms of transient phenomena, such as flares \citep[see review by][]{benz_flare_2017}, CMEs \citep[see reviews by][]{Chen2011,Webb2012}, eruptive prominences  \citep[see review by][]{parenti_solar_2014}, and shock waves \citep[see e.g.][]{Cane1985,Gopalswamy1998,Janvier2014}. Many transients directly affect the structure and dynamics of the outflowing solar wind and thereby also eventually affect Earth's magnetosphere and upper atmosphere as well as an important part of interplanetary space \citep[e.g.][]{witasse_interplanetary_2017}. The rapidly evolving discipline of space weather research is focused on understanding and ultimately predicting events that may impact the near-Earth environment \citep[see e.g.][]{eastwood_scientific_2017,koskinen_achievements_2017,bocchialini_statistical_2018}.  However, at this time, there are still many fundamental questions  about the physics of these phenomena that must be answered before we can realistically expect to be able to predict the occurrence of such events. Beyond our own Solar System, answering these questions is also relevant for our understanding of other stellar systems that exhibit transient behaviour such as flaring \citep[e.g.][]{Getman:2008aa}, \cite{2019ApJ...881..114Y}.
Solar Orbiter will observe solar transients and related changes in the heliosphere in a number of ways, and we discuss two interrelated questions below.

\subsubsection{How do CMEs evolve through the corona and inner heliosphere?}
\paragraph*{Current understanding.} Since the launch of SOHO in 1995, significant progress in understanding CMEs has been made thanks to its continuous coronagraphic observations of the Sun. This has been complemented by in-situ measurements of spacecraft such as ACE, WIND, and STEREO.
Today, with more than two full solar cycles of remote and in-situ observations of CMEs and interplanetary CMEs (ICMEs) available, the basic kinematic and morphological features of these structures have been characterised separately \citep[see e.g.][]{webb_coronal_2012,Balmaceda2018,Richardson2012,Jian2018,Nieves-Chinchilla2019}. We also have a basic understanding of CME initiation aspects \citep[see review by][]{van_driel-gesztelyi_evolution_2015} and, through modelling, have also advanced our understanding of their internal magnetic structure \citep[see e.g.][]{Chen_coronal_2011,Nieves-Chinchilla2019}. In this process, different evolutionary processes have been identified that may impact the CME kinematic, morphological, and internal magnetic structure \citep{manchester_physical_2017}.

Coronal mass ejections often appear to originate from highly sheared magnetic field regions on the Sun known as filament channels, which support colder plasma condensations known as prominences \citep[e.g.][]{zhang_launch_2019}. Eruptions are frequently impulsively accelerated in the low corona within 10-15\,minutes \citep[the initial phase can take significantly longer; see][]{Liu:2010ab,liu_geometry_2019,manchester_physical_2017}. CMEs reach speeds of up to 3000\,km/s and carry a total energy of around $\sim10^{25}$\,J ($= 10^{32}$\,erg). They can also accelerate rapidly during the very early stages of their formation, with the CME velocity being closely tied in time to the associated flare's soft X-ray light profile \citep{Zhang:2006aa,ling_peak_2020}. 
Images from SOHO's LASCO coronagraph have provided evidence for a magnetic flux rope structure in some CMEs as well as for post-CME current sheets \citep[][and references therein]{li_eruptions_2013,vourlidas_streamer-blowout_2018}. Spectroscopic data from SOHO/UVCS have been used to measure the untwisting of flux ropes \citep{1997ApJ...490L.183A,2000ApJ...529..575C}.

Both features are predicted by CME initiation models \citep[e.g.][]{Lin:2000aa,Lynch:2004aa,lynch_model_2016}. More recently, during the first perihelion of Parker Solar Probe, a pristine CME with a clear flux rope \citep{HowardNature2019,hess_wispr_2020} was imaged using the WISPR instrument \citep{vourlidas_wide-field_2016} as well as other CMEs that also included flux ropes \citep[e.g.\ a CME initiated from the blowout of a helmet streamer,][]{korreck_source_2020,nieves-chinchilla_analysis_2020} and multiple other structures \citep{zhao_identification_2020}.

STEREO observations made it possible to chart the trajectories of CMEs in the corona and heliosphere in three dimensions, thereby improving our understanding of CME evolution and propagation \citep{Patsourakos:2009aa,bemporad_uncertainties_2015,susino_determination_2016,heinzel_hot_2016,susino_hot_2018,frassati_comprehensive_2019,mancuso_three-dimensional_2019}.

Despite the advances in understanding enabled by recent space missions, very basic questions remain unanswered. These concern the source and initiation of eruptions, their early evolution, and the heliospheric propagation of CMEs. Their initiation has been a core space physics problem in recent decades. The two main paradigms are distinguished primarily by the topology of the pre-eruption magnetic field: twisted flux rope \citep{2000ApJ...529L..49A, 2004ApJ...605L..73R,prior_twisted_2016-1,owens_legs_2016} or sheared arcade \citep{1994ApJ...420L..41A,gibson_solar_2018,toriumi_flare-productive_2019}. Irrespective of the pre-eruption topology, all models predict that as a result of the flare reconnection occurring below the ejection, CMEs in the heliosphere must have a twisted flux rope topology, as commonly observed (see following paragraph for details and relevant references). If the pre-eruption topology is that of a twisted flux rope, then the innermost part of its structure should exhibit relatively undisturbed filament plasma parameters. However, if the twist forms only as a result of flare reconnection, then the whole twisted structure in the heliosphere should exhibit the properties of flare-reconnection-heated plasma, hot beamed electrons, high charge states of Fe, as well as compositional anomalies of heavy ions including He. By measuring the electron and ion properties of a CME along with its magnetic structure, we determine the pre-eruption topology and the initiation mechanism. Solar Orbiter will provide the opportunity to perform these measurements near the Sun, minimising propagation effects such as internal reconnection, which homogenises the CME structure.

The topology of ICMEs is the subject of continuing research. Various types of models predict that ICMEs have a flux rope structure. However, observations at 1\,AU find that this is only the case for less than half of all ICMEs \citep[see][for a review of ICME observations]{kilpua_coronal_2017}. The question of whether or not all ICMEs contain some kind of a flux rope structure  is therefore still open \citep{Richardson:2004aa,Richardson:2010aa,Kilpua:2011aa,2013SoPh..284..179V,demoulin_quantitative_2016,nieves-chinchilla_understanding_2018,aulanier_drifting_2019,good_selfsimilarity_2019,wang_comparison_2019,2019ApJ...885..120T}. 

\paragraph*{How Solar Orbiter will address the question.} 
Combined observations of SO/PHI, EUI, SPICE, and STIX will provide information about the boundary conditions of CME initiation, while the in-situ instruments have already started measuring physical parameters of ICMEs when these pass the spacecraft.
Because a limitation to the three-baseline remote-sensing windows per orbit would make it very challenging to observe a significant fraction of spacecraft-directed CMEs, Solar Orbiter envisages the operation of a subset of the remote-sensing instruments in synoptic modes throughout the orbit to improve the observations statistics, starting with the nominal operations phase.

\subsubsection{How do CMEs contribute to the solar magnetic flux and helicity balance?}
\label{CME_balance}

\begin{figure*}
\begin{center}
\resizebox{0.85\hsize}{!}{
\includegraphics{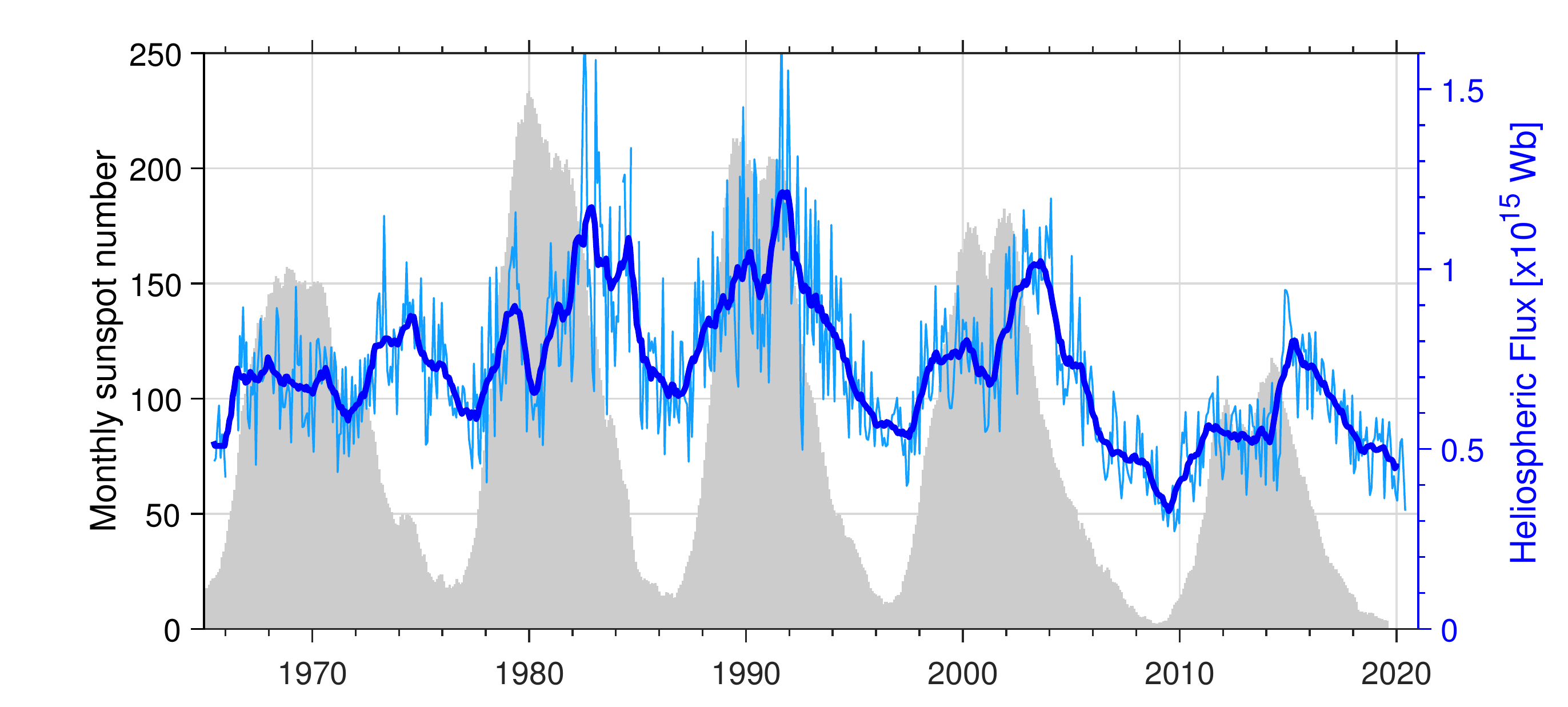}
}
\caption{Near-Earth interplanetary magnetic field strength (thick line: 1-year running mean; thin line: 27-day mean values) and sunspot number (background filled values) for the last five solar cycles. (Courtesy of M.\ Owens, University of Reading; data processed in the same way as described in \cite{Owens:2008JGR}.)}
 \label{F-HeliosphericFlux_2017}
\end{center}
\end{figure*}

\paragraph*{Current understanding.} 
The solar wind transports magnetic flux from the Sun into the heliosphere: open flux mostly through fast wind emanating from polar coronal holes, and closed flux by CMEs. Measurements of the magnetic flux content of the heliosphere from near Earth show that the total amount of magnetic flux in the Solar System changes over the solar cycle (\cite{Owens:2008aa} and Fig.~\ref{F-HeliosphericFlux_2017}).
It is still unclear what the relative contributions of the solar wind and CMEs to the heliospheric magnetic flux budget are (see e.g. \cite{linker_open_2017,owens_sunward_2017,lowder_coronal_2017,wallace_estimating_2019}. Models to explain the solar cycle variation \citep{hathaway_solar_2015} assume a background level of open magnetic flux \citep[see also][]{Owens:2008aa} to which CMEs contribute additional flux around the maxima of the solar activity cycle.

Alternatively, simple models that do not build on CMEs reproduce the solar cycle and secular change in the IMF and global solar open flux as well as concentrations of cosmogenic isotopes reasonably well \citep{2000Natur.408..445S, 2002A&A...383..706S}. These models 
require only the recorded sunspot number as input. An extension of the model also naturally reproduces the strong decrease at the end of solar cycle 23 \citep{2010A&A...509A.100V}. More sophisticated models making use of surface flux transfer \citep{2014SSRv..186..491J} reproduce more details of the open flux and the heliospheric current sheet \citep[e.g.][]{2010ApJ...709..301J} as well as the measured secular changes in the IMF \citep{2016A&A...590A..63D}.

Possibly the major shortcoming of current extrapolations of the magnetic field from measurements at the solar surface into the heliosphere is that synoptic charts of the magnetic field at the solar surface take a full solar rotation to produce. In this time the magnetic field at the Sun has evolved strongly. This can in some cases be partly compensated by making use of the results of far-side imaging, but uncertainty remains \citep{2003SoPh..212..165S,2014A&ARv..22...78W}. 

\paragraph*{How Solar Orbiter will address the question.} Knowledge of the magnetic flux related to individual CMEs forms the basis for understanding how CMEs contribute to the overall heliospheric flux budget. Along its orbit, Solar Orbiter observes CMEs in-situ and can measure their magnetic flux content directly, and doing so at different distances from the Sun will help to quantify the effect of the CMEs' evolution on their journey outwards. At some point in time, the magnetic flux that is carried outwards by CMEs must either disconnect from the Sun, or reconnect to existing open field lines. Solar Orbiter can diagnose this magnetic connectivity based on measurements of suprathermal electron and energetic particles. These particles stream rapidly along field lines and can indicate whether a magnetic flux tube is connected to the Sun at one end, at both ends, or not at all. When the magnetic field is completely disconnected from the Sun, suprathermal particles should disappear.

Solar Orbiter will also allow improvement of the computation of the heliospheric magnetic fields from photospheric measurements when combining magnetograms obtained from near-earth orbit (e.g.\ by SDO/HMI) with those recorded by SO/PHI, in particular in phases when Solar Orbiter is at a large Earth--Sun--spacecraft angle. This will allow the time it takes to produce a synoptic chart to be reduced, while also allowing us to test far-side imaging directly, a technique whose results are currently used to that end. 

\subsection{How do solar eruptions produce energetic particle radiation that fills the heliosphere?}
\label{S-Goal3}

The Sun is the Solar System's most powerful particle accelerator. Solar energetic particles (SEPs) can reach speeds close to the speed of light and their effects can even be detected on the ground, despite the protecting presence of our planet's magnetic field and atmosphere. SEP events can severely affect space hardware and disrupt radio communications, making them highly relevant manifestations of space weather.
In addition to large SEP events, which occur roughly monthly around solar activity maximum, smaller, more numerous events can occur more than once per day on average. In this section, we discuss three questions that flow down from this top-level question: how and where are SEPs accelerated? How are they released from their sources and distributed in space and time? What are the seed populations for SEPs?

\subsubsection{How and where are energetic particles accelerated at the Sun?}

\paragraph*{Current understanding.} 
Two main physical mechanisms are thought to energise SEPs: diffusive shock acceleration \citep[also known as Fermi or stochastic acceleration,][]{Jones:1991rr} and acceleration by solar flares or jets \citep{Reames:2013uq}.
Diffusive shock acceleration involves particles repeatedly interacting with moving or turbulent magnetic field, gaining small amounts of energy at each step, and is believed to operate in shock waves and in regions of high turbulence (see e.g. reviews of the mechanism physics by \cite{petrosian_stochastic_2012} and \cite{zhang_stochastic_2013}). The second mechanism involves a time-dependent magnetic field, producing an electric field which can directly accelerate particles in a single step. On the Sun, such changes take place in flares and jets (e.g.\ \cite{Aschwanden:2006aa}; \cite{Giacalone:2006aa} and review by \cite{klein_acceleration_2017}). 
In fact, multiple processes may take place even in a single SEP event, and while it is not possible to cleanly separate them, they can be split into these two broad classes.

\begin{figure}   
   \centerline{\includegraphics[width=\columnwidth,clip=true]{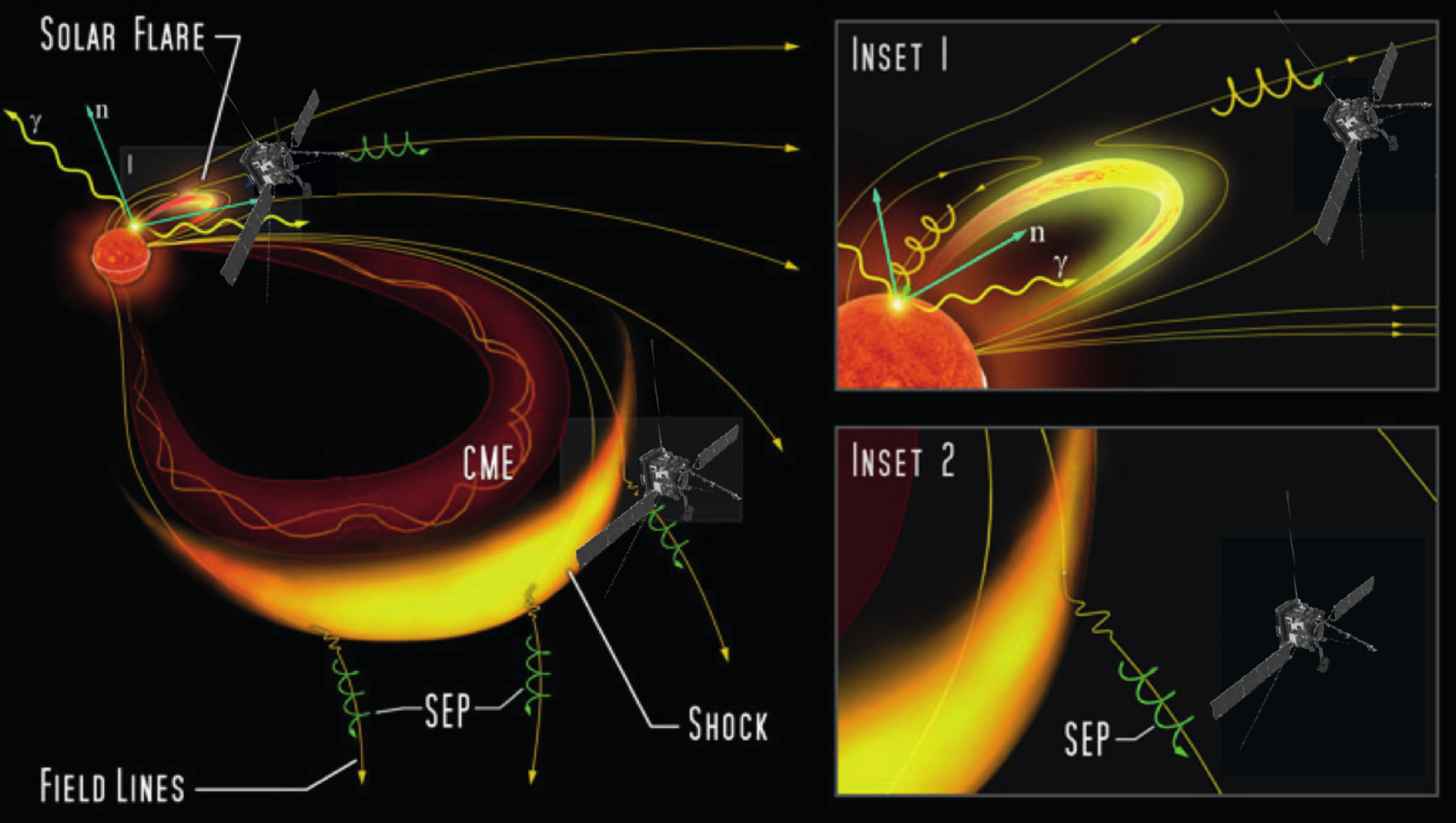}}
              \caption{Sketch showing a solar flare and CME driving an interplanetary shock. Both the flare source and shock may contribute to the interplanetary energetic particle populations. (Adapted from NASA's Solar Sentinels STDT report)}
     \label{F-SEP_gfx}
   \end{figure}

\begin{figure}   
   \centerline{\includegraphics[width=\columnwidth,clip=true]{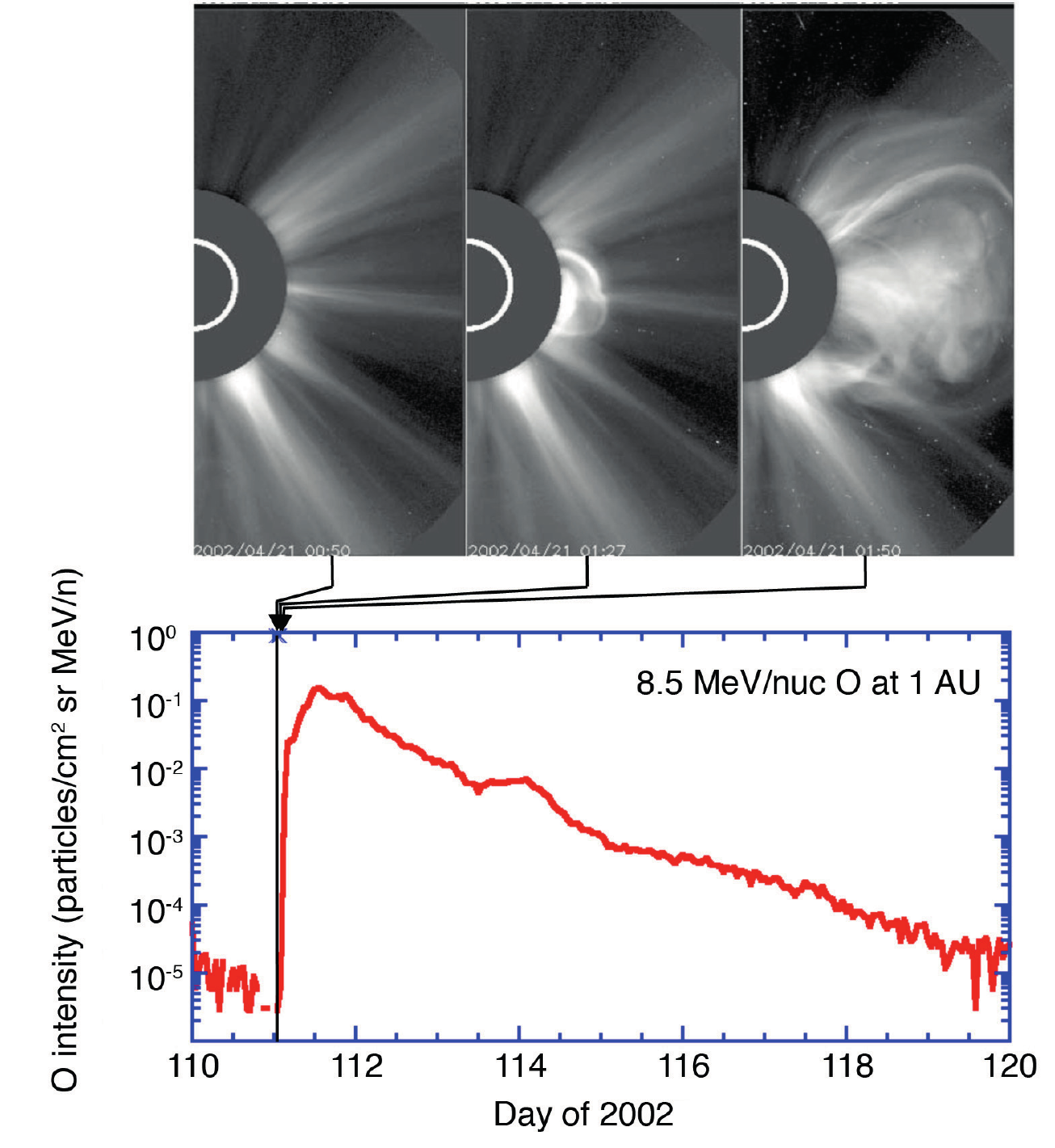}}
              \caption{Upper panel: SOHO LASCO observations of a CME erupting from the Sun's western hemisphere, with exposure times at 00:50, 01:27, and 01:50\,UT. The CME reached a speed of 2700\,km/s at 18\,R$_\sun$, and the associated interplanetary shock passed Earth around 04:15\,UT on 23 April 2002, about  51\,hours after CME lift-off. 
Lower panel: ACE observations of high-energy oxygen nuclei showing an increase in intensity of nearly five orders of magnitude beginning shortly after the CME lift-off. We note that, while the CME images are all taken near intensity onset, the ACE intensities remained elevated for days, long after the shock had passed the Earth. \citep[From][adapted from \cite{Emslie:2004aa}]{Mueller:2013a}
}
     \label{F-CME+Oxygen_in-situ}
   \end{figure}

Figure~\ref{F-SEP_gfx} (from NASA's Solar Sentinels STDT report\footnote{https://ntrs.nasa.gov/archive/nasa/casi.ntrs.nasa.gov/20090024212.pdf}) shows a sketch where an instability in coronal magnetic loops has resulted in an eruption that launches a CME. As it moves into space, it drives a shock creating turbulence that accelerates SEPs from a seed population of ions filling the interplanetary medium (inset 2). Mixed into this may be particles from an associated solar flare (inset 1). CMEs often accelerate particles for hours as they move away from the Sun, and in some cases are still accelerating particles when they reach Earth orbit in a day or two (Fig.~\ref{F-CME+Oxygen_in-situ}). For this reason, CMEs can lead to a large portion of the heliosphere being filled with SEPs. However, the correlation of the observed radiation intensities with CME properties is poor, which indicates that additional aspects such as the the seed populations of SEPs or the shock's geometry must play important but not yet fully understood roles (see \cite{Gopalswamy:2006aa,Desai:2006aa,Mewaldt:2006aa,richardson_properties_2015} and the review by \cite{desai_large_2016}).

\begin{figure}   
   \centerline{\includegraphics[width=\columnwidth,clip=true]{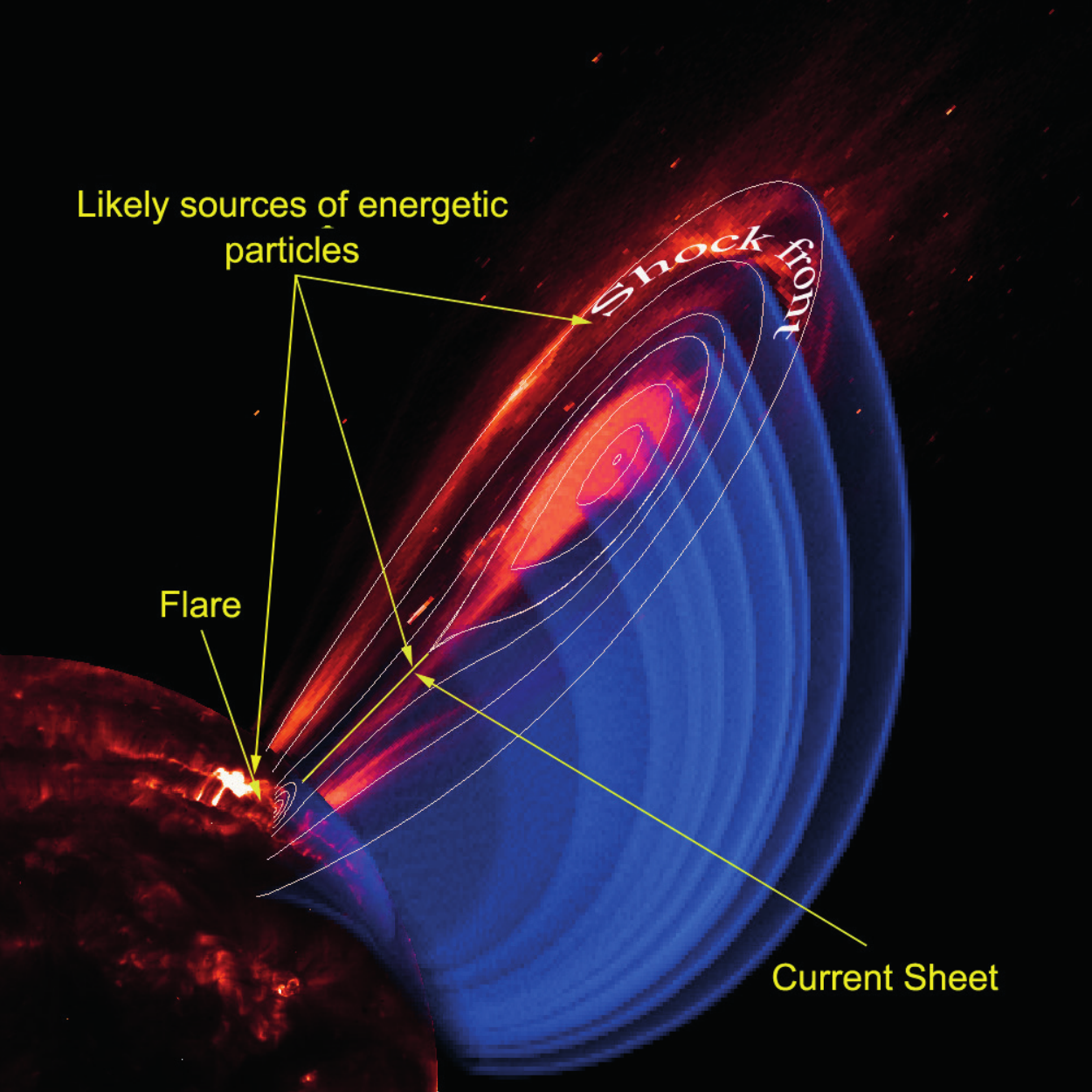}}
              \caption{Composite illustration of a unified flare--CME system showing potential solar energetic particle source regions. The coronagraph image (red image off the limb) shows the CME with a trailing current sheet seen nearly head-on. A cutaway of the modelled magnetic field structure is shown by the blue overlap. Post-flare loops are shown on the UV disk image. (From NASA's Solar Sentinels STDT report)
}
     \label{F-Flare+CME_composite}
   \end{figure}

Acceleration by solar flares or jets is associated with dynamic processes observed in solar coronal loops and active regions (see \cite{toriumi_flare-productive_2019} for a review on flare-productive active regions). Reconnecting magnetic loops, and emerging magnetic flux regions provide sites for stochastic energetic particle acceleration or acceleration by electric fields (see e.g. reviews by \cite{reames_two_2013,reames_abundances_2018}). Because these regions are relatively small, the acceleration process only takes seconds or minutes, and the resulting event is small and therefore often difficult to observe. In the relatively dense lower corona, the energised particles collide with other particles, resulting in UV and X-ray emission that makes it possible to locate the acceleration sites and infer the local plasma density. Most of these particles remain trapped by closed magnetic field lines, travelling down the legs of coronal loops to the solar surface where they deposit their energy, producing X- and $\gamma$-rays. A few SEPs escape on magnetic field lines leading to interplanetary space, and can be traced by their `type III' radio signatures, escaping electrons, and highly fractionated ion abundances (the rare $^3$He can be enhanced by $1000 - 10,000$ times more than in solar surface plasma). Figure~\ref{F-Flare+CME_composite} illustrates another site where reconnection can accelerate particles: in the current sheet behind a CME lift-off. In this case, particles can be accelerated for hours. This way, they may travel towards the sides of a CME and mix with the shock-accelerated particles \citep{Lin:2006aa,Cargill:2006aa,Drake:2009aa}.

\paragraph*{How Solar Orbiter will address the question.}

Almost the entire Solar Orbiter payload contributes to studying solar energetic particles in the wider sense. This includes visible, UV, and X-ray imaging of solar coronal loops and flares; observing radio signatures of coronal shocks and escaping electrons; in-situ measurements to determine turbulence levels and identify shock passages; inferring the SEP seed population from heavy ion abundances and suprathermal particles; and detecting accelerated energetic particles themselves: their timing, velocity distributions, scattering characteristics, and composition. 

Shocks evolve rapidly when moving from the lower corona to the interplanetary medium, because the sound speed (respectively Alfv\'en speed) drops as plasma density (respectively magnetic field strength) declines as $\sim 1/r^2$. Solar Orbiter's coronagraph will remotely identify shock front location, speed, and compression ratios through this critical region within about 10 solar radii. Combining this information with local electron densities as well as coronal ion velocities will provide constraints on shock evolution models in regions even closer to the Sun than Parker Solar Probe's minimum perihelion distance ($\sim 10$\,R$_\sun$). 

Having observed the CMEs and their radio signatures in the corona and the X-ray signatures of the energetic particles near the Sun, Solar Orbiter will then determine the arrival time of the particles \textit{in situ}. As the shock travels past the spacecraft, Solar Orbiter will measure its physical parameters. In the high-latitude phase of the mission, Solar Orbiter will be able to look down on the longitudinal extent of CMEs in visible, UV, and hard X-rays, allowing for the first ever direct observations of the longitudinal extent of the acceleration region. This will make it possible to test currently unconstrained acceleration and transport models by using measured CME size, speed, and shape to specify the accelerating shock.
We expect new insights into particle acceleration along coronal loops as photon and particle signatures will increase by $1/r^2$ as Solar Orbiter gets close to the Sun, enabling at least an order of magnitude more detections of small events than from 1\,AU. Studies of these faint coronal sources will provide important information about the location and plasma properties of suspected electron acceleration sites in the high corona. 

\vspace{3mm}
\noindent
\subsubsection{How are energetic particles released from their sources and distributed in space and time?}

\paragraph*{Current understanding.} A puzzling aspect of SEPs associated with CME-driven shocks is that they often arrive much later at 1\,AU (timescale of hours) than expected based on their velocities \citep[][see also references at the end of the paragraph for more details]{Van-Hollebeke:1975aa, Tylka:2006aa}. Two different processes have been proposed to explain this: (1) in the acceleration process, significant time may be required to energise the particles by repeated interactions with the shock to gain energy; or (2) the particle intensities near the shock may create a region of strong turbulence that traps the particles in the vicinity of the shock, and their intensity observed at 1\,AU depends on the physics of the particles escaping from this region. 
Subsequently, scattering at kinks in the IMF might further delay the arrival of these particles at 1\,AU. The amount of scattering in the interplanetary space varies depending on other activity such as recent passage of other shocks or solar wind stream interactions. By the time the particles reach 1\,AU, they are so thoroughly mixed that these effects cannot be untangled \citep[e.g.][]{Gopalswamy:2006aa,Cohen:2007aa,gomez-herrero_sunward-propagating_2017,klassen_strong_2018,dresing_long-lasting_2018,effenberger_relation_2018,he_propagation_2019}.

\begin{figure}   
   \centerline{\includegraphics[width=\columnwidth,clip=true]{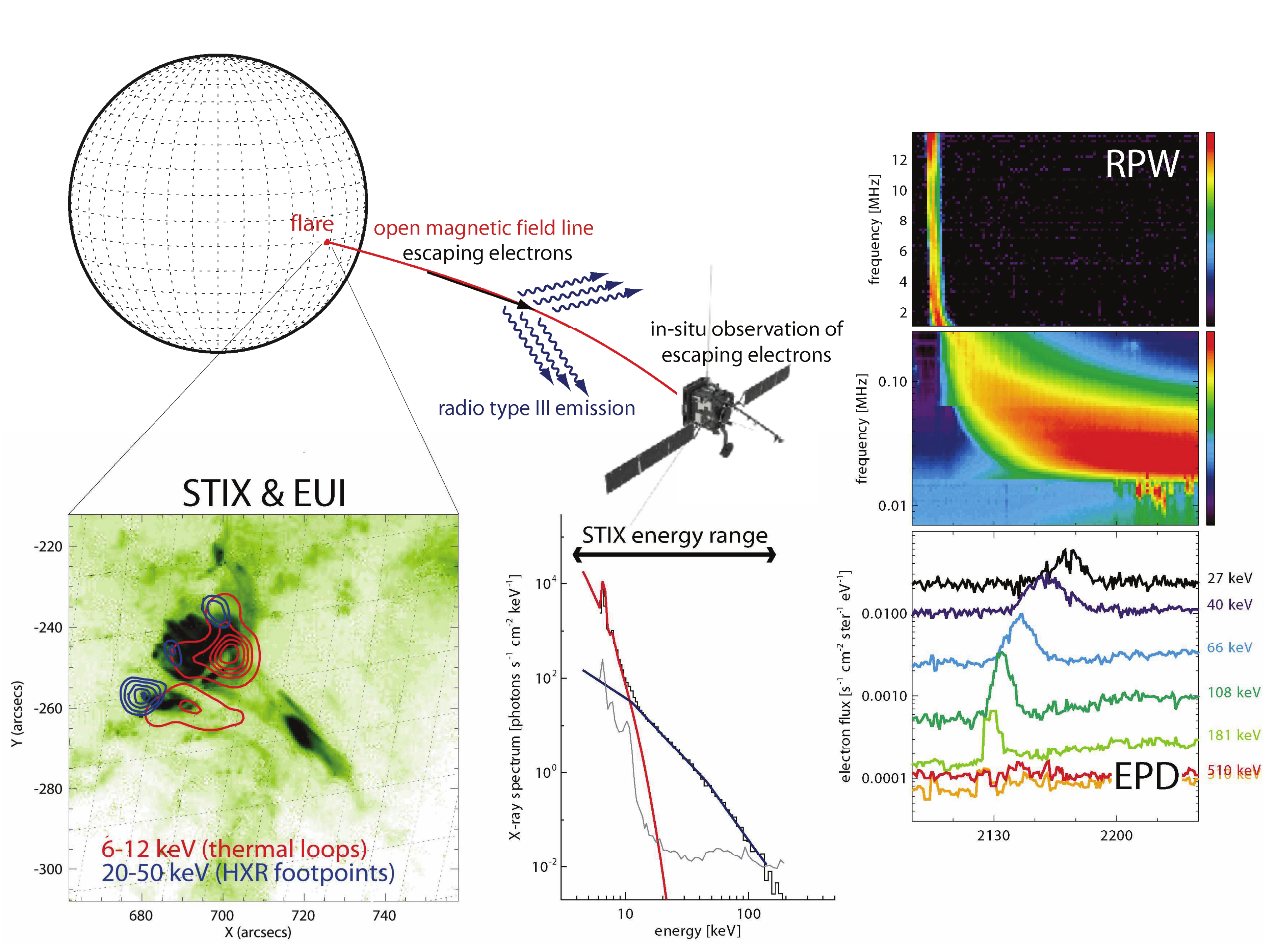}}
              \caption{Coordinated remote-sensing and in-situ observations of a flare source (lower left) producing a jet seen in UV and X-rays, which outline the loops and interactions at loop footpoints (blue). Escaping electrons produce a radio burst (upper right) whose frequency depends on the coronal height of the emitting particles. Subsequently, the energetic electrons are the first particles to arrive at the Solar Orbiter spacecraft, while the heavier energetic ions arrive later, and provide signatures of extreme fractionation produced by the acceleration mechanism.
              The prompt arrival of the particles establishes that Solar Orbiter is magnetically connected to the X-ray source, allowing comparison with coronal magnetic field models in the region of the active region. (Adapted from A.\ Benz, $3^{\rm rd}$ Solar Orbiter Workshop, Sorrento) 
}
   \label{F-Benz_coord_obs_Sorrento}
   \end{figure}

Particles accelerated along magnetic loops can reach high energies within seconds after the onset of flaring activity, and then collide with the solar surface where they emit $\gamma$ radiation \citep{Mandzhavidze+Ramaty1998}. There is a poor correlation between the intensity of the $\gamma$ radiation and the SEP intensities observed at 1\,AU, which indicates that most particles that undergo this powerful acceleration process remain trapped near the Sun. Much more common are flare events observed in UV and X-rays that produce sudden acceleration of electrons, sketched in Fig.~\ref{F-Benz_coord_obs_Sorrento}. The electrons can escape from the corona, producing non-thermal radio emission as they interact with the local plasma. Moving from higher to lower frequencies as the local plasma density decreases with altitude, the (type III) radio emission makes it possible to track the energetic electron burst into interplanetary space. Energetic ions, greatly enriched in $^3$He and heavy nuclei, accompany these electron bursts \citep{Lin:2006aa,Mason:2007aa}. Key open questions related to shock-associated events include the following: Are particle arrival delays at 1\,AU due to the length of time needed to accelerate the particles, or due to trapping in a turbulent region near an accelerating shock, or a combination of both? For particles that are accelerated along loops, are the electrons and ions accelerated at low or high altitudes? How are they related to the X- and $\gamma$-ray signatures?

Recently, Parker Solar Probe detected multiple energetic particle events \citep{mccomas_probing_2019}. Based on these observations, a variety of acceleration mechanisms has been identified including particles accelerated by stream-interaction regions \citep{desai_properties_2020,joyce_energetic_2020,cohen_energetic_2020,bandyopadhyay_observations_2020}, by a slow CME \citep{giacalone_solar_2020}, and a very small event that was not detected at 1 AU, raising the possibility
that such events are more common than one might expect \citep{leske_observations_2020}.

\paragraph*{How Solar Orbiter will address the question.}
Solar Orbiter will be able to observe how CME-driven shocks evolve, and whether they are still accelerating particles as they pass by the spacecraft. If particle arrivals are controlled by the time it takes the shock to accelerate them, then the highest energy particles will be delayed, because they require many more interactions with the shock. If trapping and release controls the timing, then the faster and slower particles will have similar intensity changes as the shock moves by. As Solar Orbiter will simultaneously measure the turbulence properties in the shock acceleration region, it will be possible to construct a complete model of the acceleration process.

For SEPs accelerated along loops or in reconnection regions, Solar Orbiter will observe their source regions in UV and X-rays, and then trace the progress of released electrons by radio emission that will drift to the plasma frequency at the spacecraft when the event passes by. In this way, magnetic connectivity from the spacecraft to the source region can be established. X-ray emission can be used to derive the energetic electron spectrum at the flare site, which in turn can be compared with the escaping population to see what fraction of the accelerated electrons escape. 

\subsubsection{What are the seed populations for energetic particles?}
\paragraph*{Current understanding.} The low-energy particles that are accelerated by CME-driven shocks to SEP energies are called the seed population. The observed ionisation states of SEP ions show temperatures typical of the corona, ruling out hot material on flare loops as the seeds. However, SEPs also show significant abundances of ions that are uncommon in the solar wind (e.g.\ $^3$He and He$^+$). The observed energetic particle abundances indicate that ions travelling at a few times the speed of the solar wind to a few tens of this value are the likely source. This is known as the suprathermal ion pool. At 1\,AU, this ion pool
 is approximately 100 times more variable in intensity than the solar wind, and varies in composition depending on solar activity and interplanetary conditions. Suprathermal ions are continuously present at 1\,AU, but it is not known whether there is a continuous solar source or these ions originate from acceleration by different processes, such as for example\ fast and slow solar wind streams. Inside 1\,AU, the suprathermal ion pool is expected to show significant radial dependence due to the different processes that contribute to the mixture, and is largely unexplored \citep{Desai:2006aa,Mewaldt:2007aa,Lee:2007aa,Fisk:2007aa,mason_possible_2018,kozarev_early-stage_2019,filwett_spectral_2019,kahler_suprathermal_2019}.

For SEPs accelerated along loops or in reconnection regions that give rise to electron and type III radio bursts, ionisation states are coronal-like at lower energies and change over to much hotter flare-like states at high energies. This may be evidence for a complex source, or more likely for energetic particle stripping as the ions escape from a low coronal source. For SEPs accelerated at reconnection sites behind CMEs (Fig.~\ref{F-Flare+CME_composite}), abundances and ionisation states are expected to be coronal \citep{Klecker:2006aa}. Recent observations by Parker Solar Probe \citep{schwadron_seed_2020, wiedenbeck_3_2020} show enhancements in energetic particle seed populations demonstrating how the early evolution of ICMEs could enhance the fluxes of energetic particle seed populations, which precondition the particle acceleration process at distances farther from the Sun where compressions can steepen into shocks.
The nature of the suprathermal ion pool in the inner heliosphere, and the mechanisms in the inner heliosphere that accelerate particles to suprathermal energies are two open questions in this area. 

\paragraph*{How Solar Orbiter will address the question.} Solar Orbiter will provide the missing seed particle data for models of SEP acceleration associated with shocks by systematically mapping the suprathermal ion pool in the inner heliosphere with spectroscopic and in-situ data. The extended mission phase will add the out-of-ecliptic dimension to this, making it possible to  further probe the solar and interplanetary origins of the seed particle populations. 

For SEPs accelerated along coronal loops or in reconnection regions, the Solar Orbiter's perihelia will be advantageous because particle properties will be accurately measured, also with much more precise information on the coronal location. This will allow us to distinguish between low coronal sources that result in stripping of escaping particles and\ higher sources. SEPs accelerated from reconnection regions behind CME lift-offs will be identified by comparing the timing of energetic particle detections with the location of the CME, and the composition of energetic particles can be compared with the composition of the source region, determined spectroscopically.

\subsection{How does the solar dynamo work and drive connections between the Sun and the heliosphere?}
\label{S-Goal4}

As stated earlier, the Sun's magnetic field plays a dominant role in the solar atmosphere: It structures the plasma of the solar corona, is responsible for most of its dynamic flows, and is the driving force behind all energetic phenomena.
On the global scale, the most evident manifestation of its magnetism is the Sun's  $11$-year activity cycle (or 22-year magnetic cycle, taking magnetic polarity into account; e.g.\ \cite{2015LRSP...12....4H}). Similar activity cycles are also observed in a broad range of stars in the right half of the Hertzsprung-Russell diagram \citep[][]{1985ARA&A..23..379B,1999ApJ...524..295S}, and the Sun is an important test case for dynamo models of stellar activity (see \cite{brun_magnetism_2017} for a review).

The Sun's global magnetic field is generated by dynamo processes long thought to be located in the tachocline, the shear layer at the base of the convection zone. However, other possibilities have been brought into play in recent years, such as dynamos located in the lower part of the convection zone, distributed dynamos, and dynamos located around the near-surface shear layer \citep{2006ASPC..354..121B}. The importance of the surface layer for the global dynamo process has been demonstrated by \cite{2015Sci...347.1333C}. This supports dynamo models that rely on magnetic flux transport (see e.g. \cite{Dikpati:2008aa} and the review by \cite{2010LRSP....7....3C}). In these models, a `conveyor belt' of meridional circulation and other near-surface flows transports magnetic flux from decaying active regions to the Sun's poles. From there, it is transported downwards into the convection zone by subduction (possibly even down to the tachocline), to be reprocessed for the subsequent solar cycle or cycles. 

Unfortunately, current dynamo models often fail to predict actual solar behaviour on a global scale. The 2009 sunspot minimum for example was significantly lower in solar activity and extended for a longer time than predicted by any model \citep{mcintosh_what_2019}. This indicates that existing models are still missing key elements. In any case, current global dynamo models (see \cite{cameron_global_2017} for a review) are insufficiently constrained, in particular regarding the meridional circulation at high solar latitudes: the exact profile and nature of the turnover from poleward flow to subduction strongly affects the behaviour of the resulting global dynamo \citep[e.g.][]{Dikpati:1999aa}. However, observing and quantitatively measuring this low-amplitude meridional surface flow near the Sun's poles is currently impossible from a location in the ecliptic plane due to geometric foreshortening. 
In addition to the global dynamo, turbulent convection may drive a small-scale turbulent dynamo \citep{2007A&A...465L..43V, 2014ApJ...789..132R} that could give rise to the observed ubiquitous, weak, small-scale internetwork field \citep{buehler_quiet_2013,2014PASJ...66S...4L,faurobert_solar-cycle_2015,2010A&A...513A...1D,danilovic_internetwork_2016}.

Solar Orbiter's magnetograph, SO/PHI, will help us measure and characterise (near-)surface flows that advect solar magnetic field \citep[e.g.][]{howe_persistent_2015}, the meridional flow \citep[e.g.][]{roth_verification_2016,komm_subsurface_2018}, and the Sun's differential rotation at all latitudes \citep[e.g.][]{a._lamb_measurements_2017,imada_effect_2018,dikpati_role_2018}. As outlined above, accurate knowledge of the magnetic flux transport near the poles \citep[e.g.][]{petrie_solar_2015,wang_surface_2017,petrovay_optimization_2019} is key to constraining solar dynamo models, and in turn improving our understanding of the Sun's activity cycle.

Observations by Hinode's Solar Optical Telescope \citep[SOT,][]{2008SoPh..249..167T} have already provided a glimpse of the Sun's high-latitude region above 70$^{\circ}$ \citep[see][]{Tsuneta:2008aa,petrie_high-resolution_2017} by making use of the the Sun's $B_0$ angle of 7$^{\circ}$. However, in general, observations from outside the ecliptic plane are required to reliably measure the Sun's  time-dependent surface magnetic field near the poles. In the following sections, we discuss three science questions that flow down from this top-level question in more detail. 

\subsubsection{How is magnetic flux transported to and reprocessed at high solar latitudes?}

\paragraph*{Current understanding.} The Sun's surface and subsurface flow fields at low and middle heliolatitudes have been mapped very successfully thanks to the large amount of high-quality data from the SOHO/MDI \citep{Scherrer:1995aa} and SDO/HMI \citep{Scherrer:2012aa} instruments. These data have provided accurate knowledge of differential rotation \citep{1998ApJ...505..390S}, and have allowed us to determine the low-latitude part of the meridional flows \citep{Gizon2020_in_press}, near-surface torsional oscillations \citep{Howe:2006aa}, and the three-dimensional structure of the shallow velocity field beneath the solar surface \citep{2005LRSP....2....6G, 2010ARA&A..48..289G}.
However, as described above the near-polar flow fields and the differential rotation at high latitudes \citep{Beck:2000aa,Thompson:2003aa} cannot be mapped well from locations near the ecliptic plane. In particular, the meridional flow, a key aspect of flux transport dynamo models, is not fully characterised \citep[e.g.][]{boning_inversions_2017,lin_solar-cycle_2018,mandal_helioseismic_2018}.
 
\paragraph*{How Solar Orbiter will address this question.} 
Solar Orbiter resolves small-scale magnetic features based on vector magnetographic measurements with the high-resolution telescope of SO/PHI. It will also be measuring magnetic flux transport near the solar surface through correlation tracking of small features, mapping Doppler shifts, and helioseismic observations. 
Mass flows in the upper convection zone can be probed by local helioseismology \citep[e.g.][]{Gizon:2005aa}. \cite{loptien_helioseismology_2015} describe in detail how helioseismic techniques can be used for Solar Orbiter data, taking into account the constraints on data downlink and temporal limitation of time series. Based on SOHO/MDI data, \cite{Jackiewicz:2008aa}  demonstrated that even with only a single day of observations, complex velocity fields can be derived.

Uniquely, Solar Orbiter's provision of magnetograms away from the Sun-Earth line will for the first time enable stereoscopic helioseismology by combining data with ground- or space-based helioseismic observations from 1\,AU. For global helioseismology, this will help by significantly reducing mode leakage in Fourier space. In addition, deeper  layers of the convection zone can be probed by considering additional acoustic ray paths. This will require observations beyond the baselined ten-day remote-sensing windows, and the mission's science operations teams have started assessing whether or not and in what ways this can be accomplished.

\begin{figure}   
   \centerline{\includegraphics[width=\columnwidth,clip=true]{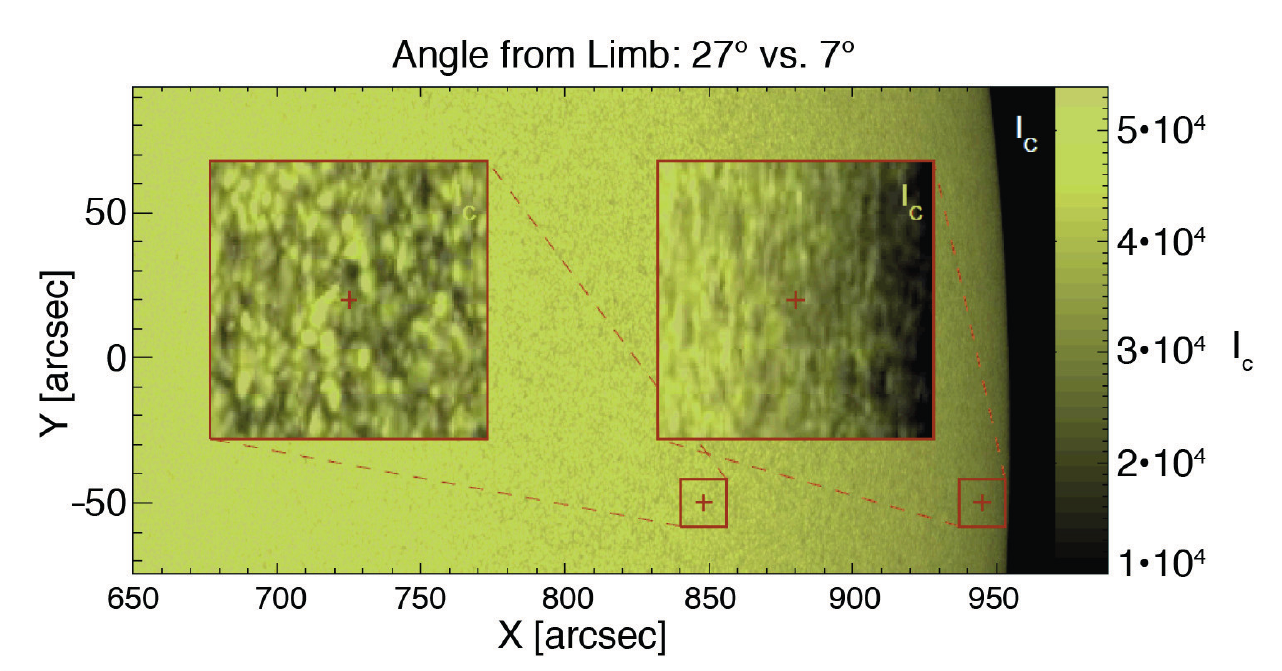}}
              \caption{Solar granulation: A comparison between images taken at different viewing angles shows that fine-scale structure can be resolved with much higher fidelity at more inclined angles. For mapping the Sun's polar regions beyond what can be achieved using the Sun's $B_0$ angle \citep{Tsuneta:2008aa}, a viewpoint outside the ecliptic plane is required.  \citep[From][]{Mueller:2013a}}
     \label{F-Granulation_angle}
   \end{figure}

\subsubsection{What are the properties of the magnetic field at high solar latitudes?}

\paragraph*{Current understanding.} 
As mentioned above, the Sun's high-latitude magnetic field cannot be properly observed from viewpoints in the ecliptic plane. This is due to (a) the directional sensitivity of the Zeeman effect and (b) magnetic polarity cancellation resulting from geometric foreshortening. By making use of the Sun's $B_0$ angle and the high-quality observations of Hinode/SOT, it has been be possible to partly   overcome the latter, but the former requires by definition out-of-ecliptic observations. 

The polar magnetic field plays a central role in the global dynamo, presumably as a source of poloidal field that is wound up by the differential rotation in the tachcline \citep{2015Sci...347.1333C, cameron_observing_2018}. Thus, the strength of the polar field is one of the best indicators of the strength of the following cycle \citep{2009ApJ...694L..11W}.

The distribution of the magnetic field near the Sun's poles also drives the formation and evolution of polar coronal holes, polar plumes, and X-ray jets. Polar coronal holes have been intensively studied from the ecliptic plane \citep[e.g.][]{chandrashekhar_characteristics_2014,gupta_observations_2014,spadaro_investigating_2017,hahn_density_2018,cho_new_2019,tei_iris_2020,peleikis_origin_2017,telloni_fast_2019}, but lack of imaging from outside the ecliptic has so far limited our understanding of the geometry of polar structures. As described in Section~\ref{CME_balance}, the magnetic flux in the heliosphere varies with the solar cycle. On the one hand, there is evidence that the heliospheric magnetic flux has increased substantially in the last hundred years; on the other hand, the interplanetary magnetic field strength during the 2009 solar minimum was lower than at any time since the beginning of the space age. While models based on the injection of flux into the heliosphere by CMEs cannot explain this reduction, models that compute the open flux using flux transport simulations show good agreement \citep[e.g.][]{2014A&A...570A..23D,2016A&A...590A..63D}.

\begin{figure}   
   \centerline{\includegraphics[width=\columnwidth,clip=true]{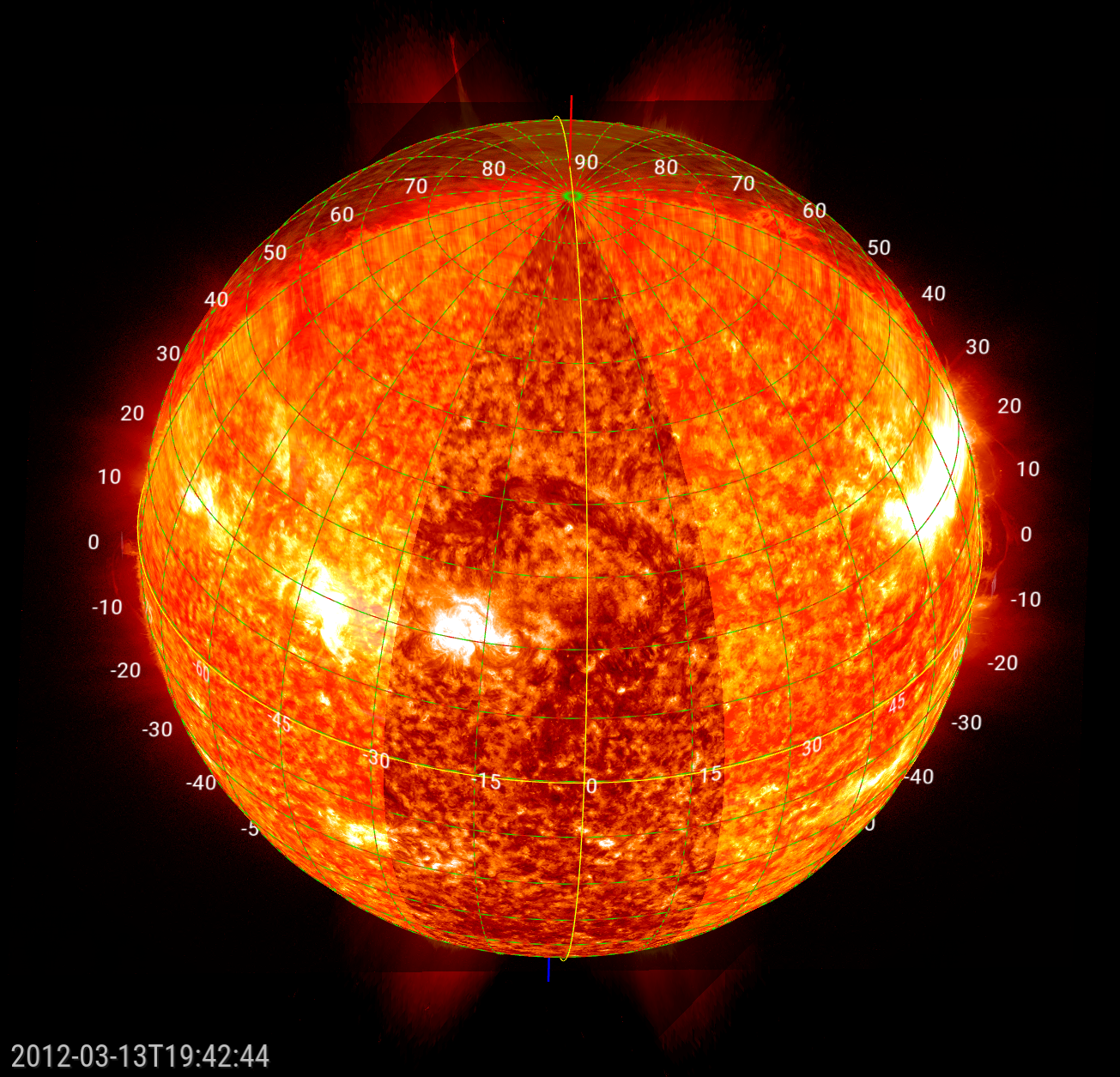}}
              \caption{Simulated view of the Sun's UV corona from 33$^\circ$ heliolatitude, composited out of three He 30.4\,nm images from SDO/AIA and STEREO/SECCHI using JHelioviewer \citep{2009CSE....11...38M,Muller:2017dq}. Solar Orbiter's remote-sensing instruments and out-of-ecliptic vantage point will enable the first simultaneous measurements of the polar magnetic field and associated structures in the corona.}
     \label{F-EUI_polar}
   \end{figure}

\paragraph*{How Solar Orbiter will address this question.} Solar Orbiter's remote-sensing instruments will characterise the Sun's polar regions for the first time (see Fig.~\ref{F-EUI_polar} for a simulated EUI image). By comparing polar magnetic flux properties from different orbits, it will provide an independent constraint on the strength and direction of the meridional flow near the pole. From higher heliographic latitudes, Solar Orbiter's SO/PHI instrument will be able to take a string of snapshots of magnetic flux from the activity belts to the poles, which drives the polarity reversal of the global magnetic field (see \cite{Wang:1989aa}; \cite{Sheeley:1991aa}; \cite{Makarov:2003aa}). In addition, SO/PHI will probe the cancellation processes that take place when flux elements of opposite polarity meet as part of the polarity reversal process. These interactions are expected to occur on comparatively short timescales, as they are driven by granular and supergranular flows once the features are sufficiently close together.   
The large overlap between Solar Orbiter's EUI Full-Sun Imager and the Metis coronagraph will allow us to trace structures from the far off-disc corona down to the limb and critically confront models of the extended coronal connectivity extrapolated from magnetograms provided by SO/PHI.

The combination of Solar Orbiter's high-latitude observations with data acquired by spacecraft in the ecliptic plane will enable investigations of the three-dimensional structure of the inner heliosphere, such as for example the heliospheric current sheet, whose inclination is commonly used as a proxy for the tilt of the solar magnetic dipole. 

\subsubsection{Are there separate dynamo processes acting in the Sun?}
\paragraph*{Current understanding.} 
It is likely that multiple physical mechanisms contribute to the generation of the Sun's magnetic field. Magnetohydrodynamics simulations indicate a local turbulent dynamo in the Sun's convection zone \citep{Brun:2004aa,strugarek_modeling_2016,whitbread_need_2019} and even in the near-surface layers \citep{Vogler:2007aa, 2014ApJ...789..132R}. Hinode's SOT and the Sunrise balloon-borne observatory \citep{2010ApJ...723L.127S} have detected ubiquitous horizontal magnetic fields in quiet regions of the Sun \citep{Lites:2007aa,2010ApJ...723L.149D,lites_are_2017}, which are possibly generated by a small-scale turbulent local dynamo \citep{Pietarila-Graham:2009aa}. These small, weak inter-network magnetic fields \citep{Zirin:1987aa} bring two orders of magnitude more magnetic flux to the solar surface than the stronger features formed by the global dynamo \citep{2011SoPh..269...13T,2017ApJS..229...17S}. Analyses show \citep{Lamb:2008aa,Lamb:2010aa,2017A&A...598A..47A} that even the smallest features observed are formed out of yet smaller features, too small to be resolved with current instrumentation. However, there is still uncertainty over whether or not the Sun indeed possesses a separate local, turbulent dynamo and how strongly this would contribute to the Sun's magnetic flux \citep[e.g.][]{borrero_solar_2017,singh_enhancement_2017,rempel_small-scale_2018}. \cite{Parnell:2009aa} showed that solar magnetic features at all spatial scales follow a power-law probability distribution function, which is scale-free. This indirectly suggests that a single turbulent mechanism may be at work.

\paragraph*{How Solar Orbiter will address this question.} By observing the distribution of small concentrations of emerging magnetic flux as a function of heliographic latitude, Solar Orbiter might be able to differentiate between the presence of a global and a local dynamo: Global dynamo action is expected to lead to the emergence of large bipolar active regions between $\approx 5^{\circ}$ and 30$^{\circ}$ latitude  and of the much smaller, ephemeral active regions over a wider latitude range, but not reaching the poles. Contrarily,  a local turbulent dynamo would be expected to lead to a much more uniform latitudinal distribution of small-scale magnetic flux concentrations. In particular, it would lead to the emergence of the small, bipolar ephemeral active regions almost independently of latitude.


\section{The Solar Orbiter spacecraft}
\label{sect-sc}
\begin{figure}
\includegraphics[width=\columnwidth]{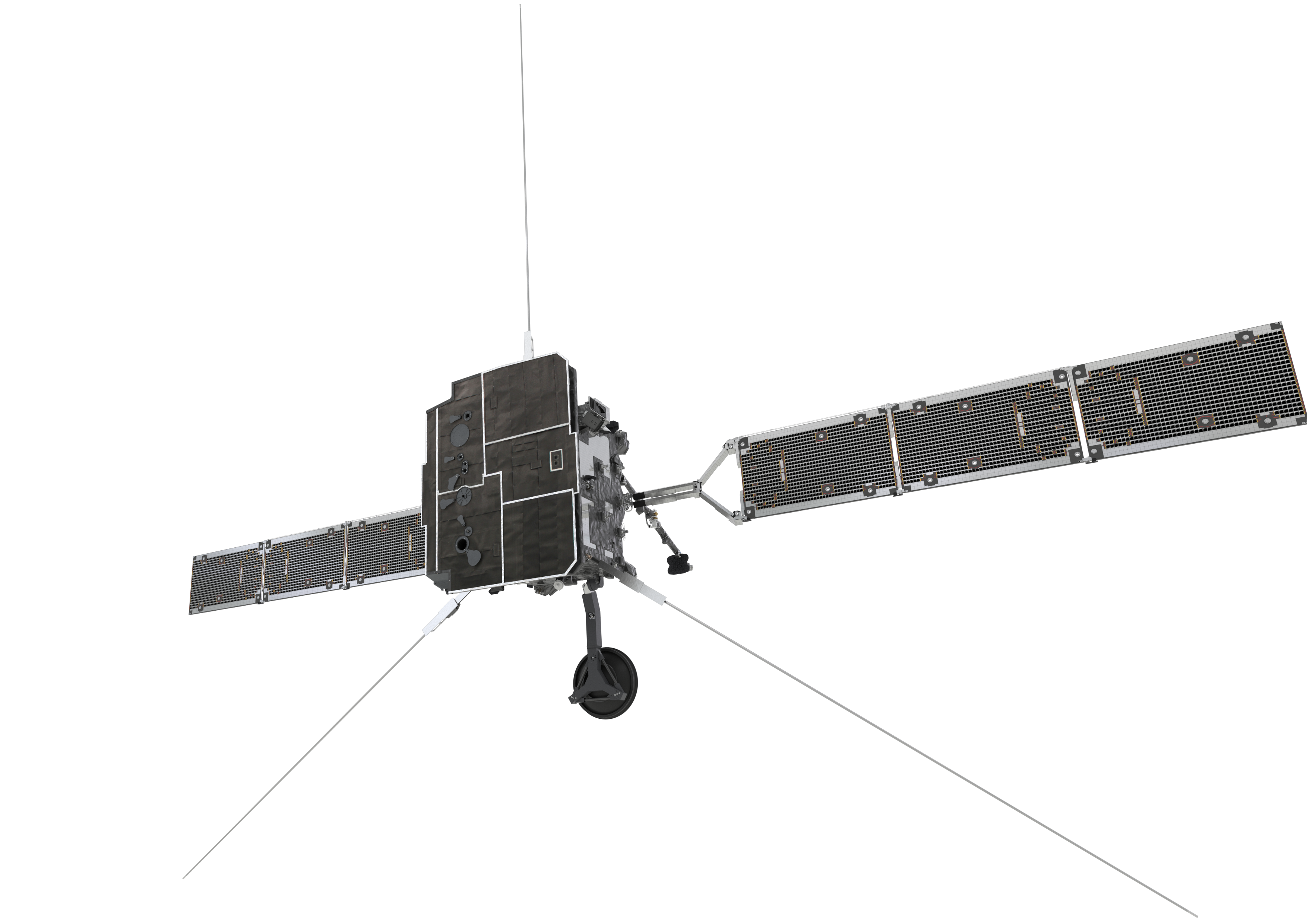}
\caption{Solar Orbiter spacecraft, front (sun-facing) view. Except for the SoloHI heliospheric imager, which is looking over the right edge of the heat shield, the remote-sensing instruments are mounted internally and view the Sun through feed-throughs in the heat shield. Most of these openings have sliding doors for additional protection. The solar arrays can be tilted around their longitudinal (yoke) axis for temperature control. The three RPW antennae are pointing radially outwards, and the high-gain antenna can be seen below the spacecraft body.}
\label{fig:so_sc_hs_view}
\end{figure}

The Solar Orbiter spacecraft was built by Airbus Defence and Space UK as prime contractor and is described in detail in \cite{Garcia2020}. 
Figure~\ref{fig:so_sc_hs_view} shows a front view of the spacecraft. 

It had a launch mass of around 1720\,kg, including\ 209\,kg of science payload, and a body size of 2.5\,m $\times$ 3.1\,m $\times$ 2.7\,m. The six solar panels of 2.1\,m $\times$ 1.2\,m each are mounted into two solar arrays that can be rotated around their longitudinal axis for temperature control. With solar arrays deployed, the spacecraft's total wingspan is 18\,m. 

Solar Orbiter is a three-axis stabilised spacecraft built around a central cylinder. Remote-sensing instruments are mounted on one side panel inside the spacecraft body while the in-situ instruments and SoloHI are mounted on external surfaces. This includes a 4.40\,m-long instrument boom to accommodate several in-situ sensors, and three 6.50\,m-long antennae of the Radio and Plasma Waves experiment (RPW). A heat shield protects the spacecraft from the intense solar flux -- up to
13 times the solar constant -- experienced during the course of the mission which will heat up the front of the heat shield to around 500$^\circ$C at perihelion.

The outer part of the heat shield, also known as the high-temperature multilayer insulation, is made of 20 thin layers of titanium foil. The sun-facing layer has a thickness of 50\,$\mu$m and its surface has been treated with the {\it SolarBlack}\footnote{\tt http://www.enbio.eu/solar-orbiter/} thermo-optical coating, which is based on black calcium phosphate. This {\it SolarBlack} `skin' has been chosen because it maintains its thermo-optical properties after years of exposure to intense radiation, whilst not shedding material or outgassing, which would risk contaminating the scientific instruments. In addition, the skin is electrically conductive, preventing the build-up of static charges which must be avoided both from scientific and engineering perspectives.

\begin{figure}[ht]
\includegraphics[width=\columnwidth]{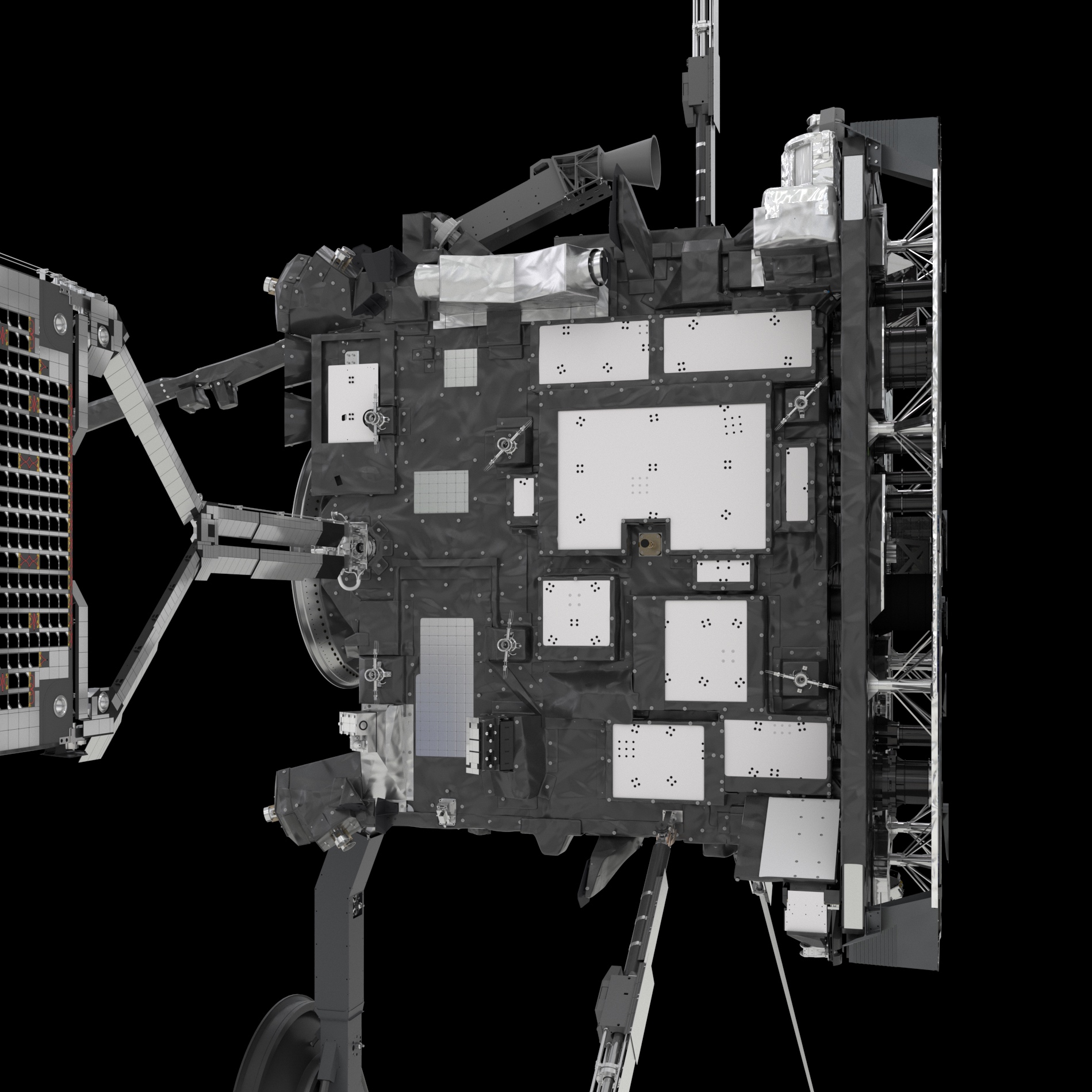}
\caption{Side view of the Solar Orbiter spacecraft. Inside the gap between the outer part of the heat shield and its base on the right-hand side, the feed-throughs of the remote-sensing instruments can be seen. Two of the four corners of the heat shield have cut-outs for the SWA/HIS (top) and SWA/PAS (bottom) sensors. The white panels on this $-Y$ side panel of the spacecraft are the spacecraft-provided payload radiators (SORAs), and the two-telescope unit near the top edge is EPD/SIS.}
\label{fig:so_sc_hs_side_view}
\end{figure}

Between the outer part of the heat shield and its base is a gap through which heat is radiated sideways and away from the spacecraft (Fig.~\ref{fig:so_sc_hs_side_view}). The white panels on the side of the spacecraft are the spacecraft-provided payload radiators (SORA).

Ten star-shaped brackets attach the top layer of the heat shield to its base. The heat shield base is composed of a 5\,cm-thick aluminium-core honeycomb support panel that is covered by 28 layers of `lower temperature' multi-layer insulation. This material can handle temperatures of up to 300$^\circ$C. The entire heat shield is then fixed to the spacecraft by ten 1.5 mm-thin titanium `blades' to minimise the transfer of heat through the spacecraft's superstructure. The remote-sensing instruments are mounted internally and view the Sun through feed-throughs in the heatshield, most of which have sliding doors for additional protection. The exception is the SoloHI heliospheric imager, which is externally mounted and is viewing the inner heliosphere over the rear edge of the heat shield, with its circular field of view offset from the centre of the Sun by about 22.5$^\circ$.

\begin{figure}
\includegraphics[width=\columnwidth]{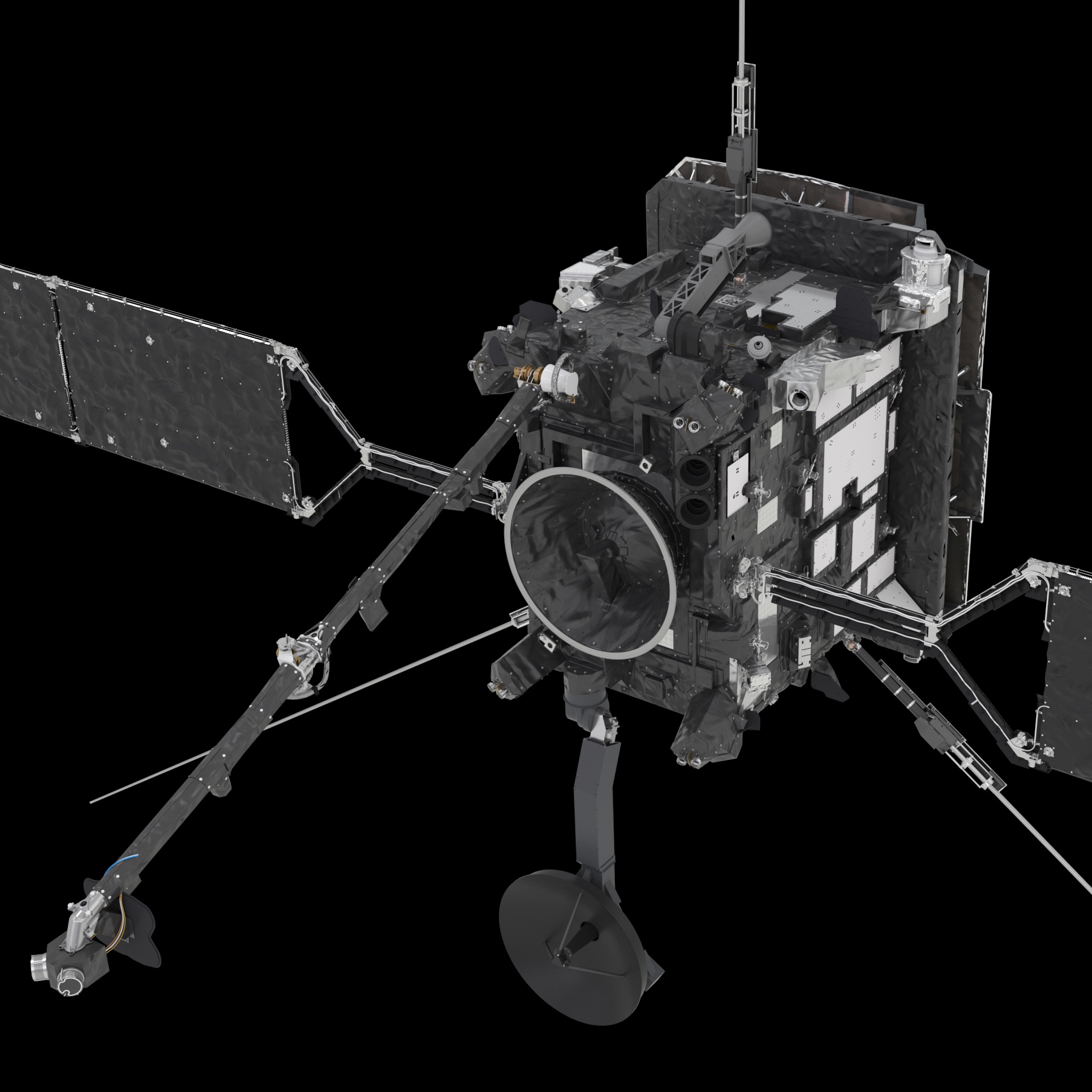}
\caption{Rear view of the Solar Orbiter spacecraft. The instrument boom hosts the SWA/EAS sensor at its tip, and the RPW/SCM sensor as well as the two MAG sensors along its length. The high-gain antenna is made out of titanium treated with {\it SolarBlack}, and is fully articulated to communicate with ground stations throughout the orbit. The rear-side of the solar arrays has been covered with conductive foil to avoid electrostatic charging.}
\label{fig:so_sc_my_view}
\end{figure}

Figure~\ref{fig:so_sc_my_view} shows the rear of the spacecraft. The rear-side of the solar arrays has been covered with conductive foil to avoid electrostatic charging, and a number of contamination-protection baffles, for example\ at the end of the instrument boom, have been added to ensure that sensitive instrumentation is shielded from any thruster-plume contaminants. The central ring on the back of the spacecraft body is the adapter to the launch vehicle.

\section{Instrument overview}
\label{sect-instruments}
\begin{figure*}
\resizebox{\hsize}{!}{
\includegraphics{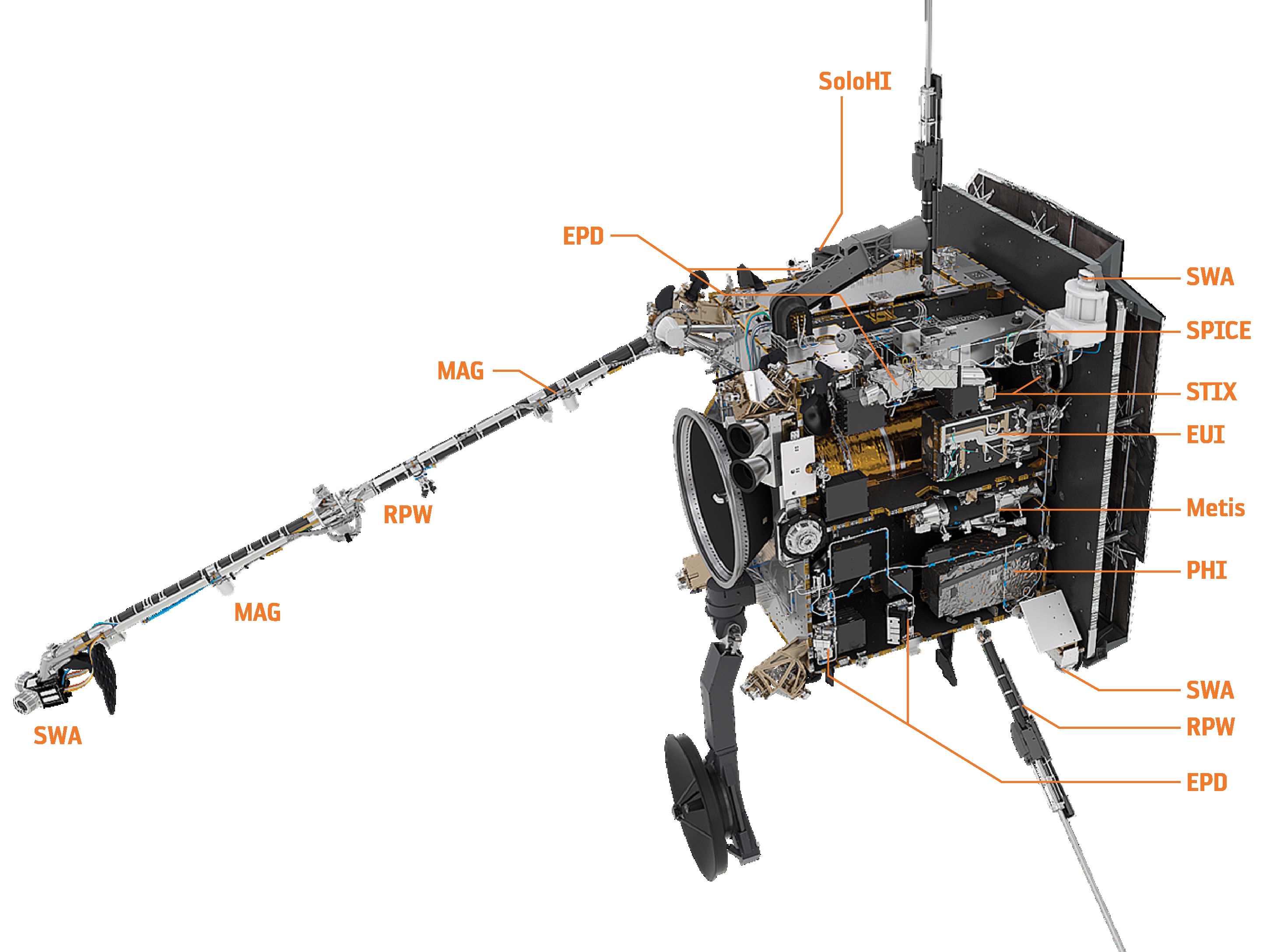}
}
\caption{Solar Orbiter payload. In this rendering, the $-Y$ side panel was removed to show the internally mounted instruments. The SWA sensor at the tip of the instrument boom is SWA/EAS, the one on the top corner of the heat shield is SWA/HIS, and the one on the bottom corner is SWA/PAS. The RPW sensor in the centre of the boom is RPW/SCM, and the lower parts of two out of three RPW antennas  (ANT) can be seen to extend radially away from the spacecraft body. The two-telescope unit visible on the outside of the removed  $-Y$ side panel is EPD/SIS. The remaining EPD sensors are STEP (close to the centre of the lower edge of the removed $-Y$ side panel) and the two EPT-HET sensors.}
\label{fig:so_sc_payload_annotated}
\end{figure*}

The scientific instruments of Solar Orbiter are provided by ESA member states, NASA, and ESA. Their accommodation on the spacecraft is shown in Fig.~\ref{fig:so_sc_payload_annotated}. Some instruments consist of several sensors and telescopes as described in the following section.

\subsection{The in-situ instruments}
\subsubsection{Energetic Particle Detector}
\label{subsect-EPD}
\begin{figure}[ht]
\includegraphics[width=\columnwidth]{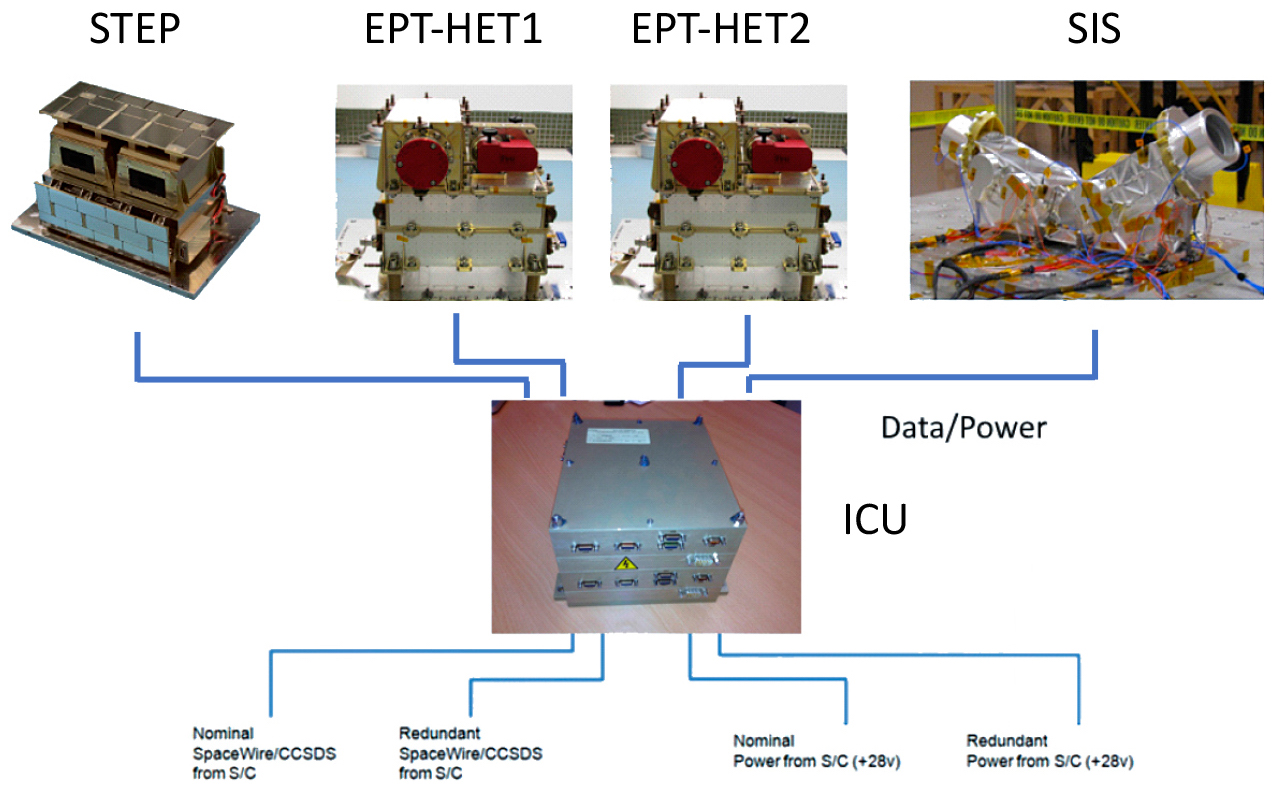}
\caption{The EPD instrument suite. The picture shows the different sensor units: The SupraThermal Electrons and Protons (STEP) sensor, the Electron Proton Telescopes (EPT), the High Energy Telescopes (HET), the Suprathermal Ion Spectrograph (SIS), and an Instrument Control Unit (ICU). The EPT and HET sensors are combined onto two almost identical sensor units: EPT-HET1 and EPT-HET2.}
\label{fig:epd}
\end{figure}

The Energetic Particle Detector (EPD; PI: J.~Rodriguez-Pacheco; \cite{Rodriguez2020a}, Fig.~\ref{fig:epd}) is an instrument suite comprising different sensors that measures the properties of suprathermal ions and energetic particles in the energy range of a few keV/n to relativistic electrons and high-energy ions. These measurements are performed over a partially overlapping energy range encompassing a few keV to 450 MeV/n (see Table~\ref{tab:EPD-sum}), with sufficient time, energy, angular, and mass resolution to achieve the mission's science goals. EPD consists of the following units: The SupraThermal Electrons and Protons (STEP) sensor, the Electron Proton Telescopes (EPT), the High Energy Telescopes (HETs), the Suprathermal Ion Spectrograph (SIS), and an Instrument Control Unit (ICU). The EPT and HET sensors are combined onto two almost identical sensor units: EPT-HET1 and EPT-HET2.

\begin{table*}
\centering
\begin{tabular}{lllllll}\hline\hline\\
& \multicolumn{1}{l}{Species} & \multicolumn{1}{l}{Energy range} & \multicolumn{1}{l}{\# of FoVs} & \multicolumn{1}{p{1.7cm}}{FoV size per aperture} & \multicolumn{1}{p{1.9cm}}{Geom. factor (cm$^2$ sr)} & \multicolumn{1}{p{1.5cm}}{Max. time resolution} \\
\hline
\\[-5pt]
STEP & e$^-$, ions & 2--80 keV & 1 (15 sectors) & $28^\circ \times 54^\circ$ & $8\cdot10^{-3}\tablefootmark{(a)}$ & 1 s \\
\\
EPT & e$^-$, H, He & 25--475 keV (e$^-$)& 4 (sunward,& $30^\circ$ & 0.01 & 1 s \\
 & & 25 keV--6.4 MeV (H) & anti-sunward, & & & \\
 & & 1.6--6.4 MeV/n (He) & north, south) & & & \\
\\
SIS & H, \element[][3]{He}, \element[][4]{He}, C, N, O,  & 14 keV/n--20.5 MeV/n & 2 (sunward, & $22^\circ$ & 0.2\tablefootmark{(b)} & 3 s\\
 & Ne, Mg, Si, S, Ca, Fe & & anti-sunward) & & & \\
\\
HET & e$^-$, H, \element[][3]{He}, \element[][4]{He}, C,  & 0.3--30 MeV (e$^-$) & 4 (sunward,& $43^\circ$ & 0.27\tablefootmark{(c)} & 1 s\\
 & N, O, Ne, Mg, Si,  & 6.8--107 MeV (H) & anti-sunward, &  & & \\
 & S, Ar, Ca, Fe, Ni & 8.1--41 MeV/n (\element[][3]{He}) & north, south) &  & & \\
 & & 6.9--105 MeV/n (\element[][4]{He}) & & \\
 & & 12--236 MeV/n (C, N, O) & & \\
 & & 16--360 MeV/n (Ne--S) & & \\
 & & 20--500 MeV/n (Ar--Ni) & & \\
\hline\hline 
\end{tabular}
\tablefoot{
    \tablefoottext{a}{can be reduced to $1.7\cdot10^{-4}$ cm$^2$ sr during high-intensity periods.}
    \tablefoottext{b}{can be reduced in several steps to a minimum of $0.002$ cm$^2$ sr using a variable aperture in front of the entrance foil.}
    \tablefoottext{c}{can be reduced to $0.01$ cm$^2$ sr during high-intensity periods.}
}
\caption{Summary of key EPD measurement capabilities.}
\label{tab:EPD-sum}
\end{table*}

\subsubsection{Magnetometer}
\label{subsect-MAG}
\begin{figure}
\includegraphics[width=\columnwidth]{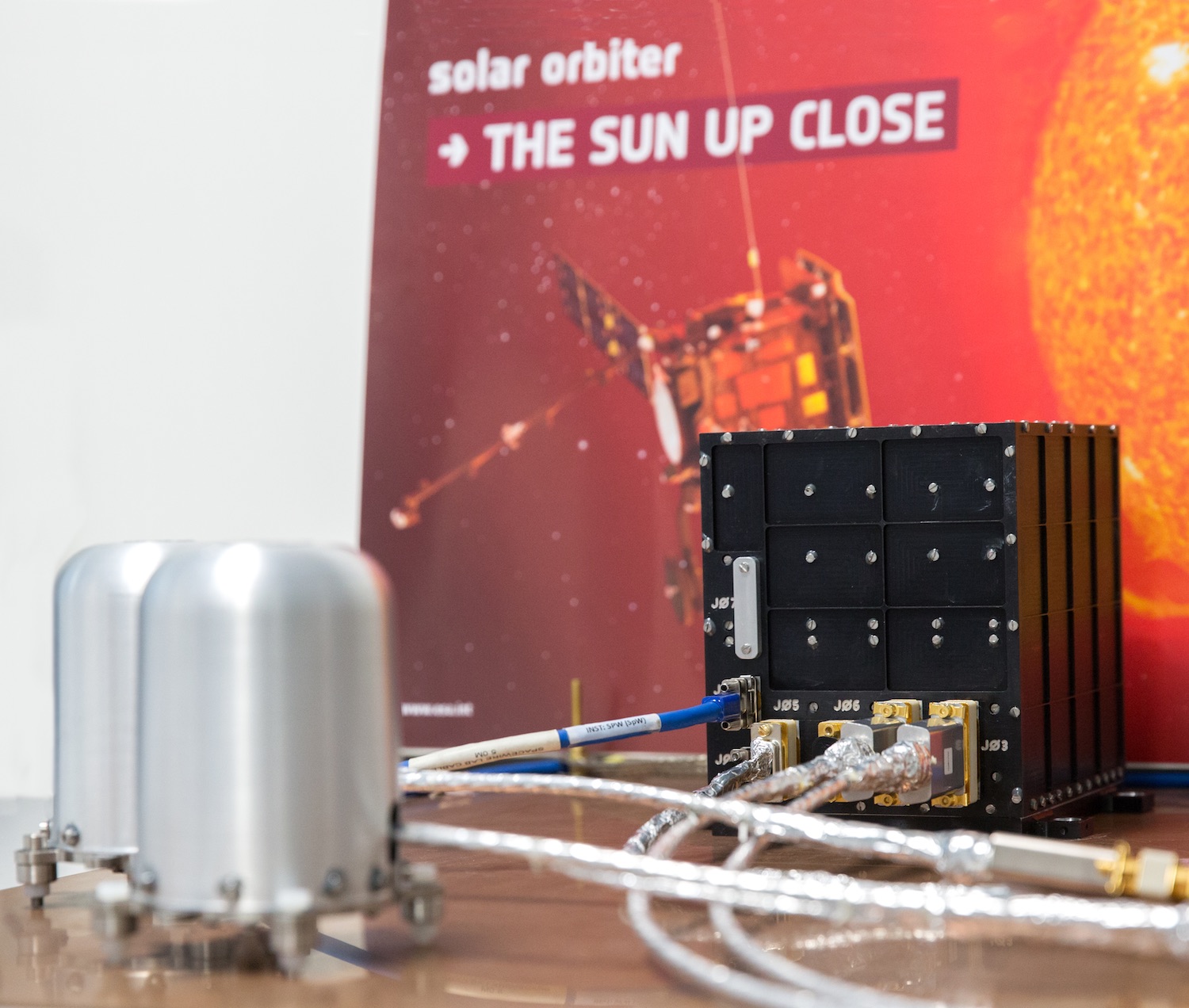}
\caption{The MAG instrument.}
\label{fig:mag}
\end{figure}

The magnetometer (MAG; PI: T.S.~Horbury; \cite{Horbury2020}, Fig.~\ref{fig:mag}) measures the \textit{in situ} magnetic field. With its dual-sensor fluxgate design, MAG operates continuously throughout the mission and records up to 16 vectors/s in its `normal' mode. A burst mode of 64 vectors/s -- and exceptionally even higher -- is recorded for an average of around an hour per day. With a precision of around 5\,pT, the magnetometer is sufficiently sensitive to record magnetic fluctuations from the largest scales associated with solar rotation, down to those below the proton gyroscale at tens of km. MAG contributes to all of the key science objectives of the mission and characterises the large-scale structure of the magnetic field in the inner heliosphere; the magnetic connectivity between the Sun and interplanetary space; dynamic plasma processes such as shock and reconnection; and the turbulence and waves that heat and accelerate the solar wind.

\subsubsection{Radio and Plasma Waves instrument}
\label{subsect-RPW}
\begin{figure}
\includegraphics[width=\columnwidth]{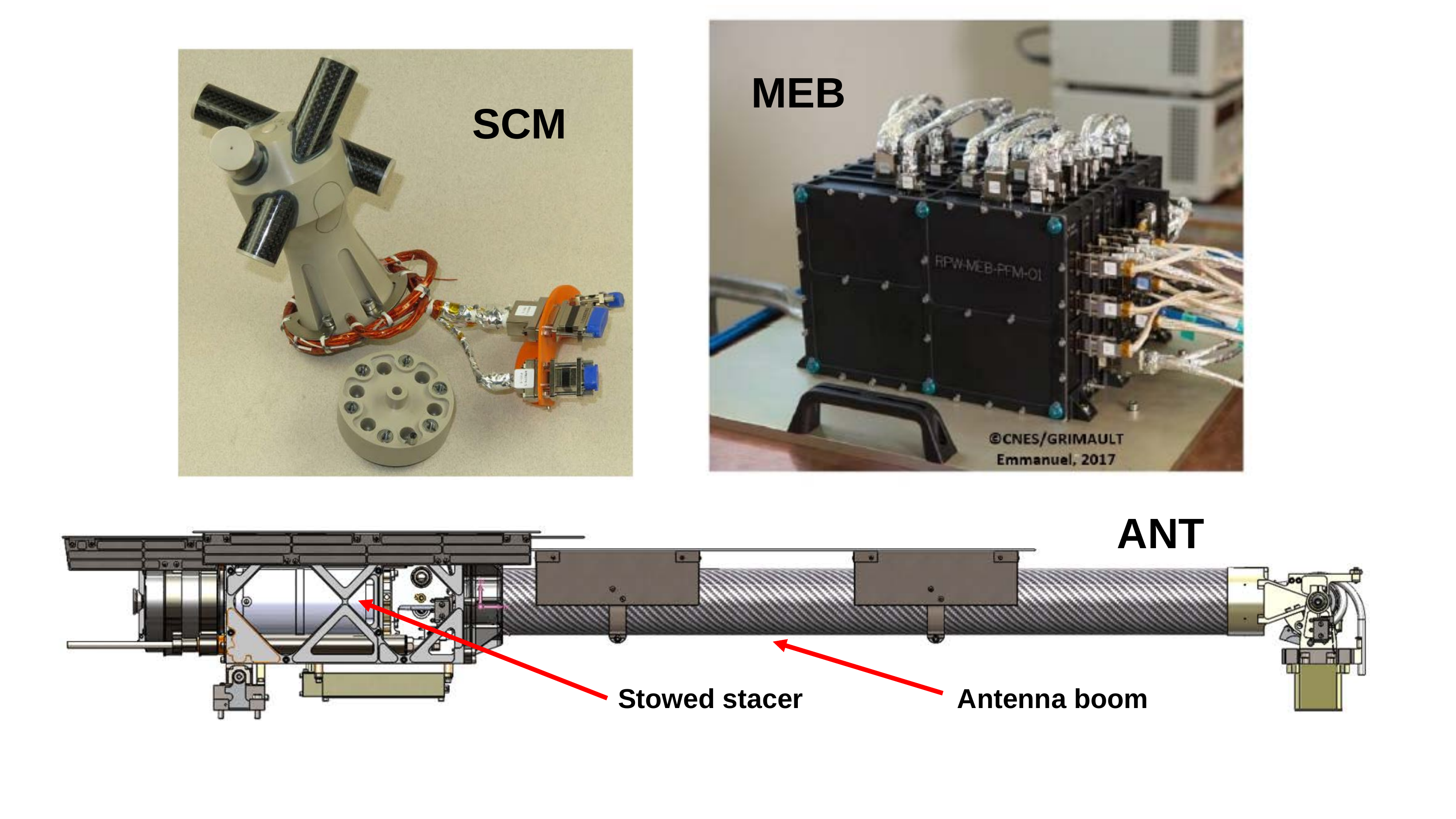}
\caption{The three main subsystems of the Radio and Plasma Waves instrument (RPW). The Search Coil Magnetometer (SCM) consists of a set of three magnetic antennas mounted orthogonally and located on the spacecraft's instrument boom. The Main Electronic Box (MEB), which collects all the signals coming from the ANT and SCM pre-amplifiers, is located inside the spacecraft. The electric Antenna system (ANT) consists of a set of three identical stacer antennas, each deployed from the tip of a rigid $\approx 1$\,m boom.}
\label{fig:rpw}
\end{figure}

The Radio and Plasma Waves instrument (RPW; PI: M.~Maksimovic; \cite{Maksimovic2020a}, Fig.~\ref{fig:rpw}) measures magnetic and electric fields, plasma wave spectra and polarisation properties, the spacecraft floating potential and radio emissions of solar origin generated in the interplanetary medium. It consists of three antenna units (ANTs) deployed in the plane perpendicular to the spacecraft--Sun direction, and a Search Coil Magnetometer (SCM) that is mounted on the spacecraft boom.

More specifically, RPW measures the three-component magnetic field fluctuations from about 10\,Hz to a few hundred kHz to fully characterise magnetised plasma waves in this range. Data from the three electric antennas are combined to retrieve the local plasma potential and to produce two components of the DC ambient electric field in the Solar Wind. RPW also observes solar radio emissions up to 16\,MHz and occasionally the associated Langmuir waves around the local plasma frequency. Finally, the instrument's radio receiver detects the local quasi-thermal noise providing accurate measurements of the in-situ absolute electron density and temperature when the ambient plasma Debye length is adequate. 

\subsubsection{Solar Wind Analyser}
\label{subsect-SWA}
\begin{figure}
\includegraphics[width=\columnwidth]{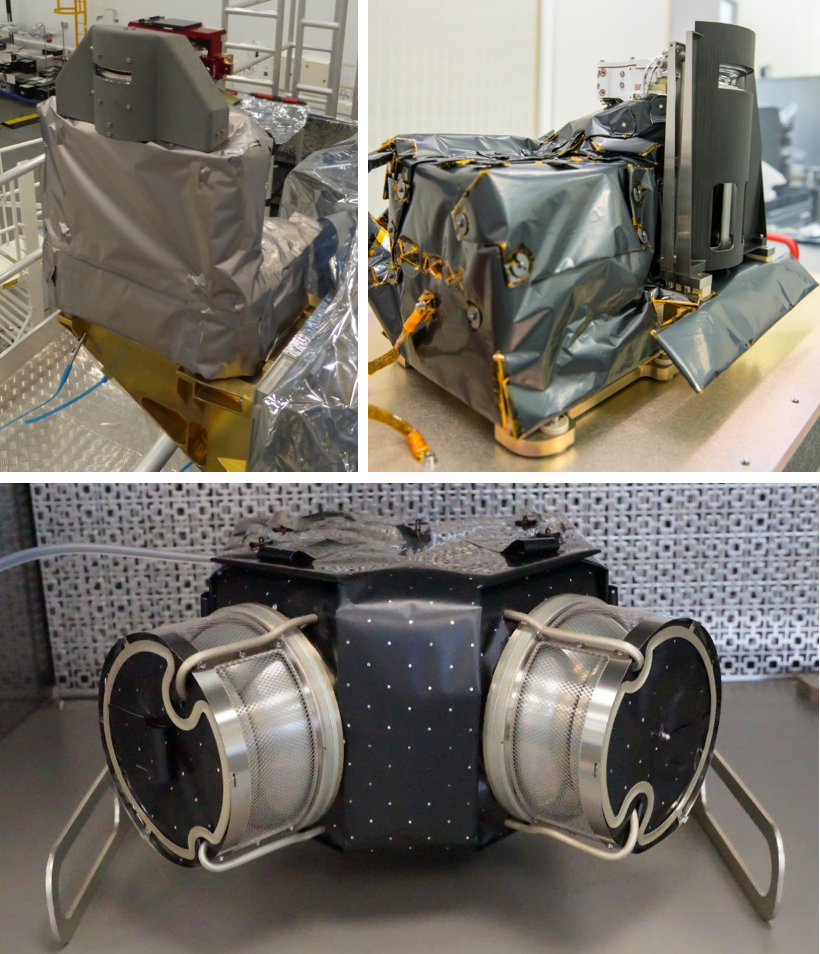}
\caption{SWA flight model sensors with their thermal blanketing.  Clockwise from top left are the Heavy Ion Sensor (SWA/HIS), the Proton and Alpha Particle Sensor (SWA/PAS), and the Electron Analyser System (SWA/EAS).}
\label{fig:swa}
\end{figure}

The Solar Wind Analyser instrument suite (SWA; PI: C.J.~Owen; \cite{Owen2020a}, Fig.~\ref{fig:swa}) consists of three sensors -- the Electron Analyser System (SWA/EAS), the Proton and Alpha Particle Sensor (SWA/PAS), the Heavy Ion Sensor (SWA/HIS) -- and the central Data Processing Unit (SWA/DPU). Between them, the sensors fully characterise the major constituents of the solar wind plasma between 0.28 and ~1\,AU.  SWA provides high-cadence measurements of 3D velocity distribution functions of solar wind electron, proton, and alpha particle populations, together with abundant heavy ions such as O$^{6+}$ and ion charge states such as Fe$^{9+}$ and Fe$^{10+}$. 
 
SWA/EAS is a dual-head, top-hat electrostatic analyser system that takes measurements of solar wind electrons at energies below ~5 keV.  Through the deployment of an aperture deflection system, each head can sample a field of view of $90^\circ \times 360^\circ$. Field-of-view blockage and the effect of spacecraft-related interference are minimised by mounting the dual-sensor unit at the end of the 4 m boom extending into the shadow of the spacecraft and its heatshield. 

SWA/PAS consists of a single electrostatic analyser head and electronics box mounted on a forward corner of the spacecraft, with a cut-out in the heat shield to allow ions arriving from the near-Sun direction to enter the sensor aperture. The SWA/PAS system deploys a set of aperture deflection plates, which steer ions into the sensor detection system while allowing sunlight to pass straight through the aperture and sensor structure and leave from the rear of the instrument without impinging on any part of the structure. Overall, the sensor is able to provide a full 3D sampling of the raw velocity distribution function of arriving ions with a cadence of once per second in normal mode, covering 32 energy bins ($0.2 - 20$\,keV/q) $\times$ 9 elevation bins ($\pm 22.5^\circ$ from the solar direction) $\times$ 11 azimuthal bins ($-24$ to $+42^\circ$). 

SWA/HIS is mounted on a second forward corner of the spacecraft, with a cut-out in the heat shield to allow ions arriving from the near-Sun direction to enter its aperture. The SWA/HIS system uses a similar electrostatic analyser/aperture deflection system design to that of the SWA/PAS described above. This allows heavy ions to be selected from a desired narrow energy-per-charge range within solar wind and suprathermal energy ranges using the time-of-flight method.
SWA/HIS electronics provide an analysis of the signals related to each incoming ion to determine its mass ($^3$He $- ^{56}$Fe), energy ($0.5 - 60$\,keV/q), charge state, and arrival direction and return results over a sampling period of 300\,seconds (30 seconds for helium) in the normal mode.

The SWA/DPU is the `heart' of the suite and is the primary SWA connection with the spacecraft, providing the data and command interfaces for the suite and the power input for SWA/EAS and SWA/PAS (SWA/HIS has a direct power connection).  

\subsection{The remote-sensing instruments}
The remote-sensing instruments, through a suite of telescopes, provide a global view of the Sun and the heliospheric environment. Some of these instruments are precisely co-aligned such that they can observe a particular area on the Sun with a common field of view of (17\,arcmin)$^2$ at very high resolution.

\subsubsection{Extreme Ultraviolet Imager}
\label{subsect-EUI}
\begin{figure}
\includegraphics[width=\columnwidth]{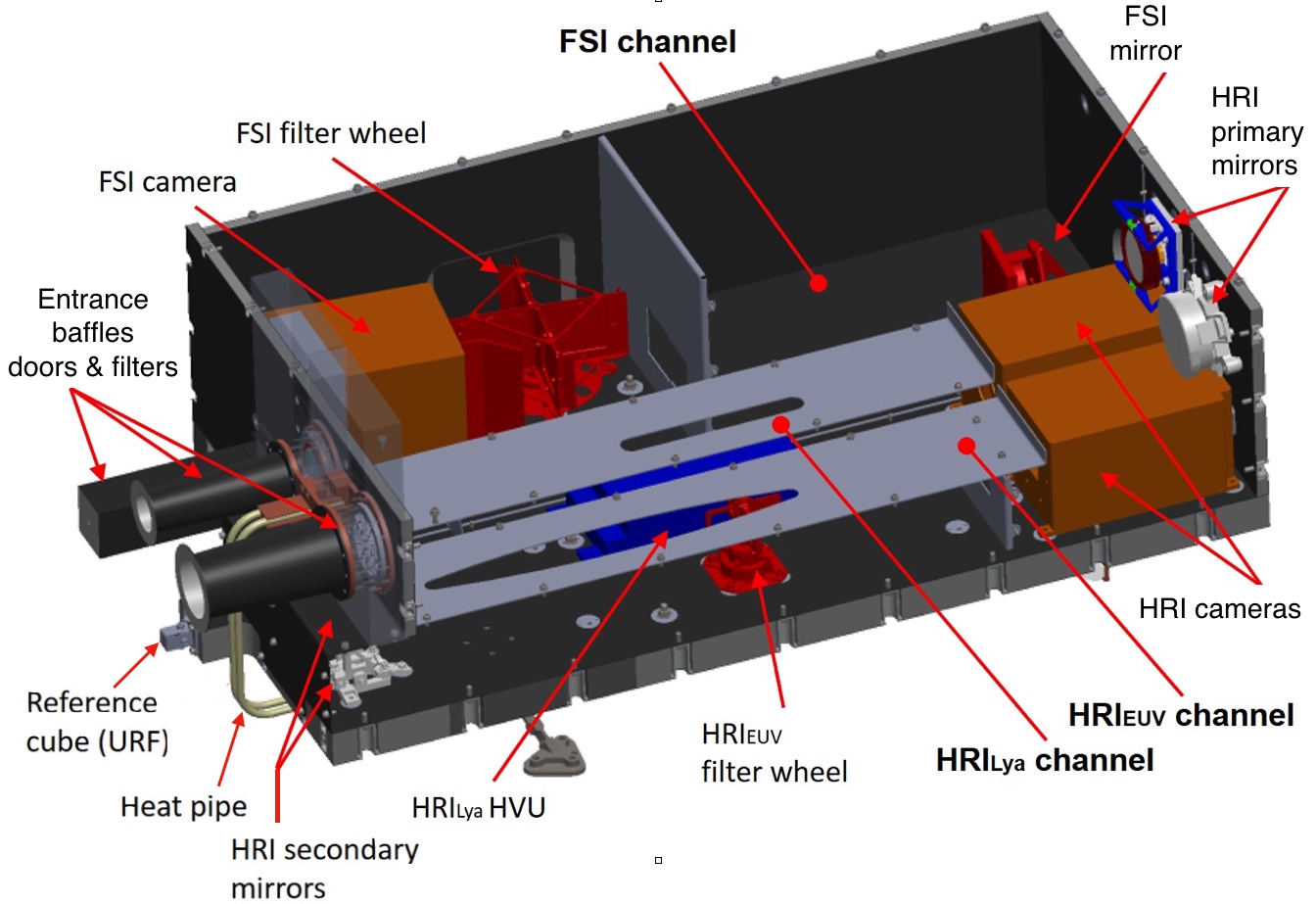}
\caption{Schematic of the EUI Optical Bench System showing its three telescopes: the Full Sun Imager (FSI), the High Resolution Imager in Lyman$-\alpha$ (HRI$_{LYA}$), and the High Resolution Imager at 17.4\,nm (HRI$_{EUV}$). \citep[From][]{Rochus2020a}}
\label{fig:eui}
\end{figure}

The Extreme Ultraviolet Imager (EUI; PI: P.~Rochus (development phase), D.~Berghmans (operations phase); \cite{Rochus2020a}, Fig.~\ref{fig:eui}) is a suite of three complementary telescopes that collectively provide image sequences of the solar atmospheric layers from the solar chromosphere into the corona. They image locally at very high resolution, as well as globally: The first two telescopes, the High Resolution Imagers (HRIs), observe features on the disc in a bandpass centred, respectively, on 17.4\,nm (HRI$_{EUV}$) and on the Lyman-$\alpha$ line (HRI$_{LYA}$) at 121.6\,nm with a pixel footprint on the Sun of down to (100 km)$^2$ and a temporal resolution of the order of a second.  The HRI fields of view of (17\,arcmin)$^2$ match that of the SO/PHI high-resolution telescope and the scanned field of view of the SPICE imaging spectrometer, and all of them are co-aligned for coordinated observations.

The Full Sun Imager (FSI) has an unprecedented field of view of 3.8$^\circ$ such that even with maximal Solar Orbiter off-pointing away from the disc centre, the full solar disc will always remain in the field of view. FSI has two bandpasses: the 17.4\,nm FSI bandpass corresponds to the 17.4\,nm bandpass of HRI$_{EUV}$ while the second FSI bandpass centred at 30.4\,nm shares the same resonance line for helium as the Lyman-$\alpha$ line of HRI$_{LYA}$ for hydrogen. FSI is designed to play an essential connection role as it images both the features studied by the high-resolution instruments on-disc, as well as the off-disc features imaged by Metis that extend further into the SoloHI field of view and are ultimately observed by the in-situ instruments.

\subsubsection{Visible light and UV coronagraph}
\label{subsect-Metis}
\begin{figure}
\includegraphics[width=\columnwidth]{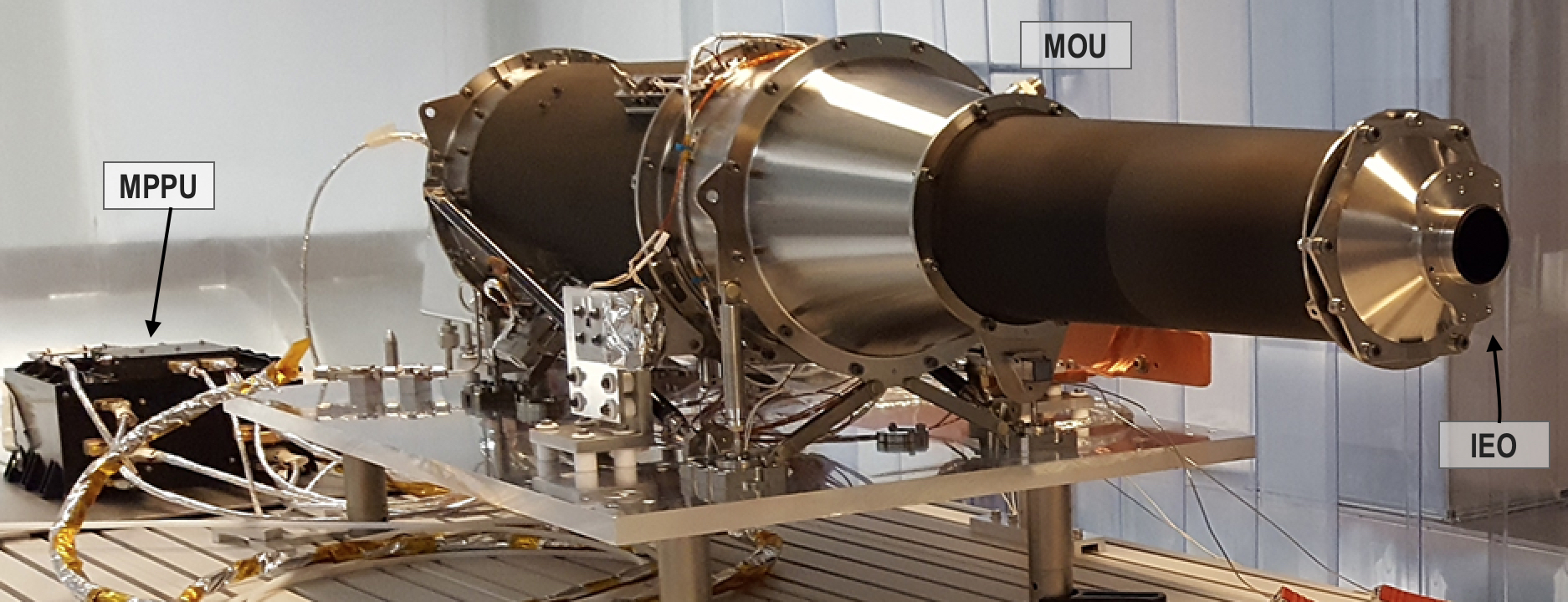}
\caption{Illustration of the Metis instrument in its flight configuration, consisting of the Metis optical unit (MOU), the camera power converter (CPC), and the Metis processing and power unit (MPPU). The high voltage unit (HVU) provides high voltage to the UV detector. \citep[From][]{Antonucci2020a}}
\label{fig:metis}
\end{figure}

Metis (PI: E. Antonucci (development phase), M. Romoli (operations phase); \cite{Antonucci2020a}, Fig.~\ref{fig:metis}) is an externally occulted coronagraph that performs broad-band and linearly polarised imaging of the corona in visible light ($580-640$\,nm bandpass), simultaneously with imaging of the UV corona in a narrow spectral range centred on the Lyman $\alpha$ line of hydrogen an 121.6\,nm. Simultaneous observations are achieved thanks to a Al/MgF$_{2}$ interference filter mounted before the focal plane of the telescope, thus reflecting the visible light to the polarimetric unit and the VL channel while selecting and transmitting Lyman $\alpha$ radiation to an intensified camera system. Metis will observe and diagnose, with unprecedented temporal and spatial resolution, the structures and dynamics of the inner corona in a square field of view of  $\pm2.9^\circ$ width, with the inner edge of the field of view starting at $1.6^\circ$, thus spanning the solar atmosphere from 1.7\,$R_{\odot}$ to about 9\,$R_{\odot}$ (varying with solar distance). The instrument will produce maps of the electron density distribution and of the outflow speed of protons on the plane of sky.

Metis will observe the hydrogen-proton and electron components of the solar wind with the aim of obtaining global maps of the outward velocity and density in the regions of the solar corona where the outflowing wind plasma is accelerated. It will investigate how the wind is channelled along the open coronal magnetic field in order to accurately establish, on an observational basis, the speed dependence on the non-radial areal divergence of the field fluxtubes. The coronagraph will trace CMEs out to 9 solar radii and measure, for the first time, their longitudinal distribution during the out-of-ecliptic phase of the mission. Metis will detect coronal fluctuations up to very high frequencies (image cadence down to 1\,s at fixed polarisation angle in the inner part of the instrument's field of view), taking advantage of the periods of reduced velocity of the spacecraft relative to the solar surface to distinguish between corotating coronal inhomogeneities and other phenomena such as turbulence and waves, which play an important role in the wind acceleration. 

\subsubsection{Polarimetric and Helioseismic Imager}
\label{subsect-PHI}

\begin{figure}
\includegraphics[width=\columnwidth]{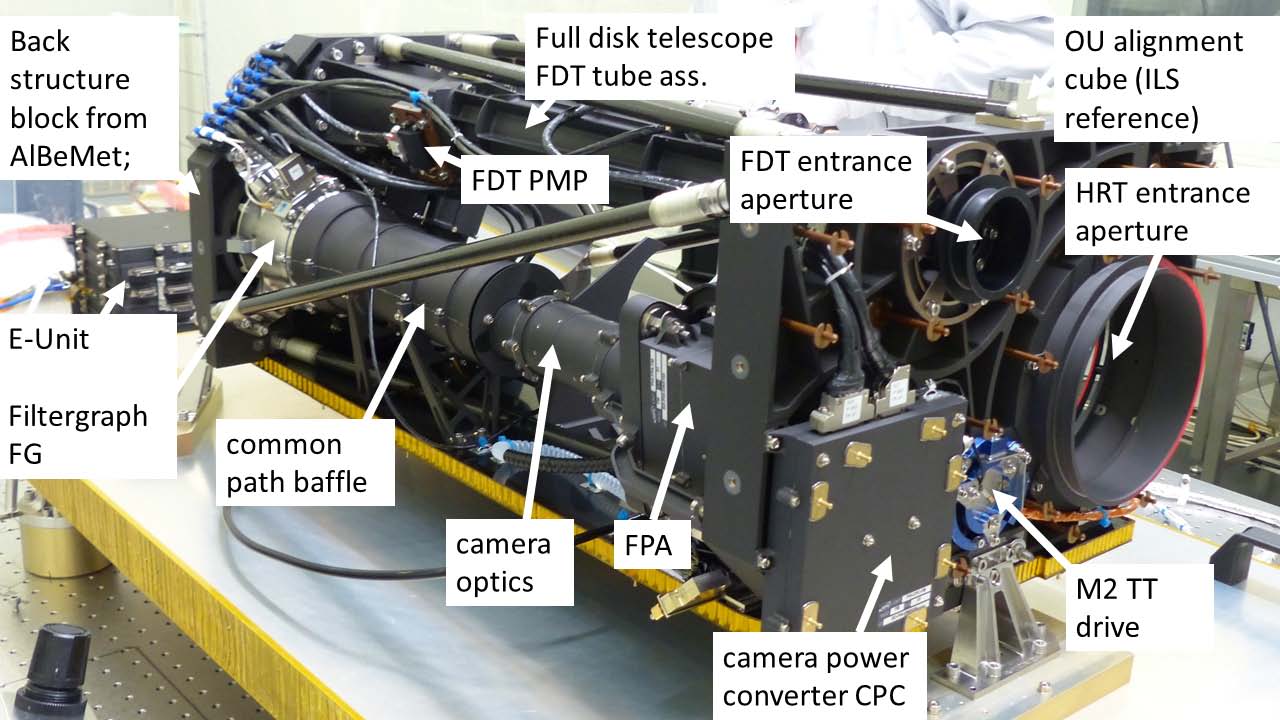}
\caption{Main subsystems of the SO/PHI optical unit \citep[From][]{Solanki2020a}}
\label{fig:phi}
\end{figure}

The Polarimetric and Helioseismic Imager (SO/PHI; PI: S.K.~Solanki; \cite{Solanki2020a}, Fig.~\ref{fig:phi}) employs two telescopes to provide high-resolution and full-disc maps of the photospheric vector magnetic field and line-of-sight velocity as well as of the continuum intensity.
Both telescopes sample the polarimetric properties of light within the strongly Zeeman-sensitive \ion{Fe}{I} line at 617.3\,nm. The spectral analysis is made with a tunable, solid LiNbO$_3$ Fabry-P\'erot etalon and the polarisation modulation is done with liquid crystal variable retarders. The quasi-monochromatic, polarimetric measurements are translated into the vector magnetic field and line-of-sight velocity by means of the Zeeman and Doppler effects.

A high-resolution telescope and a full-disc telescope feed the light into the instrument (with only one being operational at a given time). The high-resolution image is stabilised against jitter with the help of a correlation tracker and an active mirror. To reduce the data rate, the measured Stokes parameters are inverted onboard by solving the set of polarised radiative transfer differential equations and only the final physical parameters in the solar atmosphere are sent back to Earth. 
The onboard inversion is performed under the assumption of a Milne-Eddington atmosphere, although simpler reduction methods are also available \citep[for details, see][]{Solanki2020a}.

As mentioned earlier, the field of view of the SO/PHI high-resolution telescope of $17\times17$\,arcmin$^2$ matches those of the two EUI high-resolution imagers and the scanned field of view of the SPICE imaging spectrometer, and all of them are co-aligned for coordinated observations.

\subsubsection{Heliospheric Imager}

\label{subsect-SoloHI}
\begin{figure}
\includegraphics[width=\columnwidth]{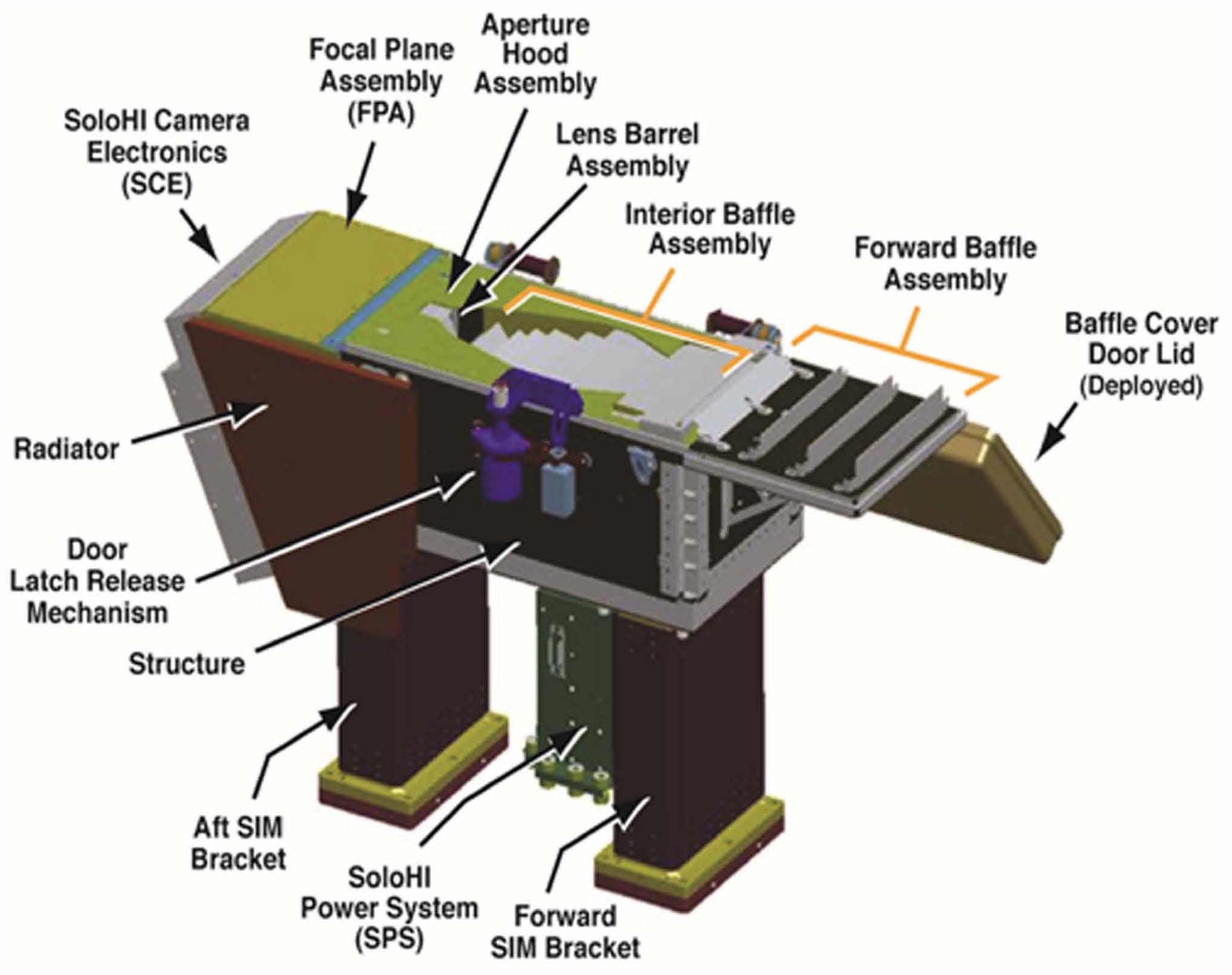}
\caption{The SoloHI instrument and power supply. \citep[From][]{Howard2020a}}
\label{fig:solohi}
\end{figure}

The Solar Orbiter Heliospheric Imager (SoloHI; PI: R.A.\ Howard; \cite{Howard2020a}, Fig.~\ref{fig:solohi}) images the inner heliosphere over a wide field of view by observing visible photospheric light scattered by electrons in the solar wind emitted by the Sun and interplanetary dust in orbit about the Sun. 
It is a single, white-light telescope of 20$^\circ$ half angle with the inner limit of the field of view at an elongation of 5$^\circ$ from Sun centre. 
As Solar Orbiter approaches the Sun, the spatial resolution will increase relative to the resolution at 1\,AU and the absolute field of view will decrease correspondingly. At perihelion, SoloHI will have the same effective resolution as the SOHO LASCO/C2 coronagraph, with a larger field of view (6-60\,$R_\sun$) than the LASCO/C3 coronagraph and a higher signal-to-noise ratio than the latter.

Baffles within the instrument combined with the edge of the heat shield reduce the light scattered by the solar disc to reveal the very faint light scattered by the solar wind electrons and the dust.  SoloHI will directly address the first three Solar Orbiter science objectives.  By observing the region between the Metis and the spacecraft, SoloHI will aid in determining the details of how the Sun and the spacecraft are connected.  Structures such as CMEs will be tracked and the interaction with the background solar wind will be observed.  As the spacecraft orbit moves out of the ecliptic plane, SoloHI will observe CMEs from a unique vantage point   above and below the ecliptic plane, enabling the observation of the effects of solar rotation on the longitudinal extent of CMEs and other structures.  SoloHI can record at a high cadence the intensities of small regions of the heliosphere which will enable us to determine the regions where density fluctuations are highest, indicating where wave energy is deposited and possibly where the corona is being heated.

\subsubsection{UV Imaging Spectrometer}
\label{subsect-SPICE}
\begin{figure}
\includegraphics[width=\columnwidth]{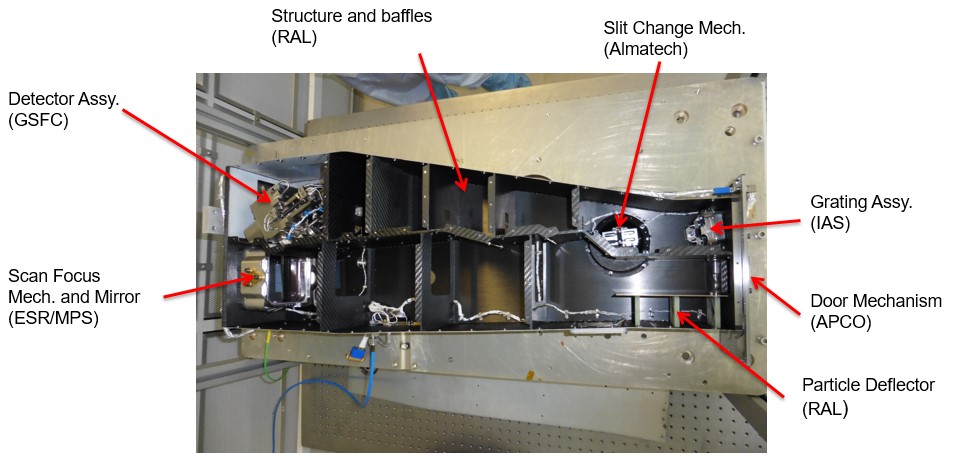}
\caption{Optics unit of the SPICE instrument. In this top view of the SPICE Optics Unit, its key components are identified along with the institutes and companies that provided them. \citep[From][]{SpiceConsortium2020}}
\label{fig:spice}
\end{figure}

The Spectral Imaging of the Coronal Environment (SPICE) instrument is a high-resolution spectral imager operating at extreme ultraviolet wavelengths (\cite{SpiceConsortium2020}, Fig.~\ref{fig:spice}).
Its wide wavelength range covers UV emission from a very large temperature regime of the solar atmosphere. By scanning the telescope mirror, it will cover a field of view that is commensurate with those of the other high-resolution instruments onboard.

SPICE will spectroscopically characterise regions at and near the Sun and provide quantitative information on the physical state and elemental composition of the observed plasma. In particular, SPICE will play a key role in investigating the source regions of outflows and ejection processes that link the solar surface and corona to the heliosphere. SPICE is of particular importance for establishing the link between remote-sensing and in-situ measurements as it is uniquely capable of remotely characterising the plasma
properties of source regions, which can be  directly compared with in-situ measurements taken by the SWA instrument suite.

\subsubsection{Spectrometer/Telescope for Imaging X-rays}

\label{subsect-STIX}
\begin{figure}
\includegraphics[width=\columnwidth]{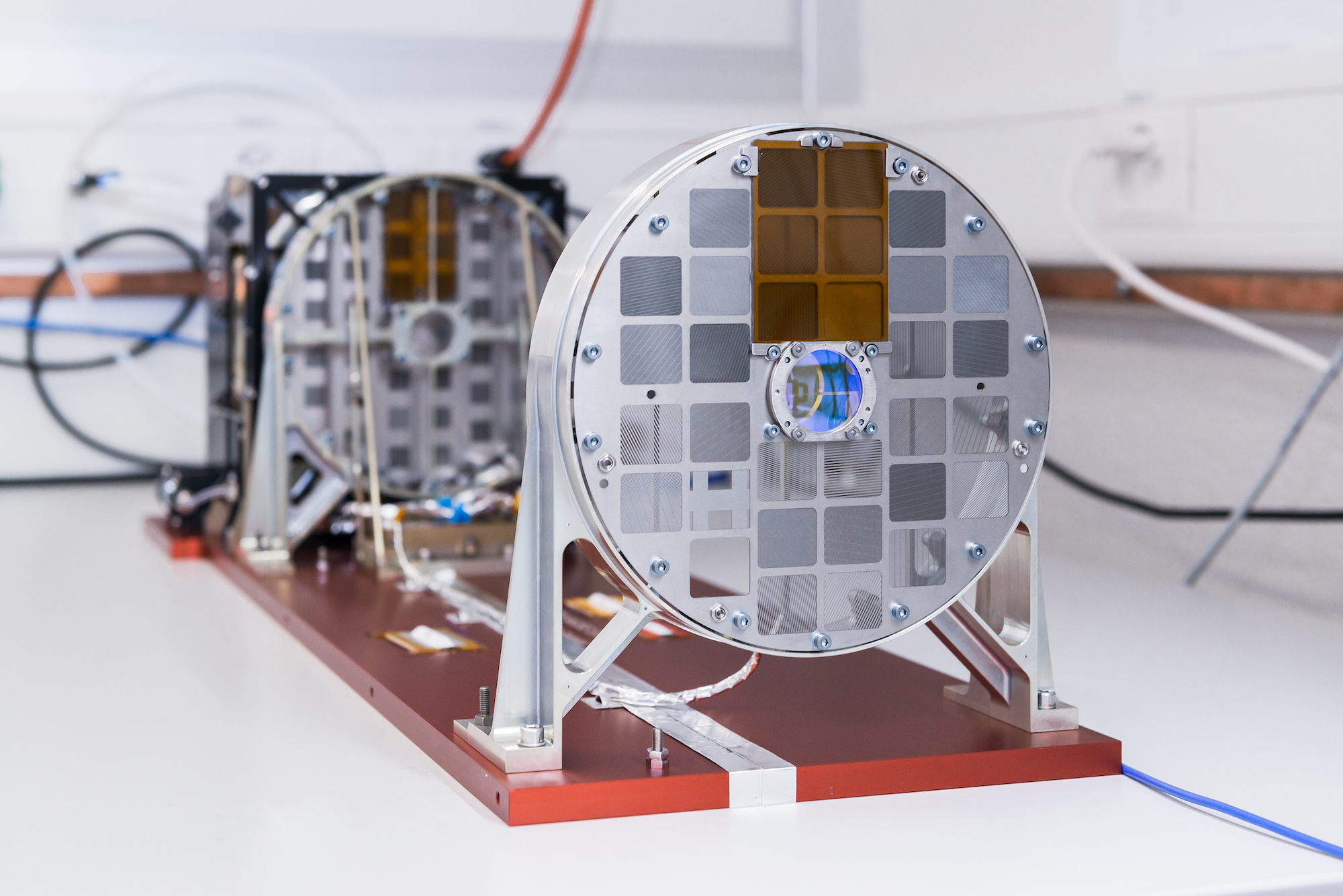}
\caption{Photograph of the STIX flight model taken from the Sun-facing side. The black box in the back is the Detector Electronics Module (DEM), which holds 32 CdTe detectors. In front of the DEM, the two grid support frames are seen, each holding 32 tungsten grids. A lens is mounted in the centre of the front grid as part of the STIX aspect system, which is used for absolute placement of the STIX hard X-ray images. The yellow rectangles are protective covers over the finest grids. }
\label{fig:stix}
\end{figure}

The Spectrometer/Telescope for Imaging X-rays  instrument (STIX; PI: S.~Krucker, Switzerland; \cite{Krucker2020a}, Fig.~\ref{fig:stix}) is a hard X-ray imaging spectrometer operating from $\sim4-150$\,keV with a spectral resolution of 1\,keV. STIX applies an indirect bi-grid Fourier imaging technique using a set of tungsten grids (at pitches from 0.038 to 1\,mm) in front of 32 coarsely pixelated CdTe detectors giving information on angular scales from 7 to 180\,arcseconds (for comparison, the spatial resolution of RHESSI reached down to 2.3\,arcseconds \citep{2002SoPh..210...61H}). With these diagnostics, STIX observations provide quantitative measurements of the hottest ($\gtrapprox$10\,MK) flare sources while quantifying the location, spectrum, and energy content of flare-accelerated non-thermal electrons.

\section{Mission design}
\label{sect-crema}
\begin{figure}
\includegraphics[width=\columnwidth]{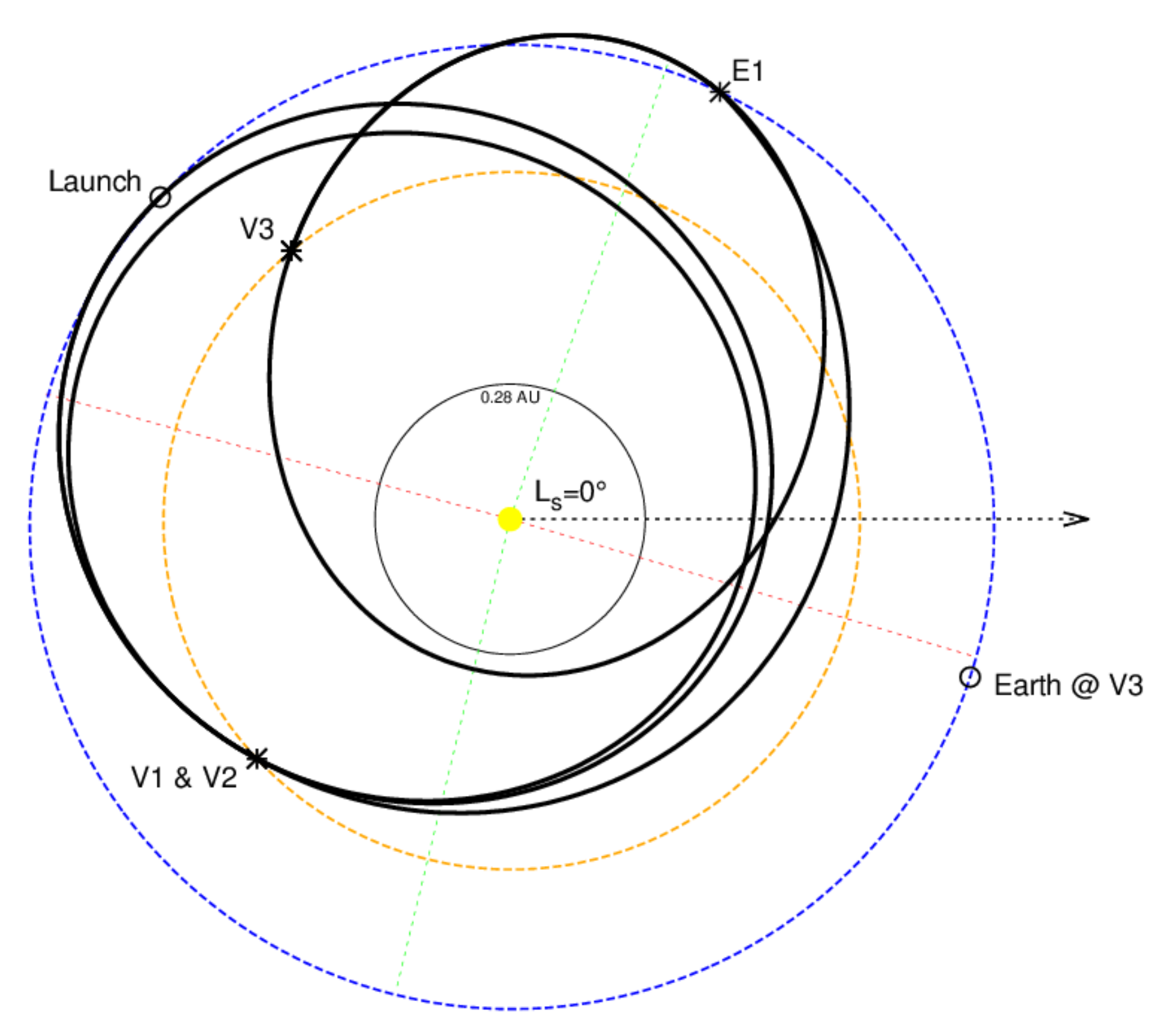}
\caption{Solar Orbiter's trajectory from launch in February 2020 to the start of the nominal mission phase after gravity assist manoeuvre (GAM) E1, viewed from above the ecliptic. The GAMs at Venus (V) and Earth (E) are indicated, along with the orbits of these two planets (Earth: blue, Venus: orange). The black inner circle indicates the minimal perihelion distance of 0.28 AU that Solar Orbiter will reach during its nominal mission phase.}
\label{fig:so_orbit_XY_plot}
\end{figure}

The spacecraft was launched on a ballistic trajectory that will be combined with planetary GAMs at Venus (V) and Earth (E). This trajectory is based on a short `EVVEV' cruise profile (Fig.~\ref{fig:so_orbit_XY_plot}), with the spacecraft having departed from Earth in February 2020, to be followed by two Venus GAMs, one Earth GAM, and another Venus GAM \citep{SO-CReMA5}.

After the near-Earth commissioning of the spacecraft and instruments, which was successfully concluded with the Mission Commissioning Results Review in June 2020, the cruise phase (CP) commenced. During cruise, the in-situ instruments are operating nominally, except for a reduced number of ground station passes compared to the nominal mission phase (NMP), while the remote-sensing instruments are only being operated during a pre-planned set of checkout windows. The first perihelion on 15 June 2020 took place at a distance of 0.51\,AU. After two Venus GAMs in December 2020 and August 2021, an Earth GAM in November 2021 will inject the spacecraft into a heliocentric science orbit with perihelion at 0.32 AU. This will mark the beginning of the NMP.

\begin{figure*}
\resizebox{\hsize}{!}{
\includegraphics{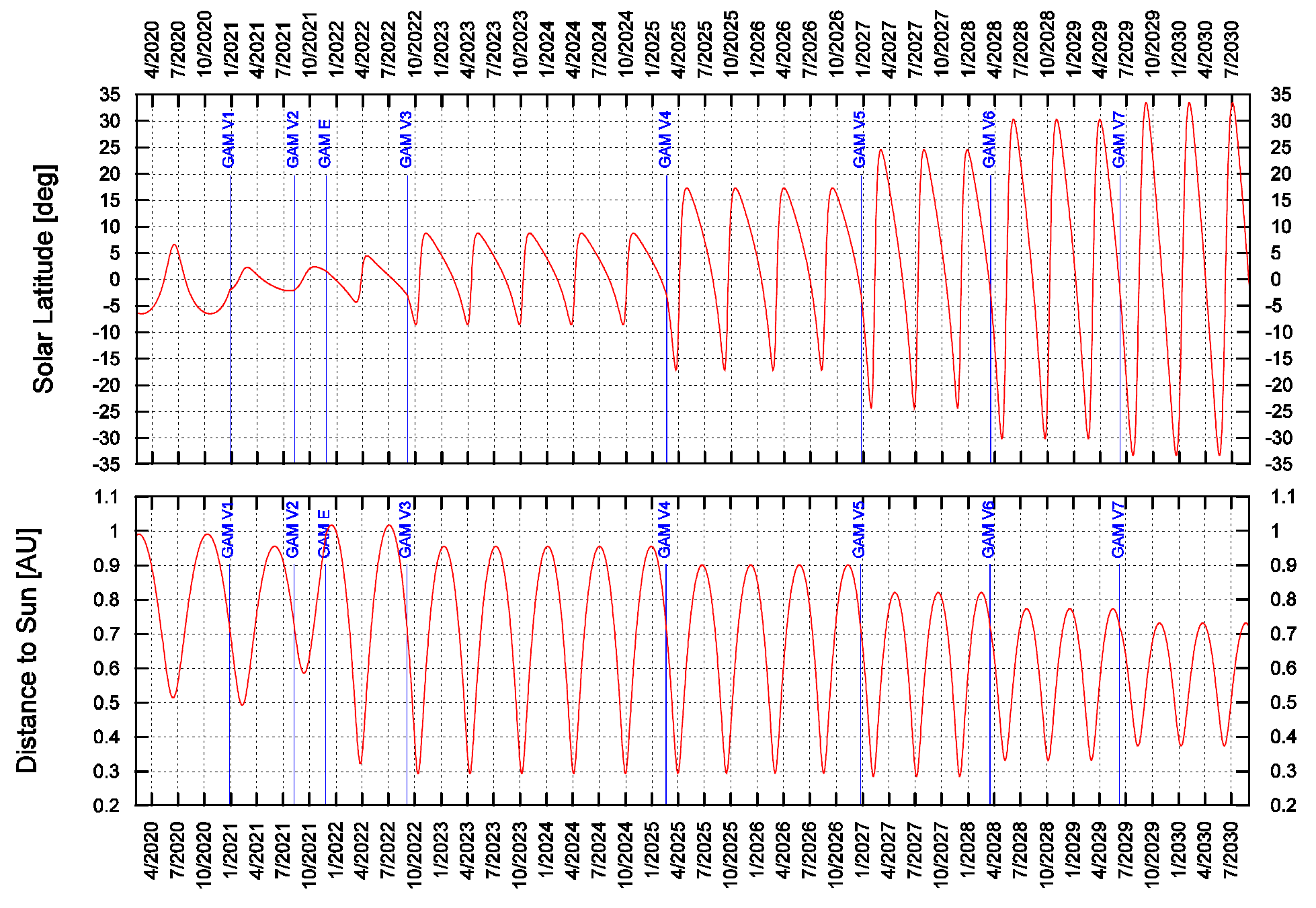}
}
\caption{
Solar Orbiter's trajectory. Heliocentric latitude (top) and distance (bottom) are plotted as a function of time. The blue vertical lines indicate the times at which gravity assist manoeuvres at Venus and Earth occur.}
\label{fig:so_orbit_dist_lat}
\end{figure*}

Venus will be encountered inbound for the third Venus GAM (V3) in September 2022, starting a sequence of resonance orbits of 5:4, 4:3, 3:2, 3:2, and 3:2 with respect to Venus' orbit. These ratios determine the different orbital periods of Solar Orbiter during the mission's lifetime, which are 180, 169, and 150\,days, respectively.
Figure~\ref{fig:so_orbit_dist_lat} displays heliocentric latitude and distance as a function of time. This sequence has been selected primarily for its excellent data downlink properties, which had to be traded off against reaching high solar inclinations more quickly.

\begin{itemize}
\item The 5:4 resonance is reached 2.57\,years after launch and reduces the perihelion to 0.292\,AU. The orbits start with an excellent phasing at an Earth--Sun--Venus angle of 148$^\circ$ which provides three aphelia gradually closer to Earth, with very high data downlink capability.
 \item The 4:3 resonance maintains perihelia at 0.294\,AU. It has a first aphelion close to inferior conjunction and provides excellent data downlink capability as well.
\item The first 3:2 resonance lowers the perihelion to 0.284 AU\,and raises the solar inclination for the first time above 20$^\circ$. This is considered the start of the extended mission phase (EMP). However, the downlink capability of this resonance is limited.
\item The second 3:2 resonance raises the perihelion to 0.331\,AU and the solar inclination for the first time above 30$^\circ$. The downlink capability is again excellent.
\item A last, Venus GAM (V7) is used for injecting the spacecraft again into a 3:2 resonance with an even larger solar inclination of 33.4$^\circ$. In parallel, this raises the perihelion distance to 0.373\,AU.
\end{itemize}

\begin{figure*}
\resizebox{\hsize}{!}{
\includegraphics{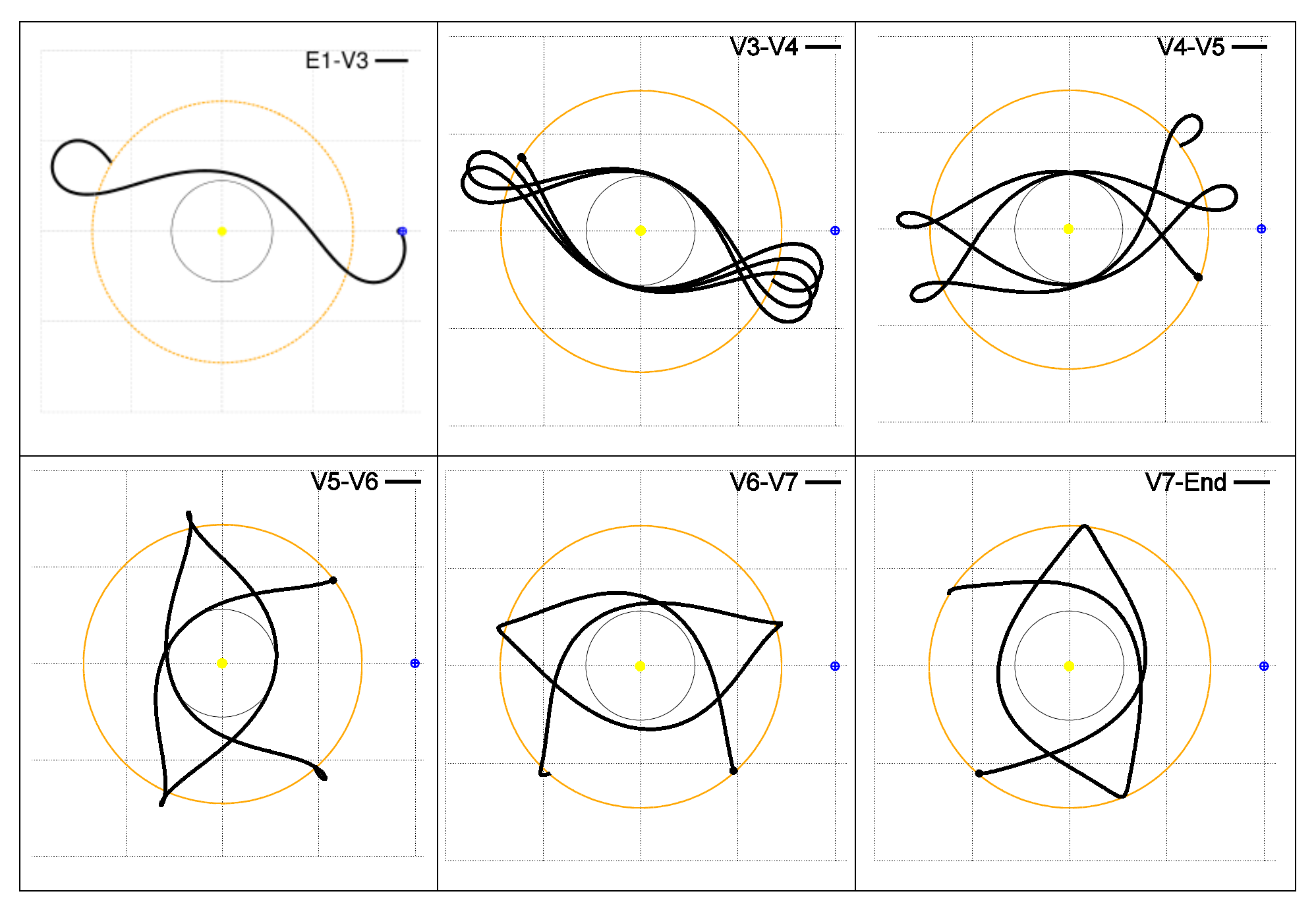}
}
\caption{Science orbits during the different parts of the nominal and extended mission phases, plotted in the Sun--Earth co-rotating frame (black lines). The Sun is at the centre of each plot (yellow dot), Earth (blue dot) is at centre-right. Venus' orbit is shown in orange. The data downlink increases quadratically with decreasing distance to Earth, which is why the V3--V4 and V4--V5 parts of the trajectory are ideal to downlink large data volumes.}
\label{fig:so_orbit_science_orbits_corot}
\end{figure*}

\noindent
Figure~\ref{fig:so_orbit_science_orbits_corot} illustrates these different parts of the mission's trajectory.

\section{Science operations}
\label{sect-sciops}
The science payload of Solar Orbiter comprises both remote-sensing and in-situ instruments. The in-situ instruments have started operating continuously since the beginning of cruise phase. As a baseline, the complete instrument suite will be operated during three ten-day windows (remote-sensing windows, or RSWs) during each orbit, nominally centred around closest approach, and at the minimum and maximum heliographic latitudes. This concept is driven by the overall constraints in telemetry. However, starting with the nominal operations phase, we envisage the operation of a subset of the remote-sensing instruments in synoptic modes throughout the orbit to improve the observations statistics of major solar eruptions and to further support collaborative science investigations with the in-situ instruments on Solar Orbiter, as well as with other missions, such as Parker Solar Probe.

\begin{figure*}
\resizebox{\hsize}{!}{
\includegraphics{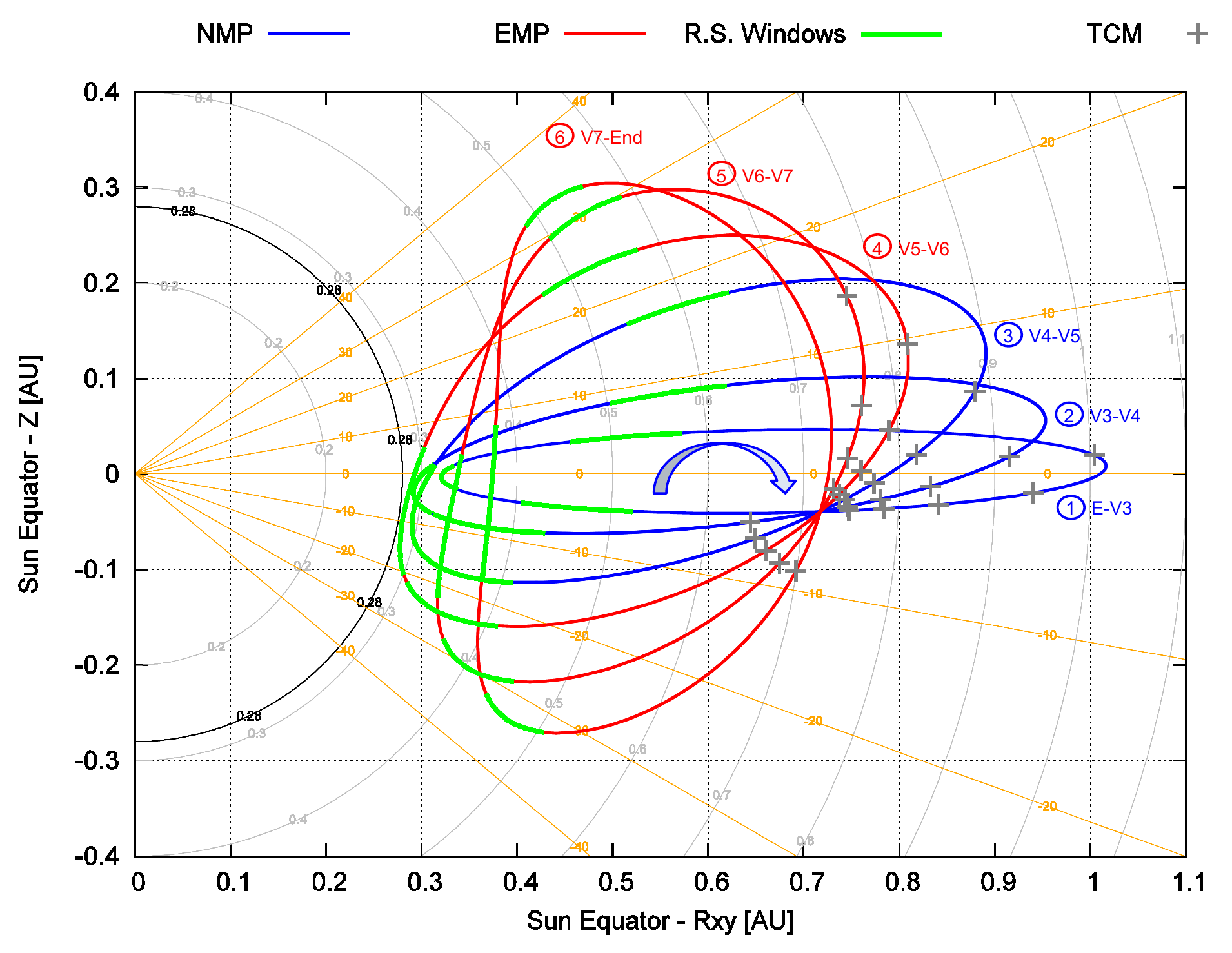}
}
\caption{Projection of science orbits. The spacecraft traces the orbits in clockwise direction. Vertical distance ($Z$) to the solar equatorial plane is plotted as a function of radial distance to the Sun in that ($X$--$Y$) plane. The blue line denotes the nominal mission (NMP), the red line the extended mission phase (EMP). For all orbits, the default remote-sensing windows are over-plotted in green. Lines of constant solar latitude are indicated in orange in increments of 10$^\circ$, and the minimum perihelion distance is plotted as a black semi-circle. The grey crosses indicate the trajectory correction manoeuvres used for the Venus GAMs.}
\label{fig:so_orbit_projection_science_orbits}
\end{figure*}

Figure~\ref{fig:so_orbit_projection_science_orbits} provides a Sun-centred representation of the six resonance orbits used during the NMP and EMP, including the second Earth-Venus leg. The plot shows vertical distance to the solar equatorial plane (projection onto $+Z$ axis, which points from the centre of the Sun in the direction of the Sun's north pole) as a function of radial distance to the Sun in that ($X$--$Y$) plane. In this figure, the spacecraft traces the orbits in a clockwise direction. Orange straight lines from the Sun represent points of constant solar latitude, while grey circles represent points with constant solar distance. The minimum perihelion constraint of 0.28\,AU is highlighted in black. This plot is useful for very quickly identifying the location of perihelion and the points of minimum and maximum solar latitude that will be used for the remote sensing windows. The grey crosses indicate the trajectory correction manoeuvres (TCMs) used for the Venus GAMs.

The mission's science operations approach is described in \cite{Sanchez2020}. Given that the mission's orbital characteristics will change over time, individual orbits will be dedicated to specific science questions. This is detailed in the mission-level science activity plan \citep[SAP,][]{Zouganelis2020a}. 

Coordinated observations will be key to the scientific success of the Solar Orbiter mission. Coordination among the in-situ instruments is described in more detail in \cite{Walsh2020}, and coordination of the remote-sensing instruments in \cite{Auchere2020a}. Coordination between Solar Orbiter, the Parker Solar Probe mission, and other space-and ground-based observatories is described in \cite{Velli2020a}.

                                                                                                                                                                                                                                                                                     
\section{Summary}

Understanding the coupling between the Sun and the heliosphere is of fundamental importance to understanding how the Solar System works and is driven by solar activity.
To address this and other fundamental questions of solar and heliospheric physics, Solar Orbiter will combine in-situ measurements as close as 0.28\,AU from the Sun with simultaneous high-resolution imaging and spectroscopic observations of the Sun. These will be acquired in and out of the ecliptic plane, and Solar Orbiter will be the first mission ever to make remote-sensing observations of the Sun's polar regions.
The combination of in-situ and remote-sensing instruments on the same spacecraft, together with the new, inner-heliospheric perspective, distinguishes Solar Orbiter from all previous and current missions. In addition to delivering ground-breaking science in its own right, Solar Orbiter has important synergies with NASA's Parker Solar Probe mission, as well as other space- and ground-based observatories. Coordinated observations from all of these perspectives will greatly contribute to  advancing our understanding of the Sun and its environment. 
\begin{acknowledgements}
Solar Orbiter is a space mission of international collaboration between ESA and NASA. The spacecraft has been developed by Airbus and is being operated by ESA from the European Space Operations Centre (ESOC) in Darmstadt, Germany. Science operations are carried out at ESA's European Space Astronomy Centre (ESAC) in Villafranca del Castillo, Spain.

Conceiving, designing and building Solar Orbiter has been an international team effort of many people. In particular, the authors would like to thank ESA's Mission Operations Centre (MOC) and Science Operations Centre (SOC) teams, Yves Langevin and Jose-Manuel S\'anchez P\'erez for their skillful optimisation of mission trajectories, the ESA and NASA Project offices, Airbus, IABG, NASA-LSP, ULA, and all national funding agencies that have enabled Solar Orbiter.
The German contribution to SO/PHI is funded by the Bundesministerium f\"ur Wirtschaft und Technologie through Deutsches Zentrum f\"ur Luft- und Raumfahrt e.V.\ (DLR), Grants No. 50 OT 1001/1201/1901 as well as 50 OT 0801/1003/1203/1703, and by the President of the Max Planck Society (MPG). The Spanish contribution has been partially funded by Ministerio de Ciencia, Innovaci/'on y Universidades through projects ESP2014-56169-C6 and ESP2016-77548-C5. IAA-CSIC acknowledges financial support from the Spanish Research Agency (AEI/MCIU) through the ``Center of Excellence Severo Ochoa" award for the Instituto de Astrof\'isica de Andaluc\'ia (SEV-2017-0709). The French contribution is funded by the Centre National d'Etudes Spatiales.
Further detailed acknowledgements regarding each instrument can be found in the individual instrument papers of this special issue.
R.A.H. is supported by the NASA Solar Orbiter Collaboration Office, under contract NNG09EK11I. 
The Spanish contribution to SO/PHI has been funded by the Spanish Ministry of Science and Innovation through several projects, the last one of which being RTI2018-096886-B-C5, and by ``Centro de Excelencia Severo Ochoa'' Programme under grant SEV-2017-0709.
The authors would like to highlight Rainer Schwenn's (1941 -- 2017) important and enthusiastic contribution to the Solar Orbiter mission in its early phase.
Portions of the text have been reproduced with permission from \cite{Mueller:2013a} copyright by Springer. The authors would like to thank John Leibacher and Bernhard Fleck for their support, and the referee for providing helpful suggestions.
\end{acknowledgements}

%

%
%

\bibliographystyle{aa}
\bibliography{aamnem99,loops,solarorbiter,fits_wcs,ral,SO_Book_references_DM,operations,SO_Book_cross_references,allzotero}

\end{document}